\documentclass[twocolumn]{aastex631}

\usepackage{bm}
\usepackage{array}
\usepackage{amsmath}

\newcommand{\pc}{\mbox{${\rm pc}$}}

\newcommand{\kms}{\mbox{${\rm km}~{\rm s}^{-1}$}}

\newcommand{\be}{\begin{equation}}
\newcommand{\ee}{\end{equation}}
\newcommand{\bea}{\begin{eqnarray}}
\newcommand{\eea}{\end{eqnarray}}
\newcommand{\dd}{\mathrm{d}}

\shorttitle{Star formation in quenched and main sequence environments}
\shortauthors{S.~M.~R.~Jeffreson et al.}

\begin{document}

\title{Learning the Universe: GalactISM simulations of resolved star formation and galactic outflows across main sequence and quenched galactic environments}

\author[0000-0002-4232-0200]{Sarah M. R. Jeffreson}
\affiliation{Center for Astrophysics, Harvard \& Smithsonian, 60 Garden Street, Cambridge MA, USA}

\author[0000-0002-0509-9113]{Eve C. Ostriker}
\affiliation{Department of Astrophysical Sciences, Princeton University, 4 Ivy Lane, Princeton, NJ 08544, USA}
\affiliation{Institute for Advanced Study, 1 Einstein Drive, Princeton, NJ 08540, USA}

\author[0000-0003-2896-3725]{Chang-Goo Kim}
\affiliation{Department of Astrophysical Sciences, Princeton University, 4 Ivy Lane, Princeton, NJ 08544, USA}

\author[0000-0001-6119-9883]{Jindra Gensior}
\affiliation{Department of Astrophysics, University of Zurich, Winterthurerstrasse 190, 8057 Zurich, Switzerland}

\author[0000-0003-2630-9228]{Greg L. Bryan}
\affiliation{Department of Astronomy, Columbia University, 550 W 120th Street, New York, NY 10027, USA}
\affiliation{Center for Computational Astrophysics, Flatiron Institute, 162 5th Ave, New York, NY 10010, USA}

\author[0000-0003-4932-9379]{Timothy A. Davis}
\affiliation{Cardiff Hub for Astrophysics Research \&\ Technology, School of Physics \&\ Astronomy, Cardiff University, Cardiff, CF24 3AA, UK}

\author[0000-0001-6950-1629]{Lars Hernquist}
\affiliation{Center for Astrophysics, Harvard \& Smithsonian, 60 Garden Street, Cambridge MA, USA}

\author[0000-0002-1050-7572]{Sultan Hassan}
\altaffiliation{NHFP Hubble fellow}
\affiliation{Center for Cosmology and Particle Physics, Department of Physics, New York University, 726 Broadway, New York, NY 10003, USA}
\affiliation{Center for Computational Astrophysics, Flatiron Institute, 162 5th Ave, New York, NY 10010, USA}
\affiliation{Department of Physics \& Astronomy, University of the Western Cape, Cape Town 7535,
South Africa}

\correspondingauthor{Sarah M. R. Jeffreson et al.}
\email{sarah.jeffreson@cfa.harvard.edu}

\begin{abstract}
We present a suite of six high-resolution chemo-dynamical simulations of isolated galaxies, spanning observed disk-dominated environments on the star-forming main sequence, as well as quenched, bulge-dominated environments. We compare and contrast the physics driving star formation and stellar feedback amongst the galaxies, with a view to modeling these processes in cosmological simulations. We find that the mass-loading of galactic outflows is coupled to the clustering of supernova explosions, which varies strongly with the rate of galactic rotation $\Omega = v_{\rm circ}/R$ via the Toomre length, leading to smoother gas disks in the bulge-dominated galaxies. This sets an equation of state in the star-forming gas that also varies strongly with $\Omega$, so that the bulge-dominated galaxies have higher mid-plane densities, lower velocity dispersions, and higher molecular gas fractions than their main sequence counterparts. The star formation rate in five out of six galaxies is independent of $\Omega$, and is consistent with regulation by the mid-plane gas pressure alone. In the sixth galaxy, which has the most centrally-concentrated bulge and thus the highest $\Omega$, we reproduce dynamical suppression of the star formation efficiency (SFE) in agreement with observations. This produces a transition away from pressure-regulated star formation.
\end{abstract}


\section{Introduction} \label{Sec::Introduction}
Since the first detections of cold gas in elliptical, early type galaxies (ETGs) at low redshift~\citep{1986A&A...164L..22W,1987ApJ...322L..73P}, the presence of star-forming gas in such galaxies has been shown to be relatively common. Molecular gas has been detected in at least 22\% of local ETGs~\citep{2003ApJ...584..260W,2007MNRAS.377.1795C,2011MNRAS.414..940Y,2019MNRAS.486.1404D}, and some of the most massive ETGs are found to have large molecular gas reservoirs between $10^9$ and $10^{11}$ solar masses~\citep[e.g.][]{2003A&A...412..657S,2016MNRAS.458.3134R,2018A&A...618A.126O,2019MNRAS.490.3025R}.

With the recent advent of high-sensitivity sub-millimeter interferometers, it has become possible to resolve these molecular gas reservoirs in great detail, and even to distinguish individual giant molecular clouds (GMCs) within them~\citep{Utomo2015,Liu2021,2023arXiv230805146W,2024MNRAS.subm}. Such studies demonstrate that the interstellar media of lenticular and elliptical galaxies display very different properties to their main sequence spiral galaxy counterparts, forming very smooth gas disks that more-closely resemble protoplanetary disks than they do galaxies~\citep{Davis2022}. A large fraction of these ETGs also display cold gas and molecular gas depletion times that are elevated by up to an order of magnitude, relative to the values measured in main sequence galaxies. Interestingly, these prolonged depletion times are not seen in all ETGs: the average increase in the cold gas depletion time across the population of observed ellipticals, relative to spiral galaxies, is just 2.5 times~\citep{2014MNRAS.444.3427D}.

This suppression of star formation in the cold gas of bulge-dominated galaxies is also a prominent feature in large galaxy surveys that directly detect atomic and molecular gas over a range of redshifts~\citep[e.g.][]{Saintonge2012,2018ApJ...853..179T,2020A&A...644A..97C}. Computing gas masses via the dust reddening of optical spectra in the SDSS sample, \cite{2022MNRAS.512.1052P} have shown a suppression of the molecular gas star formation efficiency (SFE) in quenched galaxies by two orders of magnitude, comparable to the reduction factor in their overall gas fractions. Across the EDGE-CALIFA survey~\citep{2020A&A...644A..97C}, it is found that the offset from the galactic star-forming main sequence for low gas-fraction galaxies is driven predominantly by a large drop in their SFEs per unit molecular gas mass, rather than by variation in the molecular gas fractions.

Such data indicate that the quenching of star formation occurs \textit{both} due to the removal of star-forming gas from galaxies, \textit{and} due to the quenching of star formation within the remaining cold gas. While feedback from Active Galactic Nuclei (AGN) has been shown to effectively eject gas from galaxies, and to halt the accretion of new gas by heating the surrounding intracluster medium~\citep[see][and references therein]{2012ARA&A..50..455F}, it has not been shown to suppress star formation within the remaining cold gas. A mechanism shown to produce the latter effect in numerical simulations is `dynamical suppression'~\citep{2009ApJ...707..250M,Martig2013,2020MNRAS.495..199G,2021MNRAS.500.2000G}, whereby stabilizing torques due to the rapid rate of galactic rotation in bulge-dominated environments prevents the collapse of cold gas to form new stars.

To correctly predict and therefore understand the pathways to the quenching of star formation throughout the course of galaxy evolution, it is therefore necessary to correctly model the physics driving star formation and stellar feedback in the cold interstellar media of both star-forming and quenched galaxies. Unfortunately, state-of-the-art hydrodynamical cosmological volume simulations lack the resolution to model this cold, star-forming ISM. Though substantial work has been done to model resolved star formation via cosmological zoom simulations~\citep[e.g.][]{Agertz13,2016MNRAS.460.3335C,2017MNRAS.467..179G,Hopkins18}, it is currently infeasible to extend such models to volumes containing many thousands of galaxies.

Currently, cosmological volume simulations adopt sub-grid treatments for star formation, stellar feedback and galactic winds that are typically calibrated to observed scalings in low-redshift, main sequence galaxies (in the case of star formation and stellar feedback) or that are tuned to reproduce key galaxy scaling relations (in the case of wind mass and energy loading, see~\citealt{2023arXiv230107116S} and citations therein). In particular, the depletion time is commonly calibrated to the relationship between the star formation rate and gas density in nearby spiral galaxies~\citep[e.g.][]{Springel03}: the same relationship that is shifted systematically for the interstellar media of ETGs.

Perhaps as a result of these highly-simplifed sub-grid models, hydrodynamical cosmological simulations are currently unable to accurately model the onset of star formation quenching as a function of stellar mass $M_*$, black hole mass $M_{\rm BH}$ and halo mass $M_{\rm Halo}$~\citep{2018MNRAS.475..624N,2022MNRAS.512.1052P}. Of the Illustris~\citep{2014Natur.509..177V}, Illustris-TNG~\citep{2018MNRAS.475..624N} and EAGLE~\citep{2015MNRAS.446..521S} simulations,~\cite{2022MNRAS.512.1052P} find that Illustris-TNG displays the best qualitative agreement with trends in the SDSS survey at low redshift, but over-estimates the mass at which quenching sets in, by three times in $M_*$, and around ten times in $M_{\rm Halo}$ and $M_{\rm BH}$. The discrepancy in the quenched fraction of galaxies with $M_*$ is also reproduced by~\cite{2018MNRAS.475..624N,2019MNRAS.485.4817D}, though we note that the comparison between simulations and observations may be affected by sample selection, the choice of star formation rate indicator, and a host of other complicating factors~\citep{2021MNRAS.506.4760D}.

\begin{figure}
  \includegraphics[width=\linewidth]{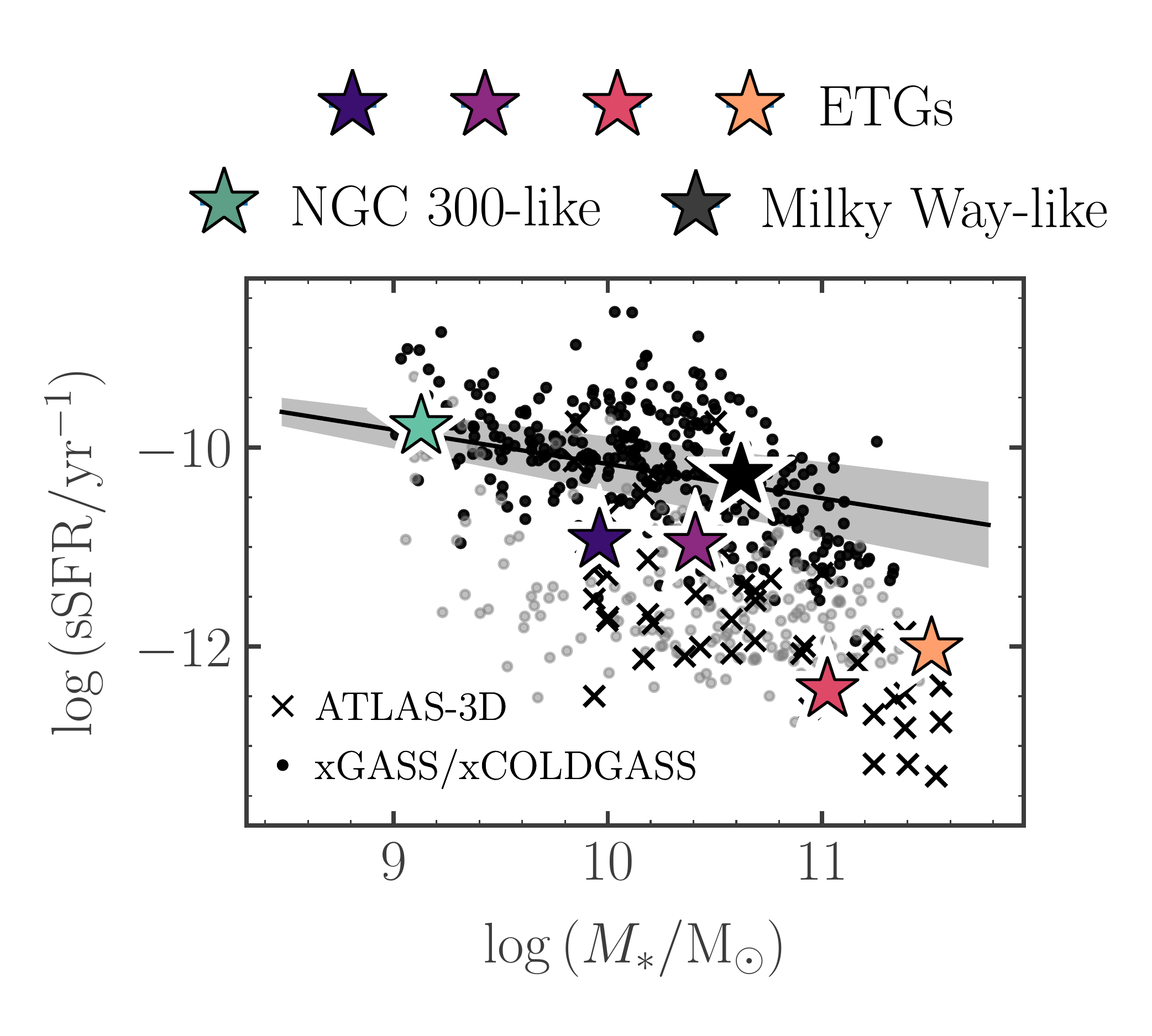}
  \caption{Each of the six isolated galaxy simulations (star symbols) in the plane of total stellar mass $M_*$ vs. specific star formation rate ${\rm sSFR}$. Black data points represent atomic and molecular gas detections, respectively, from the xGASS~\protect\citep{2018MNRAS.476..875C} and xCOLDGASS~\protect\citep{2017ApJS..233...22S} surveys, while grey data points represent non-detections. Black crosses represent data from the ATLAS$^{\rm 3D}$ survey~\citep{2011MNRAS.413..813C}. The galactic star-forming main sequence as defined in the~\protect\citep{2018MNRAS.476..875C} in is given by the black solid line and shaded region.}
  \label{Fig::gal-main-sequence}
\end{figure}

One of the key goals of the Learning the Universe Simons Collaboration (and one of its predecessors, the SMAUG collaboration) is to substitute the existing, empirically-calibrated or tuned sub-grid prescriptions in cosmological simulations with models that are calibrated based on higher resolution simulations, which capture the relevant physics on smaller scales. The collaboration therefore aims to produce cosmological simulations that no longer require empirical calibration or tuning, allowing for the predictive modeling of star formation quenching, among other physics. In this paper, we introduce the first six of the ``GalactISM'' simulations: a suite of high-resolution chemo-dynamical isolated galaxy simulations spanning observed, dynamically-diverse star-forming environments from the galactic star-forming main sequence of spiral galaxies to the population of fast-rotating quenched\footnote{We use the term `quenched' to denote galaxies with a specific SFR $\leq 1 \times 10^{-11} {\rm yr}^{-1}$.} ETGs at low redshift. We examine the star formation rate, star formation efficiency and the properties of galactic outflows across these environments, as a function of large-scale galaxy properties such as the mid-plane pressure, density and rotation rate. We therefore determine which sub-grid processes can be modeled by physically-motivated analytic theory in the form of power-law relationships, and where complex or non-linear variations arise, that might in the future be accounted for by statistical or learned modeling techniques---another facet of the Learning the Universe collaboration.

The GalactISM simulations are complementary to the ``TIGRESS'' \citep{KimCG&Ostriker17,2020ApJ...900...61K} and ``TIGRESS-NCR'' \citep{2023ApJS..264...10K,2023ApJ...946....3K} frameworks---magneto-hydrodynamic (MHD) simulations using {\sc Athena}~\citep{2008ApJS..178..137S,2009NewA...14..139S}, with star formation and supernova+radiation feedback that have the same ($2-4$pc) resolution in all ISM phases, including the low-density, hot gas.
\footnote{This higher resolution in the hot gas phase makes it possible to follow the Sedov-Taylor (energy-conserving, momentum-generating) stage of supernova remnant (SNR) evolution, in which hot gas is created in shocks, and the resolved interaction of expanding SNRs with the warm/cold gas phases, to drive turbulence on a range of scales.} The latter allows for full UV radiative transfer via adaptive ray-tracing, as well as photochemistry \citep{2023ApJS..264...10K}. By contrast, the GalactISM simulations employ adaptive refinement in the moving-mesh code {\sc Arepo}, and use the mechanical supernova and HII region feedback prescriptions based on momentum injection introduced in~\cite{2021MNRAS.505.3470J}, appropriate when it is not possible to fully resolve the Sedov-Taylor blast waves and Str\"{o}mgren spheres. At the lower resolution of GalactISM (corresponding to $\sim 1-10$~pc in the molecular gas and $\sim 30-60$~pc in the warm gas) we can model entire galaxies and the HCO chemistry of their three-phase, star-forming gas reservoirs relatively efficiently, allowing for the potential influence of inward radial mass transport~\citep[e.g.][]{2010ApJ...724..895K,Goldbaum15,Krumholz18b}, and for the later inclusion of a circumgalactic medium, necessary for the modeling of high gas-fraction, high-redshift galactic environments. We also produce a statistical sample of around $60,000$ GMCs across these six simulations.

The remainder of the paper is structured as follows. In Section~\ref{Sec::sims} we introduce the GalactISM simulation suite, along with the numerical models used for star formation, stellar feedback, chemistry and cooling. In Section~\ref{Sec::galaxy-props} we give an overview of the dynamical properties, gas phase distribution and morphology, and star-forming behavior of our galaxies, in comparison to observed ETGs from the ATLAS$^{\rm 3D}$ survey~\citep{2011MNRAS.413..813C}. Section~\ref{Sec::stability-and-clustering} provides a systematic analysis of the properties of stellar feedback-driven galactic outflows in our simulation and their dependence on the level of supernova clustering and gravitational stability in the simulated gas disks. The scaling of the star formation rate surface density with the gas surface density, the mid-plane pressure and the ISM weight are investigated in Section~\ref{Sec::SF-regulation}, along with the equation of state between the gas density and pressure. Finally, we conclude with a discussion and summary of our results in Section~\ref{Sec::conclusion}.

\section{Simulation suite} \label{Sec::sims}
The six chemo-dynamical isolated galaxy simulations presented in this work consist of one large spiral (Milky Way-like) galaxy, one dwarf spiral (NGC~300-like) galaxy, and four early type (ETG) galaxies. The physical properties of the ETG simulations are matched to the observations of elliptical galaxies from the ATLAS$^{\rm 3D}$ and MASSIVE surveys. Together, the simulated galaxies span over two orders of magnitude in total stellar mass and specific star formation rate (see Figure~\ref{Fig::gal-main-sequence}), from the galactic star-forming main sequence (black line) down to the quenched galaxy population below.

\begin{table*} \label{Tab::Params}
\begin{centering}
\vspace{0.5cm}
  \begin{tabular}{c c c c c c c c c}
  \hline
   \textbf{Property} & \textbf{Symbol} &  \multicolumn{4}{c}{\textbf{ETGs}}  & \textbf{Milky Way} & \textbf{NGC~300} \\
    \hline
    Analysis start/Myr & $t_{\rm start}$ & 100 & 100 & 100 & 100 & 300 & 500 \\
    Analysis end/Myr & $t_{\rm end}$ & 400 & 400 & 400 & 400 & 600 & 800 \\
    \hline
    Total stellar mass & $M_{\rm *}/{\rm M}_\odot$ & $10^{10}$ & $10^{10.5}$ & $10^{11}$ & $10^{11.5}$ & $4.734 \times 10^{10}$ & $1 \times 10^{9}$ \\
    \hline
    Stellar disk mass & $M_{\rm *, d}/{\rm M}_\odot$ & $2 \times 10^9$ & $6.3 \times 10^9$ & $2 \times 10^{10}$ & $6.3 \times 10^{10}$ & $4.297 \times 10^{10}$ & $1 \times 10^{9}$ \\
    Stellar bulge mass & $M_{\rm *, b}/{\rm M}_\odot$ & $8 \times 10^9$ & $2.5 \times 10^{10}$ & $8 \times 10^{10}$ & $2.5 \times 10^{11}$ & $3.437 \times 10^9$ & $0$ \\
    Gas fraction (0 Myr) & $M_{\rm gas}/M_{\rm *}$ & 0.016 & 0.016 & 0.0016 & 0.0016 & 0.18 & 0.68 \\
    Gas fraction ($t_{\rm start}$) & $M_{\rm gas}/M_{\rm *}$ & 0.012 & 0.011 & 0.0015 & 0.0012 & 0.12 & 0.57 \\
    Bulge-to-disk ratio & $M_{\rm *, b}/M_{\rm *, d}$ & 4 & 4 & 4 & 4 & 0.125 & 0 \\
    \hline
    Concentration parameter & $c$ & $8.6$ & $7.4$ & $6.7$ & 6.4 & 10 & 15.4 \\
    Virial velocity & $V_{200}/{\rm km~s}^{-1}$ & 130 & 200 & 280 & 370 & 150 & 63 \\
    Spin parameter & $\lambda$ & 0.04 & 0.04 & 0.04 & 0.04 & 0.04 & 0.04 \\
    \hline
    Stellar disk scale radius & $R_{\rm *, d}/{\rm kpc}$ & 3.96 & 6.67 & 9.64 & 13.5 & 3.43 & 1.39 \\
    Stellar bulge scale radius & $R_{\rm *, b}/{\rm kpc}$ & 1.35 & 1.85 & 2.8 & 8.75 & 0.34 & - \\
    Gas disk scale radius & $R_{\rm gas}/{\rm kpc}$ & 0.30 & 0.47 & 0.37 & 0.62 & 3.43 & 3.44 \\
    \hline
    Stellar disk scale height & $h_{\rm *, d}/{\rm pc}$ & 50 & 49 & 47 & 45 & 123 & 90 \\
    Gas disk scale height & $h_{\rm gas}/{\rm pc}$ & 30 & 25 & 20 & 30 & 82 & 110 \\
    \hline
    Gas cell resolution & $\epsilon_{\rm gas}/{\rm M}_\odot$ & $859$ & $859$ & $859$ & $859$ & $859$ & $859$ \\
    Stellar particle resolution & $\epsilon_*/{\rm M}_\odot$ & $5\times 10^3$ & $5\times 10^3$ & $5\times 10^3$ & $5\times 10^3$ & $5\times 10^3$ & $5\times 10^3$ \\
    Dark matter particle resolution & $\epsilon_{\rm halo}/{\rm M}_\odot$ & $1.25\times 10^5$ & $1.25\times 10^5$ & $1.25\times 10^5$ & $1.25\times 10^5$ & $1.25\times 10^5$ & $1.25\times 10^5$ \\
  \hline
\end{tabular}
\caption{Physical properties of the initial conditions for each galaxy simulation, along with the simulation times between which the simulation outputs are analyzed, $t_{\rm start}$ and $t_{\rm end}$. Unless otherwise stated, all values are median values between $t_{\rm start}$ and $t_{\rm end}$.}
\end{centering}
\end{table*}

\subsection{Initial conditions} \label{Sec::ICs}
We generate initial conditions for our early-type and NGC~300-like galaxy simulations using the {\sc MakeNewDisk} code~\citep{Springel05}. The physical properties of the dark matter halos, stellar disks and bulges, and gas disks for each simulation are shown in Table~\ref{Tab::Params}, along with the mass resolutions of the associated dark matter particles, stellar particles and gas cells. For the Milky Way-like simulation, we use the Agora initial condition~\citep{Kim14}, which is designed to resemble a Milky Way-like galaxy at redshift $z=0$. All six dark matter haloes are of ~\cite{Navarro97} (NFW) type, and our stellar and gas disks follow an exponential form. The stellar bulge follows a~\cite{Plummer1911} profile in the ETG initial conditions and a~\cite{Hernquist90} profile in the Milky Way-like initial condition. None of our initial conditions contain a hot halo/circumgalactic medium component, consistent with the Agora initial condition. Our median gas cell mass is $859~{\rm M}_\odot$ for all galaxies.

We note that the properties of the simulated gas and stellar disks change substantially between the beginning of the simulation (0~Myr) and the first simulation time analyzed ($t_{\rm start}$), as the gas disk settles into a state of dynamical equilibrium. Unless otherwise stated, Table~\ref{Tab::Params} therefore gives the median value of each disk parameter during the period of simulation times analyzed, during which the values do not change substantially.

\begin{figure*}
\begin{centering}
  \includegraphics[width=\linewidth]{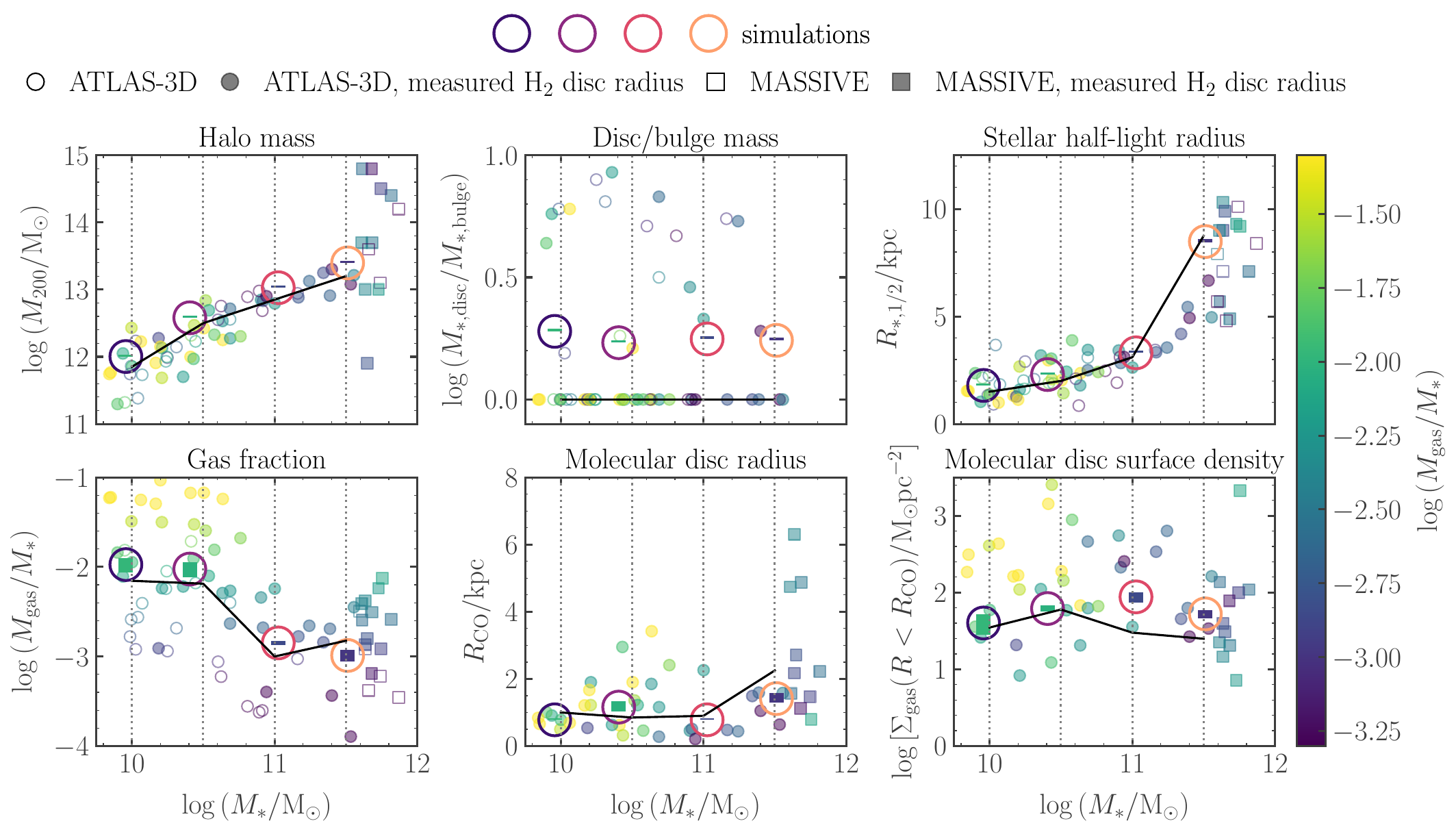}
  \caption{Physical properties of the simulated early type galaxies (circled bars), compared to observed galaxies from the ATLAS$^{\rm 3D}$ (circular transparent data points) and MASSIVE (square transparent data points) galaxy surveys. Unfilled data points represent observed galaxies with no measured values of the molecular gas disk size $R_{\rm CO}$ (center bottom panel), and so no measured values of the molecular gas surface density $\Sigma_{\rm H_2, CO}$ (right-hand bottom panel). Black lines represent the median observed values of each physical quantity in stellar mass bins centered on $M_*=10^{10}$, $10^{10.5}$, $10^{11}$ and $10^{11.5}~{\rm M}_\odot$. The vertical extent of the circled bars represents the values spanned by the simulated galaxies between the simulation times of $100$ and $400$~Myr, colored according to their gas fractions.  Our simulations roughly reproduce these median values (see Section~\protect\ref{Sec::ICs}). \textit{Observational references:} \citet{2011MNRAS.413..813C}, \citet{2013MNRAS.432.1709C}, \citet{2013MNRAS.432.1768K}, \citet{2013MNRAS.429..534D}, \citet{2014MNRAS.444.3427D}, for ATLAS$^{\rm 3D}$ and \citet{2018MNRAS.473.5446V}, \citet{2019MNRAS.486.1404D}, Timothy Davis, priv. comm. for MASSIVE.}
  \label{Fig::ICs}
\end{centering}
\end{figure*}

\subsubsection{Early type galaxies} \label{Sec::ICs-ETGs}
The physical parameters for our ETG initial conditions are designed to span the observable properties of ETGs from the MASSIVE \citep{2014ApJ.795.158M} and ATLAS$^{\rm 3D}$ \citep{2011MNRAS.413..813C} surveys (transparent data points in Figure~\ref{Fig::ICs}). We match the observed variation of the halo mass ($M_{200}$, top left), stellar half-light radius ($R_{*, 1/2}$, top right), gas fraction (bottom left), extent of the CO-luminous molecular gas disk ($R_{\rm CO}$, bottom center), and the surface density of the CO-luminous molecular gas disk ($\Sigma_{\rm gas}(R<R_{\rm CO})$, bottom right), as a function of the total stellar mass $M_*$ ($x$-axis of each panel). The black lines represent the observed median values at each stellar mass, while the circled bars represent the simulated values. The vertical extent of each bar represents the variation in value across the simulation times analyzed. The stellar masses of our galaxies span the observed samples in logarithmic intervals: our four ETGs have stellar masses of $M_* = 10^{10}, 10^{10.5}, 10^{11}$ and $10^{11.5}~{\rm M_\odot}$.

The disk-to-bulge mass ratios of our simulated galaxies are set to $M_{*, {\rm disk}}/M_{*, {\rm bulge}} = 0.2$, which will allow us, in the future, to compare our simulated molecular cloud samples to resolved observations of molecular gas in the lenticular galaxies studied by~\cite{Utomo2015,2021MNRAS.505.4048L,2023arXiv230805146W}.

According to the observable parameters presented in Figure~\ref{Fig::ICs}, we constrain the remaining physical properties of our ETG initial conditions as follows. Firstly, we determine the concentration parameters $c$ of our dark matter halos according to the halo concentration-mass relation of ~\cite{2014MNRAS.441.3359D}, for NFW fits to N-body halo density profiles in the cosmology of the \textit{Planck} satellite (their Equation 8). We use spin parameters of $\lambda = 0.04$ across our simulation suite, in accordance with the empirically-derived values of~\cite{2007MNRAS.375..163H}, across an SDSS sample of spiral and elliptical galaxies.

Within {\sc MakeNewDisk}, the stellar disk scale-length is set according to its angular momentum, which is in turn determined by the spin of the dark matter halo~\citep[see Section 2.1.1 of][and references therein]{Springel05}. We then set the stellar bulge scale-length such that the observed value of $R_{*, 1/2}$ is retrieved at each stellar mass, disk-to-bulge ratio, and gas fraction. Finally, we set the gas disk scale-length to match the sizes $R_{\rm CO}$ and surface densities $\Sigma_{\rm gas}(R<R_{\rm CO})$ of the observed molecular gas disks at each stellar mass. The CO-luminous gas fraction is computed for each simulation output using the {\sc DESPOTIC} astrochemistry and radiative transfer model~\citep{Krumholz13a,Krumholz14}, as described in Appendix~\ref{App::chem-postproc}.

The stellar disk scale-height in our initial condition is set iteratively by {\sc MakeNewDisk}, such that the stellar disk achieves a state of hydrostatic equilibrium within the full three-dimensional potential of the galaxy, assuming the density profile of a uniform isothermal sheet. The three dimensional velocity structure of the collisionless dark matter and stellar components are set according to the triaxial approximation outlined in Section 2.3 of~\cite{Springel05}, and we initialize the gas scale-height to a tenth of the stellar value, with a temperature of $10^4$~K. However, once the simulation begins, the gas and stellar disk scale-heights self-adjust to smaller values as the disks evolve towards a state of dynamical equilibrium under the influence of stellar feedback. Therefore, Table~\ref{Tab::Params} gives the median values for the stellar and gas disk scale-heights over the simulation period analyzed in this work. These values are stable once the disk has reached a state of dynamical equilibrium at $t_{\rm start}$.

\begin{figure*}
\begin{centering}
  \includegraphics[width=\linewidth]{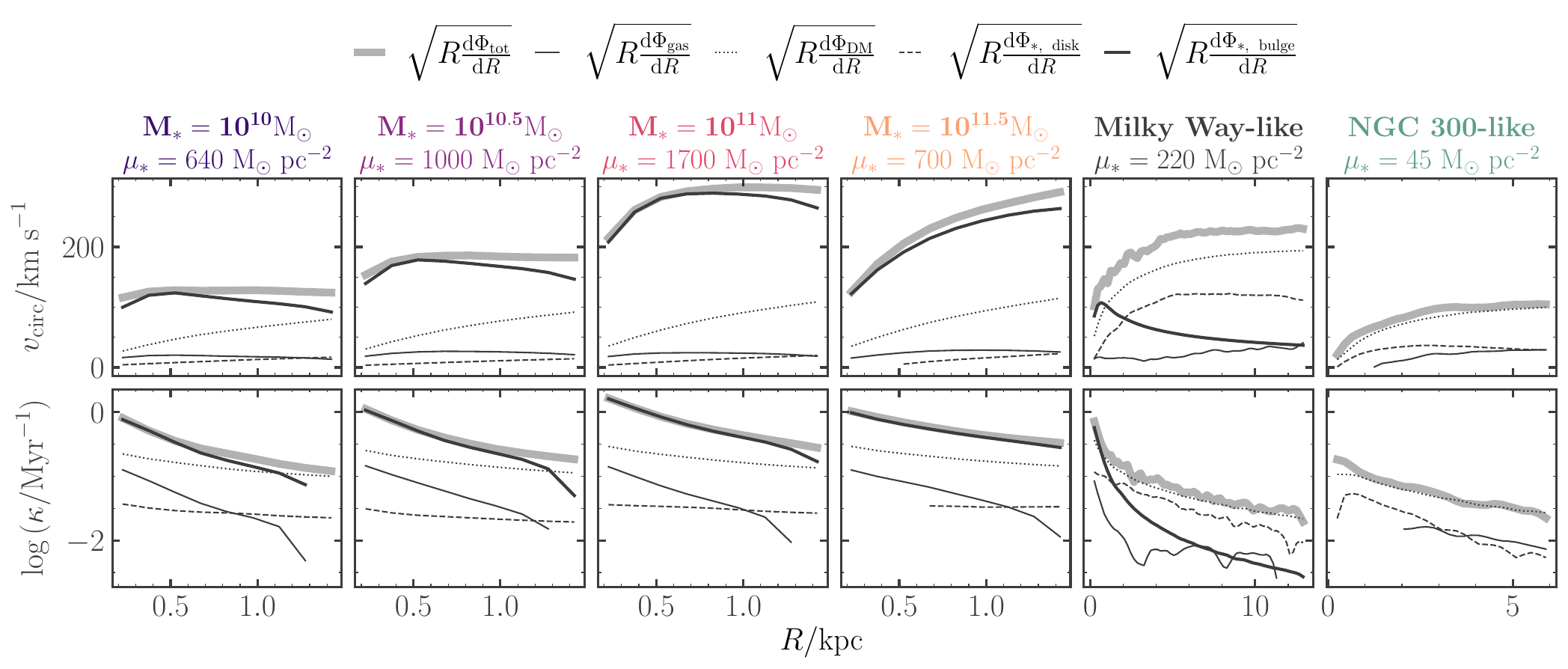}
  \includegraphics[width=.8\linewidth]{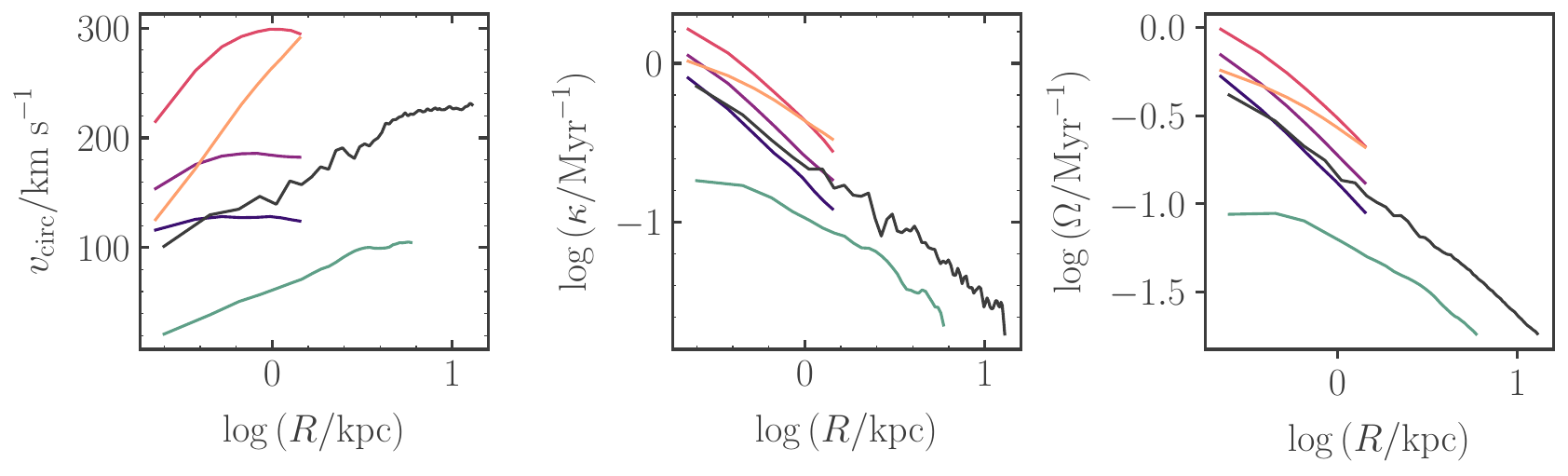}
  \caption{\textit{Top:} Mid-plane circular velocity $v_{\rm circ}$ and epicyclic frequency $\kappa$ (thick transparent lines) as a function of galactocentric radius $R$ for each of the simulated galaxies across the radial extent of the gas disk, computed directly from the gravitational potential exerted by the gas (thin lines), dark matter (dotted lines), disk stars (dashed lines) and bulge stars (bold lines). Note that the ETG gas disks are much smaller than those of the main sequence galaxies. The central stellar surface density $\mu_*$ is given below the stellar mass at the top of each column. Galactic rotation in the early type galaxies is dominated by the stellar bulge component. Each rotation profile is computed at one simulation time only, but the values change negligibly over the time period analyzed for each simulation. \textit{Bottom:} Comparison of the total circular velocities and epicyclic frequencies for each of the six galaxy simulations.}
  \label{Fig::vcircs}
\end{centering}
\end{figure*}

\subsubsection{Milky Way-like galaxy} \label{Sec::ICs-MW}
The Agora disk initial condition~\citep{Kim14} is designed to resemble a Milky Way-like galaxy at redshift $z \sim 0$. It has a dark matter halo mass of $M_{200} = 1.07 \times 10^{12} M_\odot$, a virial radius of $R_{200} = 205$~kpc, a halo concentration of $c=10$ and a spin parameter of $\lambda = 0.04$. The stellar bulge has a mass of $3.437 \times 10^9{\rm M}_\odot$, while the exponential disk has a mass of $4.297 \times 10^{10}{\rm M}_\odot$, a scale-length of $3.43$~kpc, and an initial scale-height of 343~pc.

The stellar disk scale-height equilibrates to a value of $123$~pc over the first $300$~Myr of simulation run-time, and the total gas disk scale-height to a value of $82$~pc. Specifically, the final molecular disk scale-height is around $50$~pc, the atomic disk scale-height is around $200$~pc, and the ionized gas extends to distances of $>1$~kpc above the mid-plane. We acknowledge that while these multi-phase gas disk scale-heights are in reasonable agreement with edge-on observations of external galaxies, the stellar disk scale-height is smaller than expected. The initial stellar velocity dispersion was chosen to be consistent with~\cite{Kim14}, but a larger initial vertical velocity dispersion would have resulted in a thicker equilibrium stellar disk. The bulge to stellar disk ratio is 0.125 and the initial gas fraction is 0.18.

\subsubsection{NGC300-like galaxy} \label{Sec::ICs-NGC300}
For our NGC~300-like simulation, we match the structural parameters of the dark matter halo, stellar disk and gas disk from~\cite{2011MNRAS.410.2217W}. The dark matter halo has a circular velocity of $V_{200} = 76~{\rm km~s}^{-1}$ at the virial radius, corresponding to a virial mass of $M_{200} = 8.3 \times 10^{10} {\rm M}_\odot$. We set an NFW concentration parameter of $15.4$, which gives a reasonable approximation to the observed rotation curve of the baryons in NGC~300, and choose a spin parameter of $\lambda = 0.04$, as explained in Section~\ref{Sec::ICs-ETGs}. The stellar disk has a mass of $1 \times 10^9  M_\odot$ and an initial scale-height of $0.28$~kpc, while the gas disk has a mass of $2 \times 10^9 {\rm M}_\odot$ and an isothermal temperature of $10^4$~K. Similarly to the ETG initial conditions, the stellar disk scale-length, along with the initial velocity structure of the collisionless particles and gas cells, are set according to the methods outlined in~\cite{Springel05}. The stellar disk scale-height equilibrates to a value of $90$~pc over the first $500$~Myr of the simulation run-time, and the gas disk scale-height to a value of $110$~pc.

\subsection{Galactic rotation curves} \label{Sec::rotation-curves}
In Figure~\ref{Fig::vcircs} we show the mid-plane circular velocity $v_{\rm circ}$ of each simulated galaxy as a function of galactocentric radius $R$ (first row of panels, thick transparent lines). We also show the epicyclic frequency $\kappa$ (second row, thick transparent lines), which is given by $\kappa = \Omega \sqrt{2(1+\beta)}$ for an angular velocity of $\Omega$ and a shear parameter $\beta = \dd \ln{v_{\rm circ}}/\ln{R}$. For both $v_{\rm circ}$ and $\kappa$, we show the separate components that are contributed by the gravitational potential $\Phi_{\rm gas}$ due to the gas particles in the simulation, $\Phi_{\rm DM}$ due to the dark matter particles, $\Phi_{*, {\rm disk}}$ due to the stellar disk, and $\Phi_{*, {\rm bulge}}$ due to the stellar bulge. In the lower two panels, we directly compare the values of $v_{\rm circ}$ and $\kappa$ across the entire simulation suite.

The rotation speed is largest ($\sim 300$~km/s) in the ETG of stellar mass $M_* = 10^{11}~{\rm M}_\odot$, owing to the higher concentration of its stellar bulge. While this galaxy does not have the largest stellar mass, this mass is concentrated within the smallest stellar half-light radius $R_{*, 1/2}$. Its higher bulge concentration can be quantified by its central stellar surface density $\mu_* = 1700~{\rm M}_\odot {\rm pc}^{-2}$, 70\% higher than the second most-compact bulge, with $\mu_* = 1000~{\rm M}_\odot {\rm pc}^{-2}$. The value of $\mu_*$ is measured within $R_{*, 1/2}$ for each galaxy, and listed above the top row of panels in Figure~\ref{Fig::vcircs}.

We see that the values of $\mu_*$ and the speed of galactic rotation are approximately correlated across our simulation suite: as such, the central stellar surface density can be used as a proxy for the value of $\kappa$ (and thus the degree of support provided to the gas by tidal, centrifugal and Coriolis forces) in a given galaxy. The elevated bulge-induced rate of rotation in the $M_* = 10^{11}~{\rm M}_\odot$ ETG is an important feature of this simulation, to which we will later return.

By contrast to the ETGs, the Milky Way-like and NGC300-like galaxies (right-hand panels) have much lower levels of rotation-induced shear outside the central kpc. Their rotation curves are dominated by their dark matter and stellar/gas disk components, with negligible or zero contribution to the circular velocity from a stellar bulge component. The galactic center region of the Milky Way-like galaxy (black lines, lower panels) is the most strongly rotationally-supported\footnote{Throughout this paper, we use the term `rotational support' to refer to the degree of support against gravitational collapse that is provided to the gas in a galaxy, by the tidal, centrifugal and Coriolis forces. This support depends both on the magnitude and gradient of the gas circular velocity $v_{\rm circ}$, and thus on the gravitational potential of the galaxy.} region of the disk galaxies, and matches the value of $\kappa$ in only the most weakly-supported ETG (dark purple lines, lower panels).

We note that the circular velocity of the NGC~300-like galaxy (green lines, lower panels) is around 25\% higher than the value observed by~\cite{2011MNRAS.410.2217W}, which varies from $50$ to $80$~km/s between galactocentric radii of $0.3$ and $6$~kpc, while ours has an average value of $100$~km/s. Our simulated value is more typical of other dwarf spiral galaxies, such as M33~\citep[e.g.][]{2018MNRAS.479.2505K}.

\subsection{Hydrodynamics, chemistry, star formation and feedback} \label{Sec::chem-SF-feedback}
The initial conditions described in Section~\ref{Sec::ICs} are evolved using the moving-mesh hydrodynamics code {\sc Arepo}~\citep{Springel10}. In particular, the gas reservoir is modeled using an unstructured moving mesh that is defined by a Voronoi tesselation about a discrete set of points, moving with the local gas velocity. A hybrid TreePM gravity solver is used to calculate the gravitational acceleration vectors of the Voronoi gas cells, stellar particles and dark matter particles. We employ the native adaptive gravitational softening scheme for the gas cells, with a minimum softening length of $3$~pc and a gradation of $1.5$ times the Voronoi gas cell diameter. We set the softening length of the star particles to a constant value of $3$~pc, and set the softening length of the dark matter particles to $280$~pc, according to the convergence tests presented in~\cite{2003MNRAS.338...14P}.

Our models for the temperature and chemical composition of the gas in our simulations, along with the rate of star formation in this gas, and the rate of energy and momentum injection due to stellar feedback, are identical to those described in~\cite{2024MNRAS.527.7093J}. We give a brief overview of these models below, but refer the reader to the cited works for further details.

We use the non-equilibrium network for hydrogen, carbon and oxygen chemistry described in~\cite{NelsonLanger97} and in~\cite{GloverMacLow07a,GloverMacLow07b}, coupled to the atomic and molecular cooling function of~\cite{Glover10}. This includes cooling due to fine structure emission from ${\rm C}^+$, ${\rm O}$ and ${\rm Si}^+$, Lyman-$\alpha$ emission from atomic hydrogen, ${\rm H_2}$ line emission, gas-grain cooling, and electron recombination on grain surfaces and in reaction with polycyclic aromatic hydrocarbons (PAHs). In hot gas, the chemical network additionally allows for cooling via the collisional processes of ${\rm H}_2$ dissociation, Bremsstrahlung, and the ionization of atomic hydrogen. The dominant heating mechanism is photoelectric emission from dust grains and PAHs, with lesser contributions from cosmic ray ionization and ${\rm H_2}$ photodissociation by the interstellar UV radiation field (ISRF). We assign a value of 1.7~\cite{Habing68} units for the ISRF strength according to~\cite{Mathis83}, and a value of $2 \times 10^{-16}~{\rm s}^{-1}$ to the cosmic ray ionization rate~\citep[e.g.][]{Indriolo&McCall12}.\footnote{These choices are based on the solar neighborhood; more realistically, these values would vary in time proportional to the local star formation rate per unit area.  A lower (higher) radiation field will tend to enhance (decrease) the cold-to-warm gas mass ratio, and to decrease (increase) the thermal pressure.} The dust grain number density is computed by assuming the solar value for the dust-to-gas ratio, and the dust temperature is obtained according to the procedure described in Appendix A of~\cite{Glover&Clark12}. Finally, we use the {\sc TreeCol} algorithm presented in~\cite{Clark12} to model the dust- and self-shielding of molecular hydrogen from dissociation by the interstellar radiation field, allowing us to track the non-equilibrium abundance of molecular hydrogen during the run-time of the simulation.

The star formation rate volume density in our simulation is given by
\[ 
\frac{\dd \rho_{*,i}}{\dd t} =
\begin{cases} \label{Eqn::SF1}
  &\epsilon_{\rm ff} \rho_i / t_{{\rm ff},i}, \; \rho_i \geq \rho_{\rm thresh}, T_i \leq T_{\rm thresh} \\
  &0, \; \rho_i < \rho_{\rm thresh}, T_i > T_{\rm thresh}
\end{cases}
\]
where $t_{{\rm ff}, i} = \sqrt{3\pi/(32 G\rho_i)}$ is the local free-fall time-scale for the gas cell $i$ with a mass volume density of $\rho_i$, and $\epsilon_{\rm ff}$ is the star formation efficiency per free-fall time, which follows the parametrization of~\cite{Padoan17}, such that
\begin{equation} \label{Eqn::SF2}
\epsilon_{\rm ff} = 0.4 \exp{(-1.6 \alpha_{\rm vir}^{0.5})}.
\end{equation}

\begin{figure*}
\begin{centering}
  \includegraphics[width=.9\textwidth]{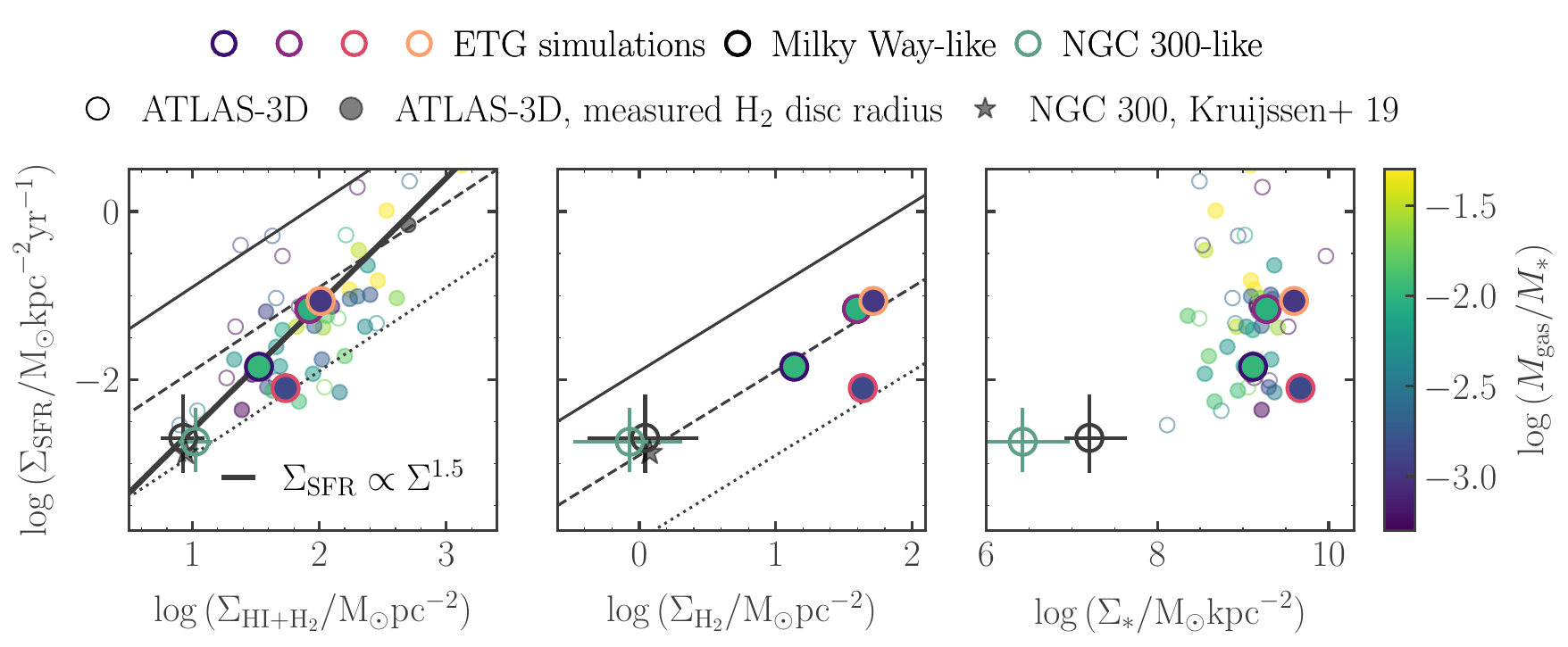}
  \caption{\textit{Left:} Median star formation rate surface density $\Sigma_{\rm SFR}$ as a function of the cold gas (atomic plus molecular) surface density $\Sigma_{\rm HI+H_2}$, integrated across each simulated galaxy (large circles), and measured for the galaxies in the ATLAS$^{\rm 3D}$ galaxy sample~\protect\citep{2013MNRAS.429..534D,2014MNRAS.444.3427D} (small transparent circles). The colors of the data points correspond to their gas fractions, and gas depletion times of $10^8$, $10^9$ and $10^{10}$ years are given by the black solid, dashed and dotted lines, respectively. Interquartile ranges over time and galactocentric radius are given by error-bars. For the early type galaxy simulations, these are too small to be shown. \textit{Center:} Similar to \textit{left} but for the molecular gas surface density $\Sigma_{\rm H_2}$. \textit{Right:} Similar to \textit{left} but for the median stellar surface density $\Sigma_*$ across the gas disk.}
  \label{Fig::KS-relations}
\end{centering}
\end{figure*}

The virial parameter $\alpha_{\rm vir}$ on cloud scales is computed during simulation run-time within overdense regions surrounding each star-forming gas cell; the scale of each overdensity is determined via a variant of the~\cite{1960mes..book.....S} approximation, as the characteristic length-scale $L=\rho/|\nabla \rho|$ for changes in the density of the surrounding gas, where $\nabla \rho = \partial \rho/\partial r$ is the density gradient with distance $r$ from the central gas cell. The algorithm is described in detail in~\cite{2020MNRAS.495..199G}. The median radius of these overdensities is $\sim 10$~pc across our simulation suite, and the average number of gas cells within each overdensity is $140$. We set an upper limit of $T_{\rm thresh} = 100{\rm K}$ on the temperature below which star formation is allowed to occur, and a lower limit of $\rho_{\rm thresh}/m_{\rm H} \mu = 100~{\rm cm}^{-3}$ on the density, where $\mu$ is the mean mass per H atom.

The star particles formed via Equations~(\ref{Eqn::SF1}) and (\ref{Eqn::SF2}) generate energy and momentum from supernova explosions and pre-supernova HII regions, via the stellar feedback prescription described in~\cite{2021MNRAS.505.3470J}. To compute the number of supernovae, ejected mass and photoionizing luminosity of each star particle, we assign a stellar population drawn stochastically from a~\cite{Chabrier03} initial stellar mass function (IMF), using the Stochastically Lighting Up Galaxies (SLUG) stellar population synthesis model~\citep{daSilva14,Krumholz15}. An energy of $10^{51}~{\rm erg}$ per supernova is assumed, and the terminal momentum from these supernovae is explicitly calculated using the few-supernovae parametrization derived from the high-resolution simulations of~\cite{Gentry17} (their Equation 17). This kinetic energy, along with the remaining thermal energy, is and injected into all gas cells surrounding each star particle.

\begin{figure*}
\begin{centering}
  \includegraphics[width=\textwidth]{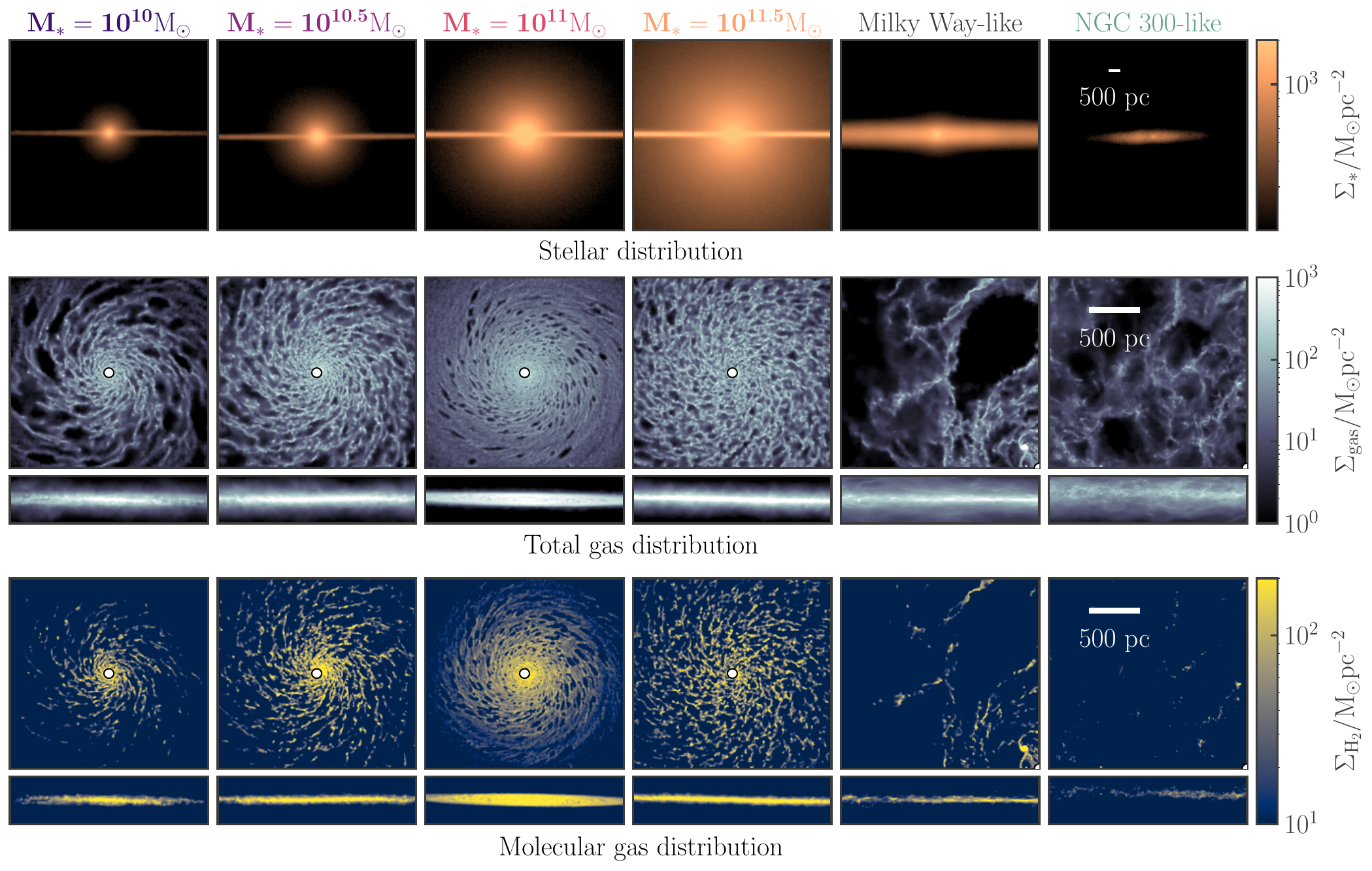}
  \caption{Surface density maps of the stellar distribution viewed parallel to the galactic mid-plane ($\Sigma_*$, upper panels), the total gas distribution viewed perpendicular to and parallel to the galactic mid-plane ($\Sigma_{\rm gas}$, center panels) and the molecular gas distribution viewed perpendicular to and parallel to the galactic mid-plane ($\Sigma_{\rm H_2}$, lower panels) for each of the simulated galaxies. All early type galaxies are shown at a simulation time of $400$~Myr, while the Milky Way-like galaxy is shown at a simulation time of $600$~Myr, and the NGC~300-like galaxy is shown at a simulation time of $800$~Myr. The small white circles denote the minimum galactocentric radius analyzed in this work. Note that only a small off-center section of the larger Milky Way-like and NGC~300-like disks is displayed.}
  \label{Fig::morphology}
\end{centering}
\end{figure*}

The photo-ionizing luminosity associated with HII regions is converted to a momentum per unit time via the model of~\cite{2021MNRAS.505.3470J}, following the analytic work of~\cite{Matzner02} and~\cite{KrumholzMatzner09} to account for both radiation pressure and the momentum injected via the `rocket effect': the ejection of warm ionized gas from cold molecular clouds. The gas cells inside the Str\"{o}mgren radii of the HII regions are fully ionized and heated to a temperature of $7000$~K.

The dense molecular gas clouds in which star formation occurs are dispersed on short time-scales by the momentum from these HII regions, as discussed at length in~\cite{2021MNRAS.505.3470J} and~\cite{2024MNRAS.527.7093J}. However, HII region momentum does not contribute substantially to the total momentum injected by stellar feedback across each simulation: in our prescription, the HII region momentum injection is less than 10\% of the supernova momentum injection for a given stellar cluster.

\section{Star formation and interstellar medium morphology} \label{Sec::galaxy-props}
Quiescent, bulge-dominated galaxies have been found to have cold gas depletion times that are, on average, substantially longer than those measured for the main sequence galaxy population~\citep[e.g.][]{Saintonge2012,2014MNRAS.444.3427D,2020A&A...644A..97C}. Such elliptical galaxies are also found to have smoother, less-fragmented interstellar media than their main sequence counterparts~\citep{Davis2022,Gensior2023}. Figures~\ref{Fig::KS-relations}-\ref{Fig::phases} examine the star-forming and gas-morphological properties of our simulated galaxies.

We note that the analysis presented in Figures~\ref{Fig::vcircs} and~\ref{Fig::phases}-\ref{Fig::rotcurves} excludes the central $50$~pc of each galactic disk. We excise this region of each simulation for two reasons. Firstly, the scale-heights of our simulated ETG gas disks become too small to be sufficiently well-resolved in the central $50$~pc. Secondly, we do not account for AGN feedback, which may affect the formation and properties of galactic centers in real galaxies. The excision does not apply to Figures~\ref{Fig::gal-main-sequence} and~\ref{Fig::ICs}, because these figures compare the global properties of each disk to observations.

\subsection{Star formation} \label{Sec::SF}
Figure~\ref{Fig::KS-relations} shows the star formation rate surface density $\Sigma_{\rm SFR}$ of our six simulated galaxies as a function of their cold gas surface densities $\Sigma_{\rm HI+H_2}$, their molecular gas surface densities $\Sigma_{\rm H_2}$, and their stellar surface densities $\Sigma_*$. Large circles represent averages over simulation time across the extent of each gas disk. Error-bars represent the corresponding interquartile ranges, where the interquartile ranges for the ETGs are too small to be displayed.

The star formation rate surface densities in Figure~\ref{Fig::KS-relations} are calculated as averages over the preceding $5$~Myr, similar to the time interval traced in observations via H$\alpha$ emission. Observed values from the ATLAS$^{\rm 3D}$ survey (small transparent and unfilled circles) are shown for comparison with the ETG simulations (large filled circles), and the observed position of NGC~300 in the left-hand and center planes (black transparent star,~\citealt{Kruijssen2019}), is shown for comparison with the late-type simulations (black and turquoise large unfilled circles). The close agreement between simulations and observations in this Figure and in Figure~\ref{Fig::gal-main-sequence} is an important validity check for our numerical models of star formation and stellar feedback, outlined in Section~\ref{Sec::sims}.

Five out of six galaxies fall along the typical powerlaw of index $\sim 1.5$ (black solid line, left-hand panel) relating the star formation rate surface density to the cold gas surface density for main sequence galaxies~\citep{Kennicutt1998,Bigiel08}. These five galaxies have molecular gas depletion times of $1$~Gyr (dashed line, second-to-left panel). However, the ETG with the highest $\kappa$ displays a suppressed star formation rate, falling substantially below the power-law, with a molecular gas depletion time of around $8$~Gyr. \textbf{That is, star formation in the ETG simulation of stellar mass $M_* = 10^{11}~{\rm M}_\odot$ (with the most concentrated stellar bulge), is dynamically suppressed by nearly an order of magnitude in depletion time.} We analyze this galaxy in detail in the second paper of this series.

\subsection{Disk morphology and fragmentation} \label{Sec::morphology}
The central and lower rows of panels in Figure~\ref{Fig::morphology} compare the total gas disk surface density $\Sigma_{\rm gas}$ and the molecular gas disk surface density $\Sigma_{\rm H_2}$ of all six galaxy simulations, within $2~{\rm kpc}$ patches at face-on and edge-on viewing angles. For the ETG simulations (left four columns), these patches cover the extent of the entire gas disk. For the Milky Way-like and NGC~300-like galaxies (right two columns), with respective gas disk diameters (by eye) of $\sim 30$ and $\sim 12$~kpc, the patch shows only a small portion of each gas disk. The excluded central $50$~pc of each gas disk, noted in Section~\ref{Sec::SF}, is marked by a white circle, and is removed from our analysis for the remainder of the paper.

The top row of panels shows the edge-on stellar surface density $\Sigma_*$ for each galaxy, highlighting the fact that the stellar distribution for the ETGs is bulge-dominated ($M_{\rm disk}/M_{\rm bulge} \sim 0.2$), whereas the stellar distribution is disk-dominated for the Milky Way-like galaxy ($M_{\rm disk}/M_{\rm bulge} \sim 0.9$) and for the NGC~300-like galaxy ($M_{\rm bulge} \sim 0$).

All four ETG simulations have much smoother gas distributions than do the main sequence galaxy simulations. While the Milky Way-like and NGC~300-like galaxies display giant feedback-driven voids of several kpc in diameter, such bubbles are reduced to diameters of $<100$~pc in the ETGs. The dynamically-suppressed ETG ($M_* = 10^{11}{\rm M}_\odot$) is particularly smooth, as expected according to its long cold-gas depletion time and thus infrequent stellar feedback. The three other ETGs are manifestly fragmented into dense gas clouds, but still remain much smoother than their main sequence counterparts.

\begin{figure*}
\begin{centering}
  \includegraphics[width=.8\linewidth]{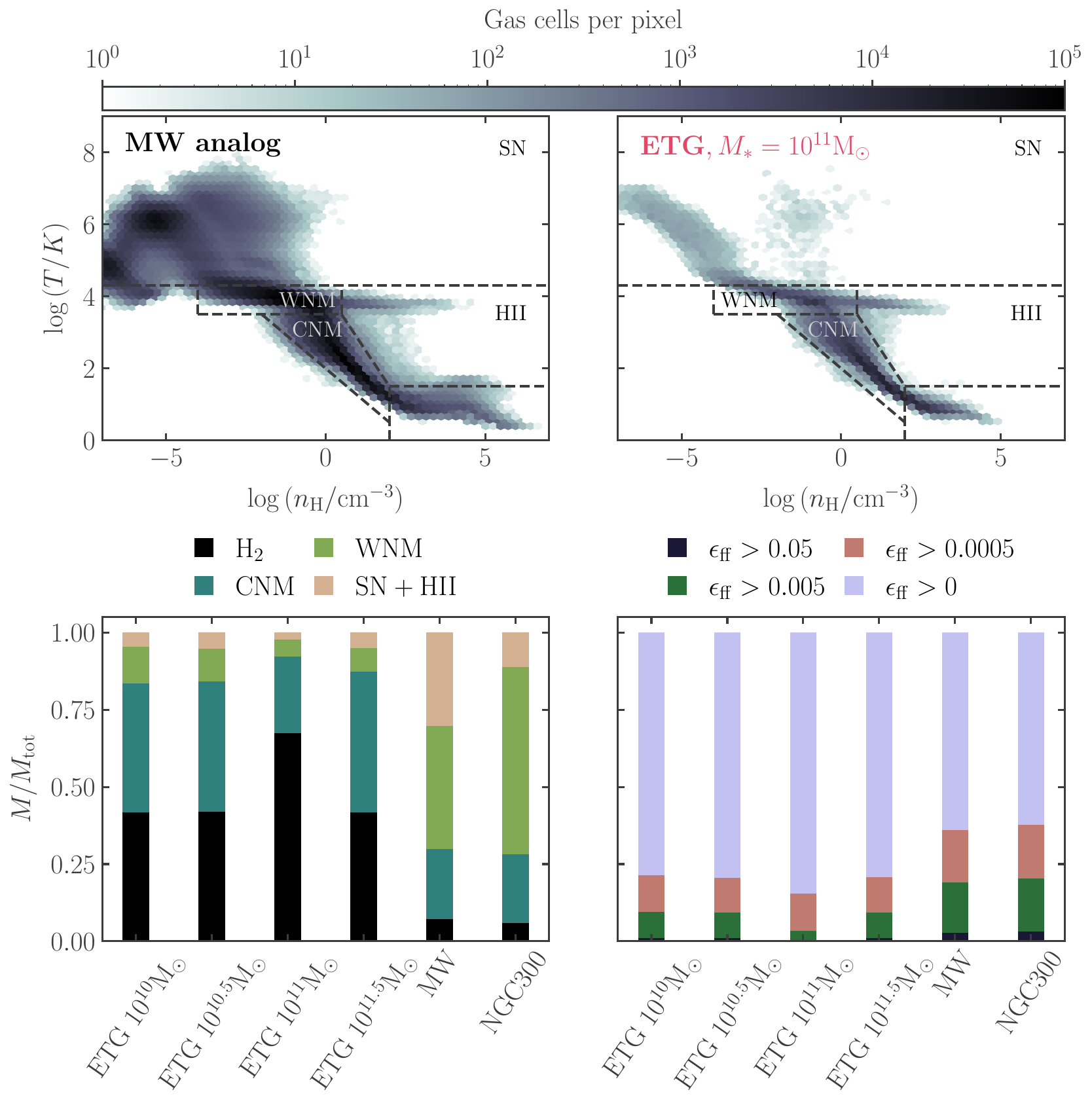}
  \caption{\textit{Top panels:} Density-temperature phase diagrams for the Milky Way-like galaxy simulation (left) and the smoothest early type galaxy ($M_*=10^{11} {\rm M}_\odot$, right). Dashed lines delineate the regions of phase space corresponding to the bar-plot in the bottom panels. \textit{Bottom left:} Partitioning of the gas mass in each simulation into four interstellar medium phases, from warmest to coolest, as a fraction of the total gas mass in the simulation: hot gas that has received thermal energy from stellar feedback ($M_{\rm SN+HII}$), the warm neutral medium ($M_{\rm WNM}$), the cold neutral medium ($M_{\rm CNM}$) and the total molecular hydrogen reservoir ($M_{\rm H_2}$). The molecular hydrogen mass is subtracted from each of the other phase-space regions to produce the bar plot. \textit{Bottom right:} Partitioning of the star-forming gas mass with density $\rho/m_{\rm H} \mu > 100~{\rm cm}^{-3}$ into fractions with differing star formation efficiencies per free-fall time, $\epsilon_{\rm ff}$.}
  \label{Fig::phases}
\end{centering}
\end{figure*}

This lower degree of gas disk fragmentation in the early type galaxy simulations, relative to the large spiral (Milky Way-like) simulation, is in qualitative agreement with the observed sample of early type and spiral galaxies in~\cite{Davis2022}. In that work, the disk clumpiness of a sample of 86 spiral galaxies, as quantified by the Gini statistic, is more than double that of a sample of 15 early type, bulge-dominated galaxies, which vary from very smooth (resembling our dynamically-suppressed ETG simulation) to manifestly fragmented (resembling our other ETG simulations). The disk smoothness is also seen to increase with the central stellar surface density in these observations, in qualitative agreement with our $10^{11} {\rm M}_\odot$ ETG, which has the most concentrated stellar bulge and the smoothest gas disk. We will discuss the physical drivers of this disk smoothness in Section~\ref{Sec::stability-and-clustering}.

Finally, we note that the gas and stellar disks of our ETG simulations develop a slight kinematic misalignment during their $400$~Myr of evolution. This misalignment likely arises due to the gravitational interaction between the gas disk and the stellar bulge. The maximum skew of 3~degrees occurs for the smoothest disk with the most compact bulge ($M_* = 10^{11}{\rm M}_\odot$). Throughout this work, the term `mid-plane' therefore refers specifically to the mid-plane of the gas disk.

\subsection{Gas phases} \label{Sec::phases}
The phase structure of the gas in each of our simulations is presented in Figure~\ref{Fig::phases}. The top two panels compare the mass-weighted distributions of gas as a function of volume density $n_{\rm H}$ and temperature $T$ (`phase diagrams') for the Milky Way-like and the dynamically-suppressed ETG simulation. The phase diagram for the NGC300-like simulation is very similar to that of the Milky Way, and the phase diagrams of the other three ETGs are relatively similar to that of the quenched ETG. The gas cells are clustered around the state of thermal equilibrium balancing the cooling rate (dominated in our simulations by line emission from ${\rm C}^+$, ${\rm O}$ and ${\rm Si}^+$) and the heating rate due to the photoelectric effect at the surfaces of PAHs and dust grains. The region of the histogram at $T \sim 7000~{\rm K}$ and high volume density corresponds to the gas that is heated by the thermal feedback from HII regions, and the gas above a temperature of $\sim 20,000$~K is heated by supernova feedback.

The lower left-hand panel of Figure~\ref{Fig::phases} shows the partitioning of the interstellar medium into the four phases that are delineated by dashed lines in the phase diagrams: feedback-heated (SN and HII), the warm neutral medium (WNM), the cold neutral medium (CNM) and the molecular hydrogen fraction (${\rm H_2}$). This partitioning is chosen by eye, with the exception of the ${\rm H_2}$ mass, which is calculated during simulation runtime using the chemical network described in Section~\ref{Sec::sims}. Any ${\rm H_2}$ mass contained in the other partitions is subtracted to produce the bar plot.

We see that the star-forming main sequence (Milky Way-like and NGC~300-like) simulations contain a much higher fraction of hot gas (salmon-colored bars, bottom left) than is present in the ETG simulations, commensurate with their much larger feedback-driven bubbles and voids, as shown in Figure~\ref{Fig::morphology}. Conversely, the ETG simulations contain a much higher fraction of cold atomic and molecular gas (black and turquoise bars, left-hand side): up to 70\% in the dynamically-suppressed ETG, and 40\% in the other ETGs, relative to $<10$\% of the gas in the main sequence galaxies. \textbf{The much larger fraction of SN-heated gas in the Milky Way-like and NGC~300-like simulations, despite their similar depletion times to three of the ETGs, points to a larger degree of supernova clustering, relative to the ETGs~\citep[see~e.g.][]{2020arXiv200911309S,2021MNRAS.505.3470J}. We will return to this point in Section~\ref{Sec::stability-and-clustering}.}

\begin{figure*}
\begin{centering}
  \includegraphics[width=.95\linewidth]{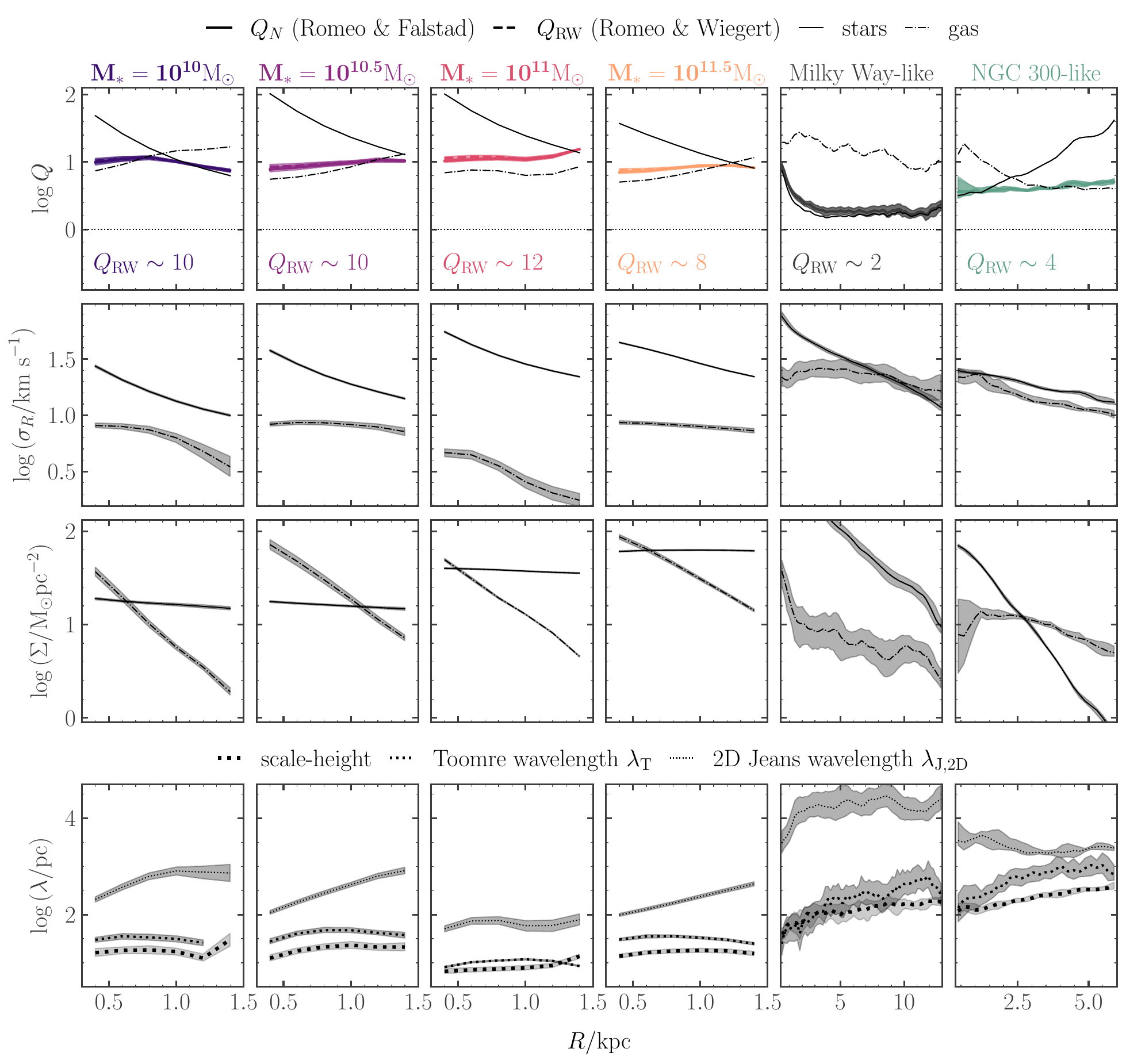}
  \caption{The median Toomre $Q$ parameter (top row), gas and stellar radial velocity dispersion (second row), and gas and stellar surface density (third row) as a function of galactocentric radius for the six galaxies in our sample. The bottom row compares the gas disk scale-height to the Toomre and 2D Jeans wavelengths. All values are computed for gas at temperatures $T<2 \times 10^4~{\rm K}$. Shaded regions represent interquartile ranges over azimuthal angle and simulation time. The Toomre $Q$ parameters of the early type galaxies are much higher than those of the main sequence (Milky Way-like and NGC~300-like) galaxies.}
  \label{Fig::Romeo-Qs}
\end{centering}
\end{figure*}

The lower right-hand panel of Figure~\ref{Fig::phases} show the partitioning of the star-forming gas (gas with $n_{\rm H}>100~{\rm cm}^{-3}$) into four logarithmic bins of star formation efficiency per free-fall time $\epsilon_{\rm ff}$, which is computed during the run-time of the simulation according to Equation~(\ref{Eqn::SF2}). Darker colors correspond to higher values of $\epsilon_{\rm ff}$. The fraction of dense gas that is forming stars, particularly at high $\epsilon_{\rm ff}$, is manifestly larger in the star-forming main-sequence galaxies than in the four ETGs. In particular, the dynamically-suppressed ETG displays a much smaller fraction of highly star-forming gas, with $\epsilon_{\rm ff}>0.5\%$ (less than half of the fraction in the other three ETGs). Despite having a much higher molecular fraction than the other ETGs, it contains no gas with $\epsilon_{\rm ff}>5\%$. \textbf{The dynamical suppression of star formation in one out of four ETG simulations therefore occurs due to the reduction of $\epsilon_{\rm ff}$ in the coldest and densest molecular gas.}

\section{Supernova clustering, galactic outflows and the equation of state} \label{Sec::stability-and-clustering}
Recent numerical work has shown that supernova clustering is likely to enhance the strength and mass-loading of galactic outflows~\citep{2017ApJ...834...25K,2018MNRAS.481.3325F,2020arXiv200911309S,2021MNRAS.505.3470J}, perhaps at the expense of turbulence-driving within the ISM~\citep{2022ApJ...932...88O}. In turn, the majority of supernova clustering occurs in the most massive giant molecular clouds~\citep{2024MNRAS.527.7093J}, which host the majority of galactic star formation~\citep{Murray&Rahman10}. These massive clouds are able to grow due to a high rate of accretion from the galactic environment, and display substantially higher lifetime star formation efficiencies than their low-mass counterparts, as they are slightly more difficult to destroy~\citep[e.g.][]{Murray10,2018MNRAS.475.3511G,2024MNRAS.527.7093J}.

In the following sections, we discuss the connection between the rotational support of the gas disk, supernova clustering and galactic outflow strength across our main sequence and quenched galaxy simulations. We then demonstrate the impact of this physics on the equation of state (pressure vs. density relation), and its implications for modeling gas in cosmological simulations.

\subsection{Disk stability and Toomre length}
The top row of panels in Figure~\ref{Fig::Romeo-Qs} demonstrates that the ETGs in our galaxy sample have a much greater level of disk stability than the main sequence galaxies. We calculate the Toomre $Q$ parameter for a multi-phase interstellar medium of finite disk thickness (as is appropriate to our simulations) via the prescriptions of \cite{2011MNRAS.416.1191R} and~\cite{2013MNRAS.433.1389R}, which are in close agreement.\footnote{The hot phase ($T>2 \times 10^4$~K) is excluded, as the majority of this gas is contained in feedback-driven bubbles or galactic outflows, and so does not contribute to the disk.} \cite{2011MNRAS.416.1191R} combine separate gas and stellar contributions to the dispersion relation, while~\cite{2013MNRAS.433.1389R} additionally consider separate contributions from the molecular, atomic and ionized gas phases. The solid and dot-dashed lines show the stellar and gaseous Toomre $Q$ parameters $Q_* = \kappa \sigma_{R, *}/3.36 G \Sigma_*$ and $Q_{\rm gas} = \kappa \sigma_{R, {\rm gas}} / \pi G \Sigma_{\rm gas}$, respectively, where $\kappa$ is the epicyclic frequency $\kappa^2 = R^{-3} \dd(\Omega^2 R^4)/\dd R$ (shown in Figure~\ref{Fig::vcircs}), $\sigma_{R,*}$ and $\sigma_{R, {\rm gas}}$ (see Appendix B) are the radial stellar and gas velocity dispersions (shown in the center row of panels), and $\Sigma_{*}$ and $\Sigma_{\rm gas}$ are the stellar and gas surface densities (shown in the lower row of panels).

\textbf{Comparing Figures~\ref{Fig::Romeo-Qs} and~\ref{Fig::vcircs}, we see that the elevated disk stability in the ETGs is driven primarily by the stellar contribution to galactic rotation.} Though the gas and stellar velocity dispersions in the ETGs are actually lower than those in the outer Milky Way, and the gas and stellar surface densities are comparable, compact stellar bulges in the four ETGs drive up their epicyclic frequencies $\kappa$ by around an order of magnitude. This translates to a substantial increase in their Toomre $Q$ values.

We caution that the Toomre $Q$ parameters in Figure~\ref{Fig::Romeo-Qs} are shown only as a comparison of the approximate level of stability between the six simulations. In turbulent media, the classical threshold, predicting fragmentation only for $Q\leq 1$, does not strictly apply. Firstly, the threshold for axisymmetric disk instability is a more complicated function of $a$ and $b$, where the gas surface density has a length-scaling relation of $\Sigma \propto \ell^a$, and the velocity dispersion has a scaling relation of $\sigma \propto \ell^b$~\citep[e.g.][]{2010MNRAS.407.1223R}. There exist non-Toomre regimes of $a$ and $b$ in which small or large scales may always be unstable~\citep[e.g.][]{2010MNRAS.407.1223R,2013ApJ...776...48H}. Furthermore, the original Toomre $Q$ criterion does not take account of non-axisymmetry or vertical dynamics, magnetic effects, non-linearity \citep[e.g.][]{2001ApJ...559...70K,2002ApJ...581.1080K} or turbulent dissipation \citep[e.g.][]{2011ApJ...737...10E}. The fact that $Q \approx 10$ for our ETGs does not imply that gravitational fragmentation is not occurring in these galaxies, and drawing a quantitative conclusion from $Q$ alone is discouraged.

The bottom row of Figure~\ref{Fig::Romeo-Qs} compares the cold+warm gas disk scale-height (thick dotted lines) for each simulation to its Toomre wavelength $\lambda_{\rm T} = (2\pi)^2 G \Sigma_{\rm gas}/\kappa^2$ (medium dotted lines) and its 2D Jeans wavelength (thin dotted lines). Due to its higher rate of galactic rotation, the Toomre wavelength associated with the dynamically-suppressed ETG simulation (stellar mass $M_* = 10^{11} {\rm M}_\odot$, pink label) is about one third the value of the other three ETGs.

\textbf{Comparing Figures~\ref{Fig::Romeo-Qs} and~\ref{Fig::morphology}, we see that smoother gas morphologies and smaller voids in the gas distribution are associated with the shorter Toomre wavelengths $\lambda_{\rm T}$ of the ETGs.} That is, angular momentum imposes much stronger constraints on the scale of self-gravitating condensation for the ETGs. Because the variation in the ratio $\sigma_{\rm R}/\Sigma$ is much smaller than that of $\lambda_{\rm T}$, this also means that smoother gas distributions are associated with higher Toomre $Q$ values, as observed by~\cite{Davis2022}. In the next section, we will show that this smoother gas distribution is associated with a lower level of supernova clustering and weaker galactic outflows.

\begin{figure}
  \includegraphics[width=.95\linewidth]{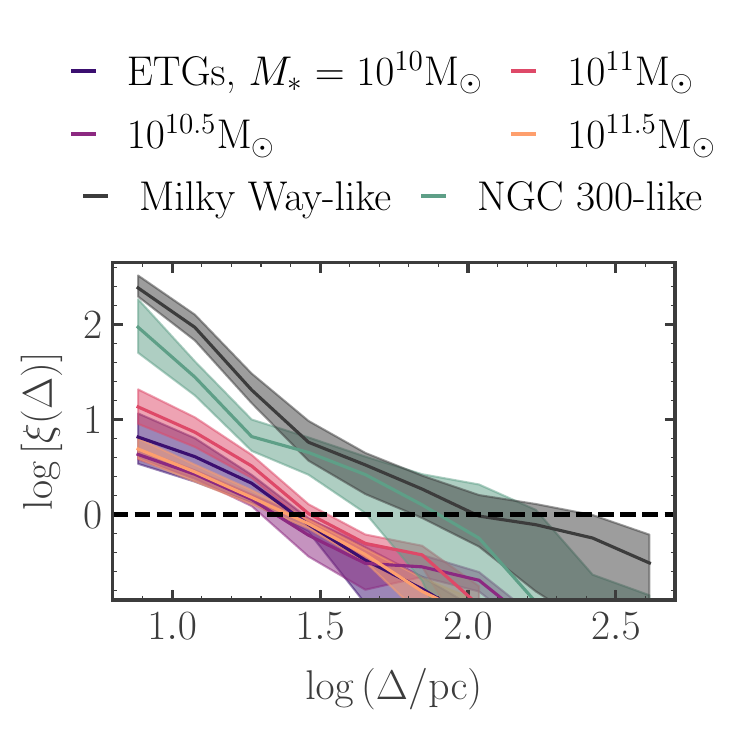}
  \caption{The two-point correlation function $\xi(\Delta)$ for supernova explosions (quantifying the degree of supernova clustering) as a function of their separation $\Delta$ over time intervals of $1$~Myr, averaged over all times throughout each simulation (solid lines). The black dashed line indicates the profile for a uniform distribution of objects across the galactic mid-plane ($\xi(\Delta) = 1$). The shaded regions give the interquartile ranges over these times. The level of supernova clustering is higher on all scales in the Milky Way-like and NGC~300-like simulations, relative to the early type galaxy simulations.}
  \label{Fig::SN-clustering}
\end{figure}

\subsection{Supernova clustering}
Figure~\ref{Fig::SN-clustering} shows the level of supernova clustering in each of our simulated galaxies, quantified by the two-point correlation function $\xi(\Delta)$ of supernova explosions as a function of spatial scale $\Delta$. If $\xi>1$, then the supernovae are \textit{more clustered} than would be expected for a uniform distribution of objects across the galactic mid-plane; if $\xi<1$ then they are \textit{less clustered}. The supernovae in the star-forming main sequence galaxy simulations display much stronger clustering on all scales than do the ETG simulations (up to an order of magnitude in $\xi$). The Milky Way-like and NGC~300-like simulations display substantial supernova clustering at all scales below $\Delta \sim 100$~pc, while the ETG simulations display supernova clustering only on much smaller scales, below $\Delta \sim 25$~pc.

The level of supernova clustering is clearly associated with the length-scale of gravitational instability $\lambda_{\rm T}$ in each disk (Figure~\ref{Fig::Romeo-Qs}), and thus by the epicyclic frequency $\kappa$, dependent on the rotational shear $\dd \Omega/\dd \ln{R}$ (Figure~\ref{Fig::vcircs}). In particular, the onset of clustering, indicated by the intercept of the black dashed line in Figure~\ref{Fig::SN-clustering}, occurs at approximately the Toomre length for each simulated disk. As seen in bottom row of Figure~\ref{Fig::Romeo-Qs}, this Toomre length-scale is substantially smaller in the ETG models, and this stricter limit from angular momentum on the outer scale of self-gravitating condensation is reflected in the lower level of SN clustering seen in Figure~\ref{Fig::SN-clustering}. A similar trend of increasing outflow strength with increasing instability scale (due to increasing box size) is also noted in Appendix A of~\cite{2020ApJ...900...61K}. \textbf{In other words, these galactic dynamics influence the clumpiness of the ISM (the free-fall times of the most massive GMCs, see Figure~14 of~\citealt{2024MNRAS.527.7093J}) and therefore the clumpiness of supernova explosions.}

\subsection{Galactic outflows}
\begin{figure}
\includegraphics[width=.95\linewidth]{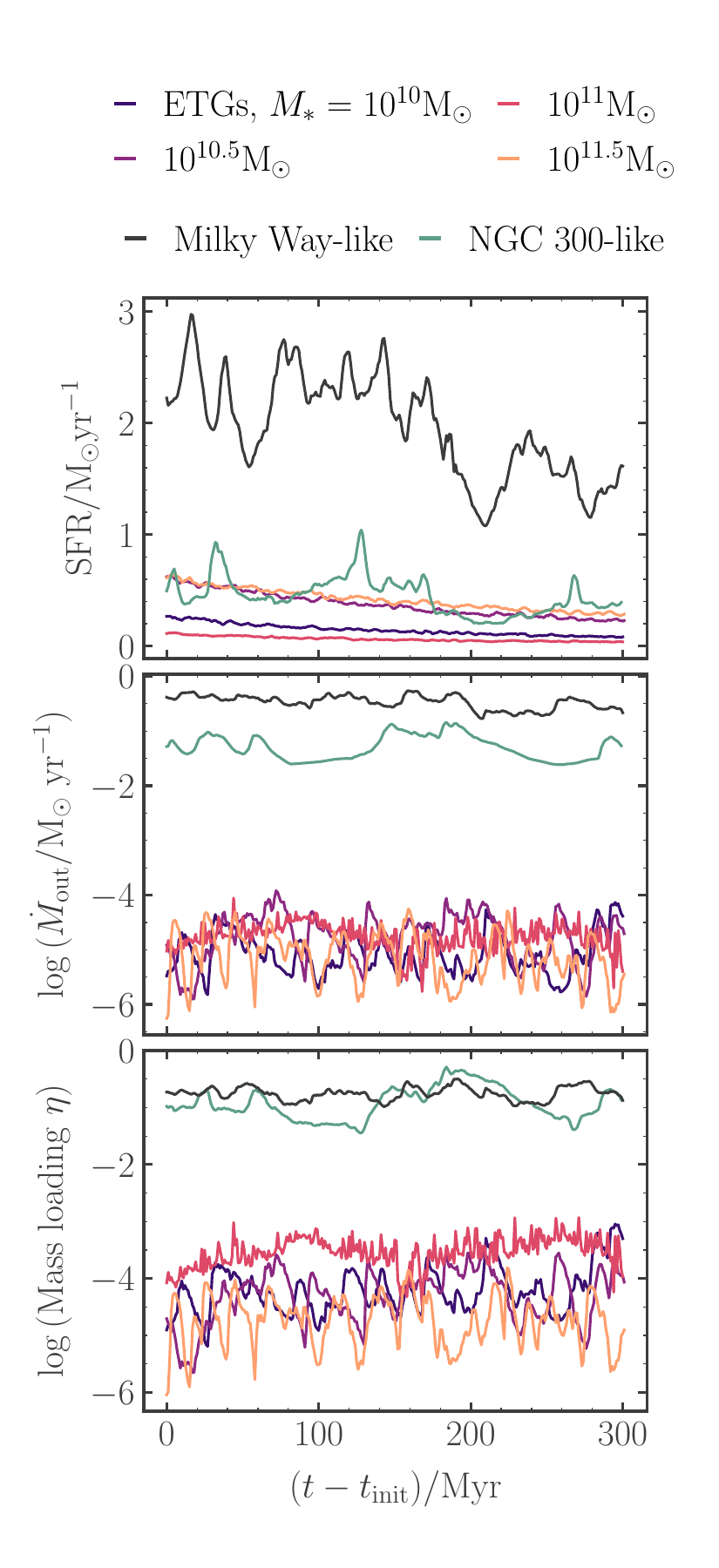}
  \caption{Global galactic star formation rate (upper panels), gas outflow rate (central panels) and mass-loading of outflows (lower panels) as a function of time. We begin tracking each property only when the gas disks have reached a state of dynamical equilibrium: $100$~Myr onwards for the ETG simulations, $300$~Myr onwards for the Milky Way-like simulation, and $500$~Myr onwards for the NGC~300-like simulation.}
  \label{Fig::outflow-rate}
\end{figure}

Figure~\ref{Fig::outflow-rate} shows the total galactic star formation rate (${\rm SFR}$, top panel), the rate of gas outflow ($\dot{M}_{\rm out}$, center panel) and the mass-loading $\eta$ of the galactic outflows in each of our simulated galaxies, as a function of simulation time. We begin tracking each property only after each disk has reached a state of dynamical equilibrium. The outflow rates are calculated as the total momentum (volume) density of the gas moving away from the disk, integrated over the area of two planar slabs of thickness 500 pc, located at $\pm 1$~kpc above and below the galactic disk, i.e. the mass flux $\dot{M}_{\rm out} = \int \dd A \: \rho v_z$. The mass-loading divides this outflow rate by the star formation rate.

\begin{figure*}
  \includegraphics[width=\linewidth]{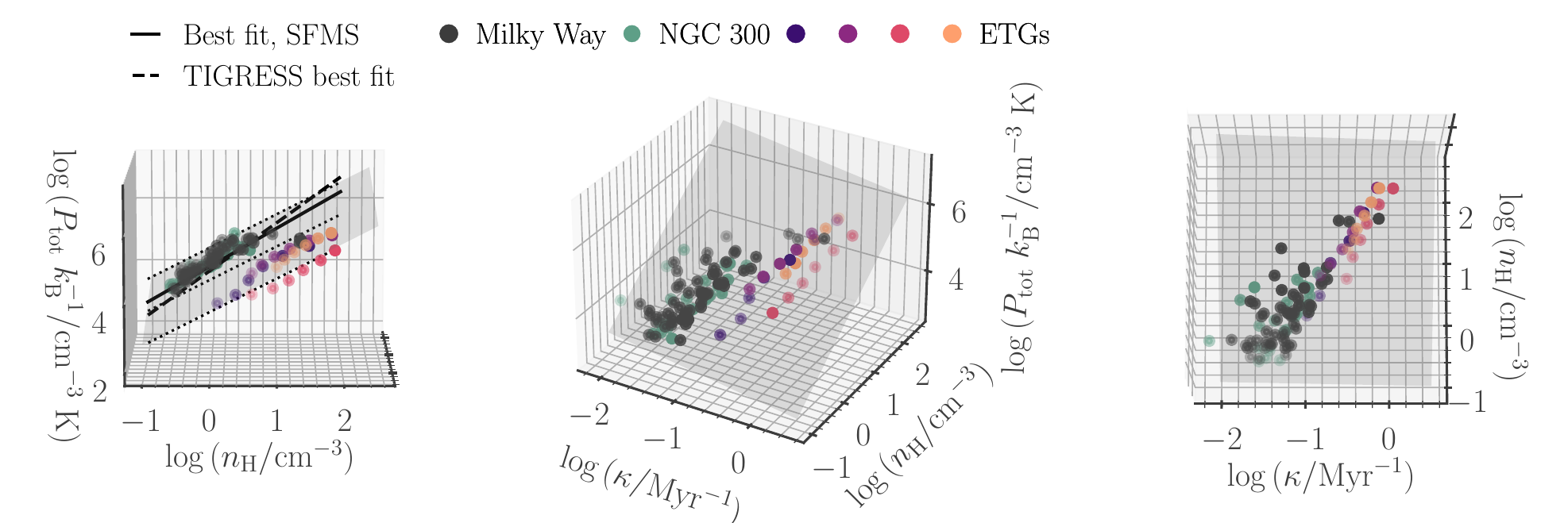}
  \caption{Total mid-plane pressure $P_{\rm tot}$ as a function of the mid-plane gas volume density $n_{\rm H}$ and the epicyclic frequency $\kappa$ for the cold+warm gas ($T<2 \times 10^4$~K) in our simulations, viewed in three different planes. The left-hand panel shows the usual equation of state ($P_{\rm tot}$ vs. $n_{\rm H}$) for our simulated galaxies. The best fit line to the Milky Way and NGC300 (star-forming main sequence) galaxy simulations is given by the solid black line, and the best-fit to the TIGRESS simulations is indicated by the thick dashed line. Three isotherms of constant $T = P_{\rm tot}/n_{\rm H} k_{\rm B}$ are given by grey dotted lines. Filled data points represent median values over time and azimuthal angle for each simulated galaxy, measured within overlapping radial annuli of width $500$~pc. The grey transparent plane shows the best fit to all data, excluding the dynamically-suppressed galaxy (pink).}
  \label{Fig::eEoS}
\end{figure*}

The strength and mass-loading of the outflows displays a very large difference of around 3-4 orders of magnitude between the star-forming main sequence galaxy simulations and the ETG simulations. This difference could be attributed to two factors: (1) to the the increased levels of supernova clustering in the Milky Way-like and NGC300-like simulations (Figure~\ref{Fig::SN-clustering}) and (2) to the shallower gravitational potential wells of these galaxies. A shallower gravitational potential well, measured perpendicular to the galactic mid-plane, decreases the escape speed perpendicular to the mid-plane.

The role of supernova clustering \textit{alone} in driving strong galactic outflows can be quantified by comparing the Milky Way-like galaxy and the ETG of stellar mass $M_* = 10^{10.5}~{\rm M}_\odot$, which have similar average values of the gravitational potential (and thus escape speed) across the extents of their respective gas disks, at the distance of $1$~kpc from the galactic mid-plane at which the outflow is measured. Figure~\ref{Fig::outflow-rate} demonstrates that these two galaxies nevertheless have very different values of the outflow rate and mass-loading, which are therefore attributable solely to their very different levels of supernova clustering, mediated by the level of galactic rotation and the associated level of disk gravitational stability $Q$ and the Toomre wavelength.

\subsection{The effective equation of state of the cold+warm gas distribution}
The turbulent and thermodynamic state of the cold+warm gas reservoir ($T<2 \times 10^4$~K) in each galaxy can be described by an equation of state relating the total turbulent plus thermal gas mid-plane pressure $P_{\rm tot}$ to the mid-plane gas density $n_{\rm H} = \rho/(\mu m_p)$, where $\mu$ is the mean molecular weight. This equation of state is equivalent to a statement of the combined turbulent and thermal gas velocity dispersions, as $P_{\rm tot} \equiv P_{\rm th} + P_{z, {\rm turb}} \sim \rho (c_s^2 + \sigma_{z, {\rm turb}}^2 )$, where $c_s$  is the isothermal sound speed. The equation of state therefore depends on the momentum and energy injected by stellar feedback, which is a key driver of turbulence.

Because this cold+warm reservoir of star-forming gas is unresolved in cosmological simulations, it is necessary to parameterize the equation of state in terms of its properties on large, resolved scales. This fixed, `effective' equation of state (eEoS) therefore provides an effective pressure that accounts for the unresolved stellar feedback in the simulation.

In Figure~\ref{Fig::eEoS}, we demonstrate how such an eEoS can be derived from our high-resolution GalactISM simulations, with star-forming gas reservoirs resolved on scales of $1$ to $100$~pc in the warm-cold gas reservoir. Filled data points represent median values of the total mid-plane pressure $P_{\rm tot}$ and density $n_{\rm H}$, as well as the epicyclic frequency $\kappa$, within overlapping radial annuli of width $500$~pc. Our method for calculating $P_{\rm th}$, $P_{\rm turb}$ and $\rho$ is outlined in Appendix~\ref{App::P-and-W}.

\begin{figure*}
\begin{centering}
  \includegraphics[width=.95\linewidth]{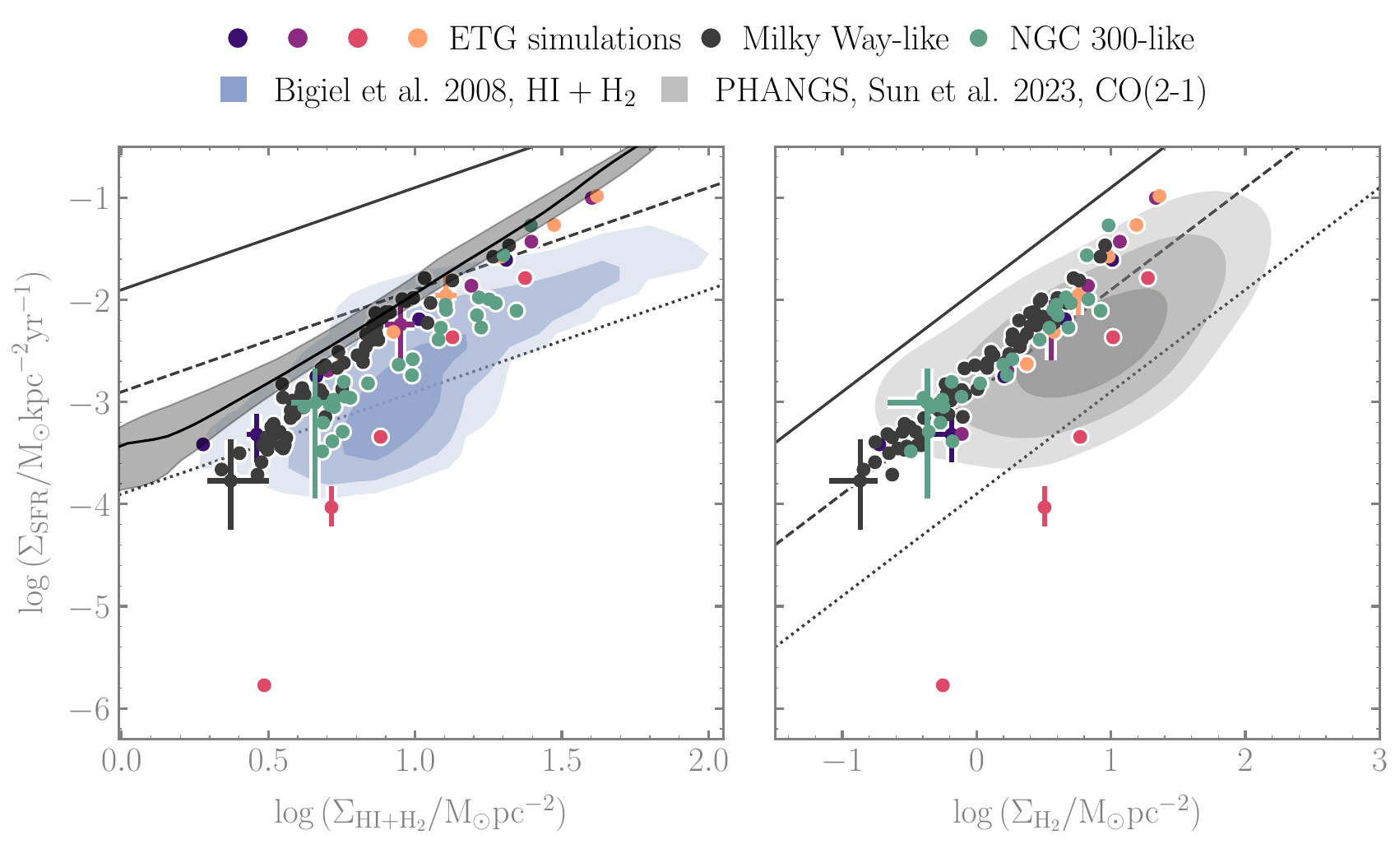}
  \caption{Star formation rate surface density $\Sigma_{\rm SFR}$ as a function of the cold gas surface density $\Sigma_{\rm HI+H_2}$ (left) and as a function of the molecular gas surface density $\Sigma_{\rm H_2}$ (right). Solid, dashed and dotted lines represent depletion times of $10^8$, $10^9$ and $10^{10}$~Gyr, respectively. Filled data points represent median values over time for each simulated galaxy, measured within overlapping radial annuli of width $500$~pc. For visual clarity, the corresponding interquartile ranges are shown at just one representative radius in each galaxy. The solid black line and shaded region represents the median and interquartile range of the star formation rate predicted by the model of~\protect\cite{Springel03}. The model is applied to all of our high-resolution snapshots at once, where each snapshot is degraded to a 3D resolution of $80$~pc: equal to the softening length used in TNG-50 (see Section~\protect\ref{Sec::Sigmagas-vs-SigmaSFR}). The blue and grey contours represent the 40\%-80\%-95\% levels of the observed observed galaxy samples from~\protect\cite{Bigiel08} and~\protect\cite{2023ApJ...945L..19S}, respectively, which assume a constant CO-to-H$_2$ conversion factor.}
  \label{Fig::obs-comparison-Sigmagas}
\end{centering}
\end{figure*}

In the left-hand panel of Figure~\ref{Fig::eEoS}, we show the standard two-dimensional eEoS, characterized by the variables $n_{\rm H}$ and $P_{\rm tot}$. The best fit to our two star-forming main sequence galaxies (the Milky Way-like and NGC~300-like simulations in grey and green, respectively) is given by a thick black line, and takes the form
\begin{equation} \label{Eqn::eEoS_2D}
\log{(P_{\rm tot}/k_{\rm B})} = 1.14\log{(n_{\rm H}/{\rm cm}^{-3})} + 4.42.
\end{equation}
This can be compared to the best fit eEoS in star-forming main sequence environments 
reported in \citet{2022ApJ...936..137O} from TIGRESS simulations, given by $\log{(P_{\rm tot}/k_{\rm B})} = 1.43\log{(n_{\rm H}/{\rm cm}^{-3})} + 4.30$, and denoted by the black dashed line. The two fits are in relatively good agreement, albeit with Equation~(\ref{Eqn::eEoS_2D}) having a slightly shallower slope than the TIGRESS fit. This difference reflects detailed differences between the feedback models, and is not unexpected.  In fact, with the updated TIGRESS-NCR framework, a shallower slope is obtained than with the original TIGRESS framework~\citep{2023ApJ...946....3K}.

In the central and right-hand panels of Figure~\ref{Fig::eEoS}, we show that the ETG simulations in our sample have different equations of state to the Milky Way-like and NGC~300-like simulations, overlapping only with the center-most regions of the main-sequence galaxies. As the level of galactic rotation is increased, the eEoS is shifted systematically towards lower gas velocity dispersions. That is, the support against gravitational collapse that is provided by galactic rotation allows gas to remain gravitationally-stable at higher densities and lower velocity dispersions, commensurate with the smaller Toomre wavelengths $\lambda_{\rm T}$, smaller scale-heights, and lower levels of supernova clustering, as reported in Figures~\ref{Fig::Romeo-Qs} and~\ref{Fig::SN-clustering}.

\textbf{Our simulations provide evidence that rotationally-supported regions in galaxies (including the central regions of the Milky Way-like galaxy) have reduced thermal and turbulent pressure, and potentially require an adjusted effective equation of state in cosmological simulations.} The grey transparent plane in Figure~\ref{Fig::eEoS} shows such a three-dimensional eEoS, characterized by the variables $n_{\rm H}$,  $P_{\rm tot}$ and $\kappa$, and fitted to all galaxies in the simulation suite, excluding the ETG with dynamically-suppressed star formation (pink data points). This best fit takes the form
\begin{equation} \label{Eqn::eEoS_3D}
\begin{split}
\log{(P_{\rm tot}/k_{\rm B})} &= -0.33\log{(\kappa/{\rm Myr}^{-1})} \\
&+ 1.03\log{(n_{\rm H}/{\rm cm}^{-3})} + 3.93.
\end{split}
\end{equation}
If $\kappa$ displays only a small variation between galaxies, then we obtain $\log{P_{\rm tot}} \propto 1.03 \log{n_{\rm H}}$, which is close to the best-fit two-dimensional eEoS reported in Equation~(\ref{Eqn::eEoS_2D}).

Finally, we note that the dynamically-suppressed ETG is offset from the plane characterizing the star-forming gas in the other five galaxies. In a future paper, we will investigate in detail how this behavior is related to the rate of galactic rotation, leading to a transition to a longer depletion time, lower gas velocity dispersion and smaller gas disk scale-height, that is non-linear in $\kappa$.

\section{Star formation regulation} \label{Sec::SF-regulation}
Power-law relationships between the star formation rate surface density $\Sigma_{\rm SFR}$ and other large-scale properties of galaxies provide important constraints for theories of galactic star formation. They also function as subgrid models for star formation in cosmological simulations, in which the cold+warm interstellar medium cannot be resolved. Most commonly, such sub-grid models are underpinned by the empirical power-law relationship between $\Sigma_{\rm SFR}$ and either the neutral gas surface density $\Sigma_{\rm HI+H_2}$, or the molecular gas surface density $\Sigma_{\rm H_2}$. Crucially, their slopes and normalizations are calibrated to observed samples of galaxies, limiting their predictive power and applicability to a diverse set of galactic environments. In this section we test a new, predictive sub-grid model for star formation in disk galaxies.

\begin{figure*}
\begin{centering}
  \includegraphics[width=.95\linewidth]{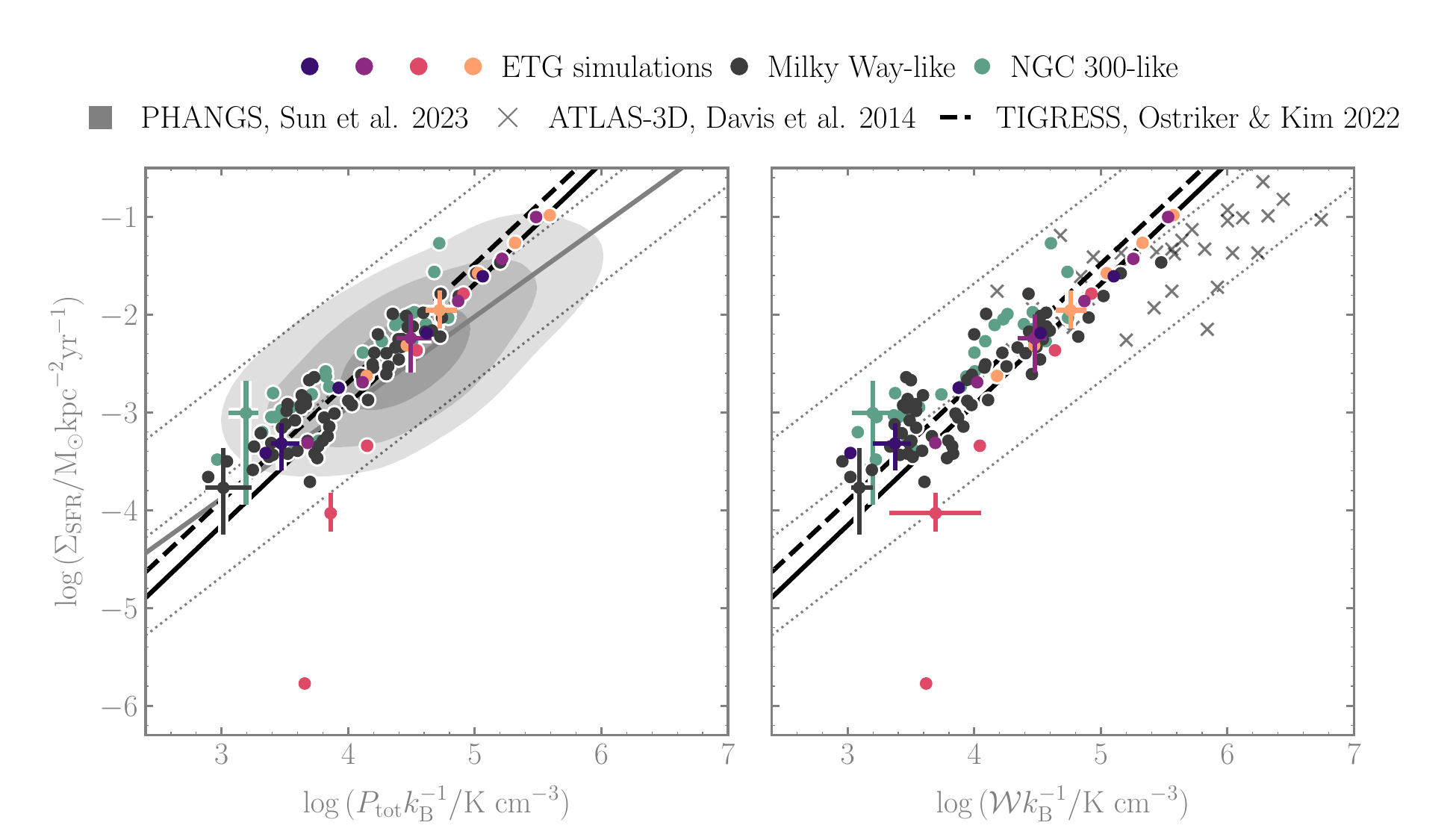}
  \caption{Star formation rate surface density $\Sigma_{\rm SFR}$ as a function of the total mid-plane pressure $P_{\rm tot}$ (left) and as a function of the interstellar medium weight $\mathcal{W}$ (right, see Section~\protect\ref{Sec::P-vs-SigmaSFR}). Black dotted lines represent constant ratios of $\Sigma_{\rm SFR}/P_{\rm tot} = 10^2$, $10^3$ and $10^4~{\rm km~s}^{-1}$. Only the cool-warm gas ($T<2 \times 10^4$~K) gas that is gravitationally unbound ($\alpha_{\rm vir}>2$) is included in the calculation. Filled data points represent median values over time for each simulated galaxy, measured within overlapping radial annuli of width $500$~pc. For visual clarity, the corresponding interquartile ranges are shown at just one representative radius in each galaxy. The thick black line represents the best linear regression fit to these data points, excluding the dynamically-suppressed galaxy (pink). The thick dashed line represents the corresponding best fit from the TIGRESS simulations~\protect\citep{2022ApJ...936..137O}. The grey contours represent the 40\%-80\%-95\% levels of the observed galaxy sample from~\protect\cite{2023ApJ...945L..19S}, and the grey crosses represent average values across the gas disks of the ATLAS$^{\rm 3D}$ galaxy sample from~\protect\cite{2014MNRAS.444.3427D}, also shown in Figure~\ref{Fig::ICs}. We note that~\protect\cite{2023ApJ...945L..19S} strictly measure $\mathcal{W}$ to obtain $P_{\rm tot}$, but we include this data in the left-hand panel for readability.}
  \label{Fig::obs-comparison-pressures}
\end{centering}
\end{figure*}

\subsection{Gas surface density vs. star formation rate surface density} \label{Sec::Sigmagas-vs-SigmaSFR}
The most common subgrid model for star formation in cosmological simulations sets a depletion time of $\tau_{\rm dep} = m_{\rm cell}/{\rm SFR}_{\rm cell} = \tau_{\rm dep, 0} (\rho_{\rm thresh}/\rho_{\rm gas})^{0.5}$ per star-forming gas cell of mass $m_{\rm cell}$, where $\rho_{\rm gas}$ is the volume density of the gas, $\rho_{\rm thresh}$ is the density above which star formation is allowed to occur, and $\tau_{\rm dep, 0}$ is the gas depletion time at this threshold. The resulting inverse proportionality between the star formation rate (SFR) and the gas free-fall time $\sqrt{3\pi/32G\rho_{\rm gas}}$ is in rough agreement with a sample of 21 observed spiral galaxies at low redshift from~\cite{Kennicutt1998}, which follow the power-law $\Sigma_{\rm SFR} \propto \Sigma_{\rm HI+H_2}^{1.4}$, averaged across galactic disks (the `Schmidt-Kennicutt relation'). This model is used in Illustris-TNG~\citep{Springel03,2013MNRAS.436.3031V}, and is qualitatively very similar to the models used in other large cosmological simulations, such as EAGLE~\citep{2015MNRAS.446..521S}.\footnote{We note that the EAGLE simulations set ${\rm SFR}_{\rm cell}/m_{\rm gas}$ according to a power-law in the gas pressure, but in combination with their effective equation of state $P\propto \rho_{\rm gas}^{4/3}$ above $n_H=0.1~{\rm cm}^{-3}$, the resulting relation is ${\rm SFR}_{\rm cell}/m_{\rm cell} \propto \rho_{\rm gas}^{0.3}$, and the normalization of this relationship is again set according to~\protect\cite{Kennicutt1998}.}

The solid black line and shaded region on the left-hand side of Figure~\ref{Fig::obs-comparison-Sigmagas} represent the median and interquartile range of this subgrid model, when applied to the outputs of our GalactISM simulations, degraded to the TNG-50 resolution of 80 pc. For comparison, the filled data points represent the true median values of $\Sigma_{\rm HI+H_2}$ and $\Sigma_{\rm SFR}$ in each simulation, within overlapping radial annuli of width $500$~pc. The blue contours in the left-hand panel represent the sample of 18 galaxies observed at $750$~pc by~\cite{Bigiel08}, and the grey contours in the right-hand panel represent the sample of 80 galaxies observed at $1.5$~kpc resolution by~\cite{2023ApJ...945L..19S}.

We note that the slight ($\approx 0.3$~dex) overestimate of the star formation rate surface density $\Sigma_{\rm SFR}$ for our Milky Way-like simulation, relative to the 95\% confidence level of the~\cite{Bigiel08} values, is likely due to differences in the stellar gravity for given $\Sigma_{\rm HI+H_2}$. Firstly, the {\sc Agora} initial condition has a smaller stellar disk scale-height than the Milky Way-like galaxies observed by~\cite{Bigiel08}, as we mentioned in Section~\ref{Sec::ICs-MW}. Secondly, the fraction of ionized gas is enhanced, as seen in Figure~\ref{Fig::phases}, and thus $\Sigma_*/\Sigma_{\rm HI+H_2}$ is slightly increased, relative to these observed galaxies.

At a given $\Sigma_{\rm HI+H_2}$, the effect of decreasing the stellar disk scale-height and increasing $\Sigma_*$ is to vertically compress the gas at the galactic mid-plane, increasing its volume density, pressure and star formation rate. In addition, the deviation between simulations and observations at high gas surface densities may be increased by the uncertainty in the CO-to-$H_2$ conversion factor used in~\cite{Bigiel08}, since a constant CO-to-H$_2$ conversion factor tends to  overestimate $\Sigma_{\rm H_2}$ in dense galactic centers (see below).

We see that the~\cite{Springel03} sub-grid model provides a reasonable approximation to the median resolved star formation rate surface density across the inner regions of the Milky Way-like galaxy simulation (black data points), but over-estimates the star formation rate in the outer regions, and fails to capture the variation in depletion time across the NGC~300-like and ETG simulations. Similarly, it does not capture the spread of resolved depletion times in the~\cite{Bigiel08} observations. This is perhaps unsurprising, considering that the slope and normalization of the power-law sub-grid model are calibrated to galaxy-averaged values in nearby spirals, with physical properties closest to those of our Milky Way-like simulation.

An alternative empirical sub-grid model for star formation in cosmological simulations sets the depletion time according to the molecular gas volume density, which in turn is computed via the sub-grid model of~\cite{2011ApJ...729...36K}. Similarly to the Illustris-TNG model, the slope and normalization of the power-law are set according to the~\cite{Kennicutt1998} galaxy sample. This approach is used in the MUFASA~\cite{2016MNRAS.462.3265D} and SIMBA~\citep{2019MNRAS.486.2827D} simulations, and takes advantage of the relatively constant slope of the relationship between the molecular gas surface density $\Sigma_{\rm H_2}$ and star formation rate surface density, shown on the right-hand side of our Figure~\ref{Fig::obs-comparison-Sigmagas}. The trend across five out of six of our simulated galaxies can be well modeled by a linear relationship between $\Sigma_{\rm SFR}$ and $\Sigma_{\rm H_2}$, with a roughly constant molecular gas depletion time of $1$~Gyr. Aside from a slight upturn in $\Sigma_{\rm SFR}$ relative to $\Sigma_{\rm H_2}$ at high surface densities, the simulated data are in good agreement with recent observations of the molecular gas distribution across a sample of 80 nearby galaxies at $1.5$~kpc resolution~\citep[][green contours]{2023ApJ...945L..19S}.

However, although $\Sigma_{\rm H_2}$ appears to be a better predictor of $\Sigma_{\rm SFR}$ than $\Sigma_{\rm HI+H_2}$~\citep[see also e.g.][]{Bigiel11}, the proportionality between these variables is still an empirical relationship that is calibrated to a set of observations. This relationship is not derived from first principles, so is not predictive in new galactic environments. Furthermore, new observations of CO isotopologues across nearby galaxies are now revealing substantial variations in the CO-to-${\rm H}_2$ conversion factor $\alpha_{\rm CO}$ (used to derive $\Sigma_{\rm H_2}$) between galaxy disks and galaxy centers~\citep{2023A&A...676A..93D,2023ApJ...950..119T}. These results strongly imply that the proportionality between $\Sigma_{\rm SFR}$ and $\Sigma_{\rm H_2}$ is not as universal as previously thought, even across the population of nearby galaxies. In fact, if the variation in the CO-to-${\rm H}_2$ conversion factor is taken into account, the upturn in $\Sigma_{\rm SFR}$ relative to $\Sigma_{\rm H_2}$ that is seen in our simulations at high surface densities would be retrieved in observations~\citep{2023ApJ...950..119T}. More generally, observational evidence has demonstrated that the correlation of star formation with chemical composition (such as CO or HCN) is much weaker when environmental variation is taken into account~\citep[e.g.][]{2018ApJ...858...90G}.

An alternative approach to using empirical relations for sub-grid star formation rates is to predict these star formation rates via a theoretical model. In the next sub-section, we compare our results to the predictions of the pressure-regulated feedback-modulated star formation theory, calibrated to the TIGRESS simulations~\cite{2022ApJ...936..137O}.

\subsection{Mid-plane pressure vs. star formation rate surface density} \label{Sec::P-vs-SigmaSFR}
Recent analyses of a large sample of main sequence galaxies from the PHANGS-ALMA sample~\citep{Leroy2021a} have demonstrated a close correlation between the kpc-scale mid-plane pressure $P_{\rm tot}$ of gas disks in dynamical equilibrium, and the galactic star formation rate surface density $\Sigma_{\rm SFR}$~\citep{2023ApJ...945L..19S}, as well as the fraction of dense and self-gravitating molecular gas~\citep{Sun2020}.

Such a relationship between the star formation rate and the mid-plane pressure is a central tenet of `pressure-regulated' theories of star formation~\citep{Ostriker+10,OstrikerShetty2011,2011ApJ...743...25K}. In this theoretical framework \citep[see][]{2022ApJ...936..137O}, the thermal, turbulent, and magnetic pressures in the diffuse ISM are driven by stellar feedback, as offset by dissipation and cooling~\citep[see also related work by][on the balance between momentum injection by feedback, and turbulent dissipation]{2005ApJ...630..167T,Hopkins11,2013MNRAS.433.1970F}. That is, there is a causal relationship whereby higher star formation rates $\Sigma_{\rm SFR}$, associated with higher fractions of dense, gravitationally-bound gas, produces more feedback and this leads to higher total pressures $P_{\rm tot}.$ In the Ostriker-Kim theory, the sum of the pressures $P_{\rm tot}$ must also satisfy vertical dynamical equilibrium by balancing the ISM weight at the midplane. As a result, $\Sigma_{\rm SFR}$ is expected to be directly proportional to the diffuse ISM weight, and inversely proportional to the total `feedback yield.'

The predicted relationship between $P_{\rm tot}$ and $\Sigma_{\rm SFR}$ is reproduced in simulations of stratified boxes representing a range of observable galactic environments~\citep[e.g.][]{2013ApJ...776....1K,KimCG&Ostriker15b,2022ApJ...936..137O}. The required condition of vertical dynamical equilibrium is also demonstrated in these simulations, and additionally in the Milky Way-like galaxy of~\cite{Benincasa16}, and FIRE cosmological zoom simulations analyzed by \citet{2020MNRAS.498.3664G}. The $P_{\rm tot}$-$\Sigma_{\rm SFR}$ relation has not yet been investigated in high-resolution isolated galaxy simulations spanning diverse galactic environments outside of the star-forming main sequence.

\begin{figure*}
\begin{centering}
  \includegraphics[width=1.05\linewidth]{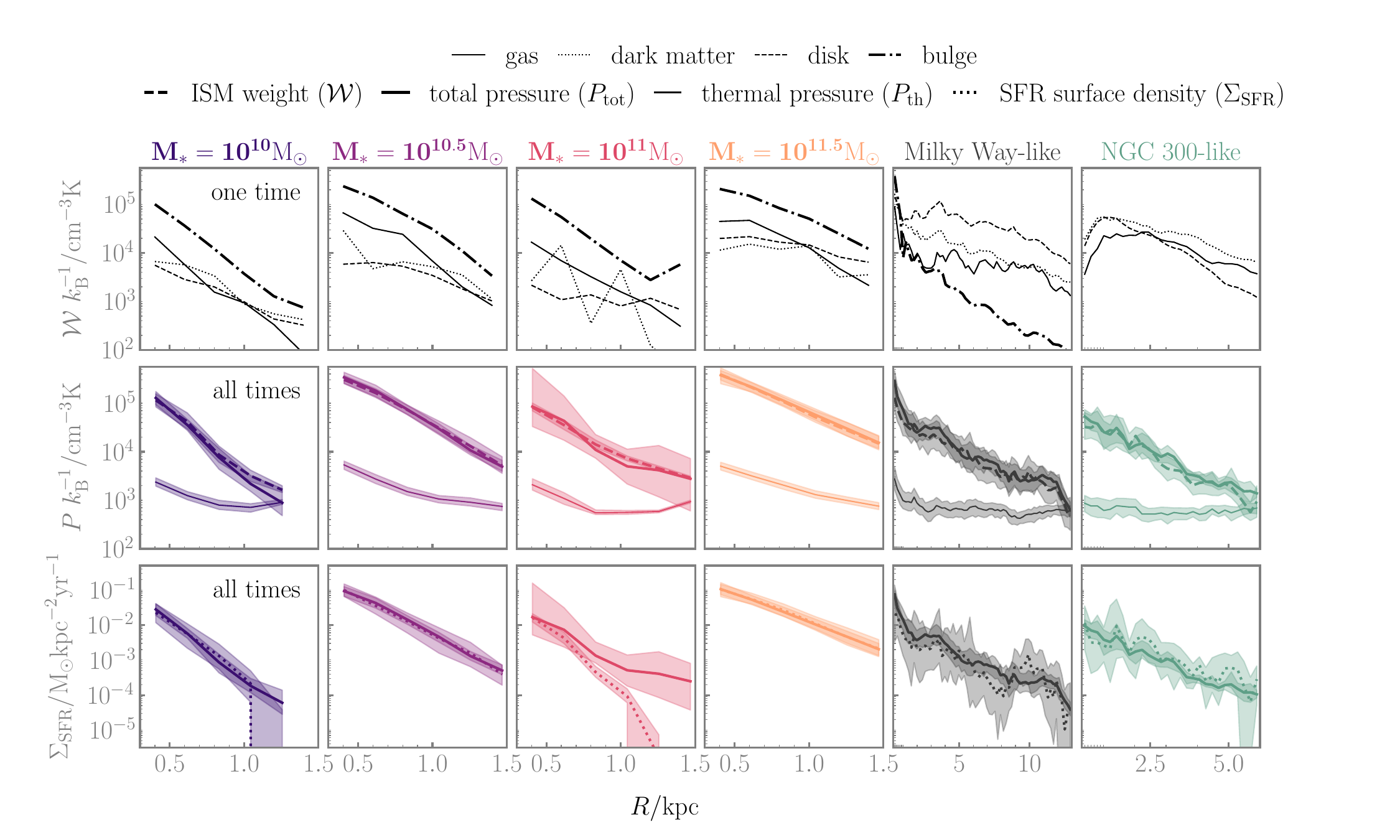}
  \caption{\textit{Top:} The ISM weight $\mathcal{W}$ due to each component of the gravitational potential, as a function of the galactocentric radius, at the final analyzed time of each simulation. \textit{Center:} Comparison of the total mid-plane pressure (solid lines) and the total ISM weight (dashed lines), along with median values of the thermal mid-plane pressure (thin lines) as a function of galactocentric radius. Only the cool-warm gas ($T< 2 \times 10^4$~K) gas that is gravitationally unbound ($\alpha_{\rm vir}>2$) is included in the calculation of the pressures and star formation rates. \textit{Bottom:} Comparison of the true star formation rate surface density $\Sigma_{\rm SFR}$ (dotted lines) and the star formation rate surface density predicted by Equation~\protect\ref{Eqn::Upsilon-fit}. The bottom two rows are median values over time and azimuthal angle, and all shaded areas are the corresponding interquartile ranges.}
  \label{Fig::rotcurves}
\end{centering}
\end{figure*}

The left-hand side of Figure~\ref{Fig::obs-comparison-pressures} shows the relationship between $P_{\rm tot}$ and $\Sigma_{\rm SFR}$ within overlapping radial annuli of width $\sim 500$~pc, for diffuse gas in our simulated galaxies.\footnote{`Diffuse' is equivalent to the cold+warm, gravitationally-unbound gas reservoir with $T\la 10^4$~K and $\alpha_{\rm vir}>2$.  This gas accounts for the majority of the interstellar medium by mass, and in our simulations we calculate $P_{\rm tot}$ for this cold+warm gas reservoir.} Our method for calculating the total (turbulent plus thermal) mid-plane pressure is described in Appendix~\ref{App::P-and-W}. The best fit to these data (thick black line), excluding the dynamically-suppressed galaxy (pink data points) is given by
\begin{equation} \label{Eqn::Upsilon-fit}
\log_{10}{\Big(\frac{\Sigma_{\rm SFR}}{{\rm M}_\odot {\rm kpc}^{-2} {\rm yr}^{-1}}\Big)} = 1.235 \log{\Big(\frac{P_{\rm tot}k_{\rm B}^{-1}}{{\rm cm}^{-3}{\rm K}}\Big) - 7.86}.
\end{equation}
This best fit shows good agreement with the observed relationship between $\Sigma_{\rm SFR}$ with $P_{\rm tot}$ across a sample of 80 nearby galaxies at $1.5$~kpc scales~\citep{2023ApJ...945L..19S}, in which star formation rates are traced by ${\rm H}\alpha+22\mu{\rm m}$ emission and gas is traced by 21 cm + CO(2-1) emission, with a CO(2-1)/(1-0) conversion factor of 0.65 and a constant CO-to-${\rm H_2}$ factor $\alpha_{\rm CO}= 4.35 M_\odot \ \pc^{-2} ({\rm K}\ \kms)^{-1}$.\footnote{In practice,~\cite{2023ApJ...945L..19S} calculate the ISM weight $\mathcal{W}$, as opposed to the pressure $P_{\rm tot}$, but we have placed the grey contours on the left-hand side of Figure~\ref{Fig::obs-comparison-pressures} for visual clarity.} The resulting logarithmic correlation between $P_{\rm tot}$ and $\Sigma_{\rm SFR}$ for the observed galaxies has a slope of $0.93$ and a normalizing coefficient of $-6.95$, with estimated upper-limit uncertainties of around $25$~per~cent and $0.20$~dex, respectively.

The slightly steeper slope in our simulations may be due a number of effects. One is that the value of the CO-to-${\rm H_2}$ conversion factor $\alpha_{\rm CO}$ is known to decrease in  high-pressure (and high-surface density) regions of galaxies~\citep[e.g.][and references therein]{2023ApJ...950..119T}, and is thus not well represented by a constant $\alpha_{\rm CO}$. In fact, using the variable $\alpha_{\rm CO}$ of~\citet{Bolatto13}, \citet{2023ApJ...945L..19S} find the slope increases to 1.08. Other possible reasons for the small discrepancy in slopes are (1) differences in the gas and star formation reservoirs that we have analyzed, relative to those traced by CO and ${\rm H}\alpha$ emission in the observations, (2) a small under-estimate of the momentum provided by feedback at high pressures/densities, in our simulated galaxies, and (3), variation in the supernova feedback momentum yield, according to differences in the galactic environments we have modeled, as noted by~\cite{Martizzi15} and~\cite{2017MNRAS.465.1682H}, among others.

We obtain a similar slope and normalization to that of the TIGRESS simulations, which are 1.21 and -7.66 respectively, and given by the dashed line in the figure. The slightly reduced normalization factor may be attributable to the different feedback model used in our simulations, or to the presence of radial mass transport, which is not present in stratified box simulations.

Excluding the dynamically-suppressed galaxy, the correlation between $\Sigma_{\rm SFR}$ and $P_{\rm tot}$ in our simulations is tight, indicating that \textbf{the mid-plane pressure is strongly correlated with the star formation rate, in line with theoretical expectations.} We show explicitly in Appendix B and Figure~\ref{Fig::pressures-vs-weight} that the mid-plane pressure can be approximated by the interstellar medium weight $\mathcal{W}$ across our simulation suite, given by
\begin{equation}
P_{\rm tot} \sim \mathcal{W} = \int^{z_{\rm max}}_0 {\rho_{\rm gas}(z) \frac{\partial \Phi}{\partial z} \dd z},
\end{equation}
where $z_{\rm max}$ is the maximum extent of the gas disk, $\rho_{\rm gas}$ is the gas volume density, and $\Phi$ is the gravitational potential due to the entire distribution of gas, stars and dark matter. The right-hand panel of~\ref{Fig::obs-comparison-pressures} shows $\Sigma_{\rm SFR}$ vs $\mathcal{W}$, since $\mathcal{W}$ is more readily accessible in observations than a direct measure of $P_{\rm tot}$. The filled circles again represent measured values within annular radial bins, and grey crosses represent estimated values across the gas disks of the ATLAS$^{\rm 3D}$ sample, with a similar spread of $\Sigma_{\rm SFR}/P_{\rm tot}$ values. Our method for calculating $\mathcal{W}$ in both the simulations and observations is described in Appendix~\ref{App::P-and-W}.  

It is worth noting that the ATLAS$^{\rm 3D}$ values of $\Sigma_{\rm SFR}/P_{\rm tot}$ in the right-hand panel of Figure~\ref{Fig::obs-comparison-pressures} display relatively good agreement with all four ETG simulations, including the dynamically-suppressed galaxy simulation. That is, they extend to lower disk-averaged star formation rates per unit ISM weight than do the spiral galaxies in the left-hand panel of Figure~\ref{Fig::obs-comparison-pressures}. This finding is consistent with the suppression of the star formation efficiency in observed ETGs~\citep[e.g.][]{2014MNRAS.444.3427D} by dynamical suppression~\citep[e.g.][]{2009ApJ...707..250M,2020MNRAS.495..199G}.

Figure~\ref{Fig::obs-comparison-pressures} therefore demonstrates that, averaged over time, and with the exception of the dynamically-suppressed galaxy, the gas disks of our simulated galaxies have $\Sigma_{\rm SFR}$ strongly correlated with $P_{\rm tot}\approx\mathcal{W}$. Because the ISM weight $\mathcal{W}$ can be calculated in terms of the large-scale properties of galaxies~\citep[see Section 3 of][for details]{Hassan2023}, Equation~(\ref{Eqn::Upsilon-fit}) can therefore be used to model star formation in cosmological simulations, so long as the assumption of dynamical equilibrium holds, and dynamical suppression of the star formation efficiency is not present.

In Figure~\ref{Fig::rotcurves}, we show the same data as is presented in Figure~\ref{Fig::obs-comparison-pressures}, but in greater detail for each simulated galaxy, with interquartile ranges over time and azimuthal angle at each galactocentric radius (transparent shaded regions). The separate contributions to the ISM weight $\mathcal{W}$ made by the gas disk (solid lines), dark matter halo (dotted lines), stellar disk (dashed lines) and stellar bulge (thick dot-dashed lines), at one single simulation time, are shown in the top row, clearly demonstrating that $\mathcal{W}$ is dominated by the stellar bulge in the ETG simulations, by the stellar disk in the Milky Way-like simulation, and by the disk and dark matter halo in the NGC~300-like simulation. We note that the strong dominance of the stellar disk weight contribution over the gas disk contribution in the Milky Way-like disk may not reflect the true balance of weights in the Milky Way, due to the small scale-height of the stellar disk in the Agora initial condition ($\sim 100$~pc), relative to the true value in the Milky Way ($\sim 300$~pc).

In the center row of Figure~\ref{Fig::rotcurves}, we show that a state of dynamical equilibrium is maintained across all of the galaxy simulations, with close overlap between the mid-plane pressure (thick solid lines) and the ISM weight (thick dashed lines). The thermal pressure (thin lines) is also shown, for comparison. The lack of an increase in $P_{\rm th}$ towards the inner parts of the Milky Way and NGC~300 models is likely due to the spatially- and temporally-constant radiation field we have adopted in our simulations. A more realistic model would have a radiation field that increases at higher $\Sigma_{\rm SFR}$, but would still produce a value of $P_{\rm th}$ that is sub-dominant to $P_{\rm turb}$ in all but very low-weight environments~\citep[see][]{2022ApJ...936..137O}. Since $P_{\rm th} \ll P_{\rm turb}$ in the star-forming gas of the supersonic interstellar medium, this should not affect the robustness of our results.

Finally, the bottom row of Figure~\ref{Fig::rotcurves} shows the correspondence between the measured star formation rate surface density $\Sigma_{\rm SFR}$ for each simulation as a function of galactocentric radius, and the value predicted via Equation~(\ref{Eqn::Upsilon-fit}). The same powerlaw relation between $P_{\rm tot}$ and $\Sigma_{\rm SFR}$ manifestly holds across five out of six galaxies, excluding the dynamically-suppressed ETG.

\textbf{We therefore find that Equation~(\ref{Eqn::Upsilon-fit}) holds promise as a predictive, first-principles} subgrid model for star formation in cosmological simulations, across a range of star-forming main sequence and bulge-dominated, quenched galaxy environments. This model provides an estimate for the star formation rate based on the theoretical prediction of~\cite{2022ApJ...936..137O}, rather than a fit to a relatively small sample of nearby main sequence spirals, as is the current state-of-the-art in cosmological simulations. It therefore has the potential to make reliable predictions of the star formation rate across a larger range of environments, particularly at high redshift.

\textbf{By contrast, the star formation rate in the dynamically-suppressed galaxy is decoupled from pressure regulation, as both the mid-plane turbulent pressure and galactic rotation appear to provide substantial support against gravitational collapse of the cold, star-forming gas.} It is likely that the star formation rate in this galaxy is determined not just by the mid-plane gas pressure, but also by the rate of galactic rotation. We will discuss this galaxy in detail in the second paper of this series, along with ways to incorporate the role of multiple physical mechanisms in driving (and thus predicting) the star formation rate.

\section{Discussion and summary} \label{Sec::conclusion}
In this work we have presented six high-resolution chemo-dynamical simulations of galaxies spanning the dynamical environments of star-forming main sequence and quenched (low specific SFR) galaxies, with a wide range of bulge-to-disk ratios, and so a wide range of epicyclic frequencies. We have investigated their global properties: the regulation of star formation, the gas phase distribution and the gas phase morphology, related to the gravitational potential, the clustering of supernovae and the driving of galactic outflows. We have found that varying the gravitational potential produces a large range of gas morphologies, phase structures, and star formation rates, which are broadly in agreement with observations. We have quantified these variations with a view to modeling such environments in cosmological simulations. We can summarize our results as follows:
\begin{enumerate}
  \item The level of supernova clustering, and thus the mass-loading $\eta$ of galactic outflows, is strongly-coupled to the rate of galactic rotation $\kappa$, via the Toomre length-scale for each disk. That is, higher rotation rates prevent large clusters from forming, such that star formation occurs in larger numbers of smaller clouds, which do not break out of the disk. The result is that the bulge-dominated galaxies have galactic outflows with mass-loadings reduced by four orders of magnitude, relative to the star-forming main sequence galaxies.
  \item The equation of state (density vs.~pressure) of the star-forming gas depends on its turbulent velocity dispersion, so also varies strongly with the level of supernova clustering and the rate of galactic rotation. The cold+warm gas reservoir in the ETGs has a higher density, lower velocity dispersion, and higher molecular gas fraction than in the main sequence galaxies.
  \item Aside from one dynamically-suppressed ETG, with the highest epicyclic frequency $\kappa$, the mid-plane pressure is strongly associated with the star formation rate surface density $\Sigma_{\rm SFR}$ across both main sequence and early type environments, in agreement with~\cite{OstrikerShetty2011,2022ApJ...936..137O}. The relationship is tighter than that between $\Sigma_{\rm SFR}$ and gas surface density.
\end{enumerate}

Our conclusions have important implications for the modeling of star formation and stellar feedback in cosmological simulations:
\begin{enumerate}
  \item The relationship between the mid-plane gas pressure $P_{\rm tot}$ and the star formation rate surface density $\Sigma_{\rm SFR}$ represents an improved model for star formation across star-forming main sequence and quenched galactic environments, relative to the Schmidt-Kennicutt relation.
  \item Across galactic environments with widely-varying levels of galactic rotation and epicyclic frequency $\kappa$, as seen across our GalactISM simulations, the equation of state between the gas density and the pressure varies strongly with $\kappa$. The value of $\kappa$ should therefore be taken into account when modeling the pressure of the star-forming gas (via the effective equation of state), and the mass-loading of galactic outflows, in cosmological simulations.
  \item The onset of dynamical suppression introduces a transition away from pressure-regulated star formation, which depends non-linearly on the rate of galactic rotation $\kappa$. This behavior represents the greatest challenge for parameterization in terms of galaxy properties that are resolved in cosmological simulations.
\end{enumerate}
We are optimistic that, like the effective equation of state, the mass-loading $\eta$ of galactic outflows, and even the onset of dynamical suppression of the SFE, can be parameterized systematically in terms of large-scale galaxy properties such as the interstellar medium weight $\mathcal{W}$, and $\kappa$. Across our galaxy sample, the variations in $\eta$, the degree of supernova clustering and the SFE are not described by simple power-laws. However, given a large number of high-resolution galaxy simulations across diverse galactic environments, from simulation suites such as GalactISM and TIGRESS, it might be possible to constrain these quantities via data-driven or machine learning techniques. An investigation of this possibility will in the future be enabled by the expertise in statistical modeling present in the Learning the Universe collaboration.

\section*{Acknowledgements}
We thank an anonymous referee for their thorough and attentive reading of the paper, which improved the clarity of the text and conclusions. SMRJ is supported by Harvard University through an Institute of Theory and Computation Fellowship. The work of ECO and CGK is supported by grant 10013948 from the Simons Foundation to Princeton University, sponsoring the Learning the Universe collaboration. JG gratefully acknowledges financial support from the Swiss National Science Foundation (grant no CRSII5\_193826). G.L.B. acknowledges support from the NSF (AST-2108470, ACCESS), a NASA TCAN award, and the Simons Foundation. We thank Volker Springel for providing us access to Arepo, and we thank Mark Krumholz and Romain Teyssier for helpful discussions. We thank Jiayi Sun for the use of his script for plotting beautiful density contours for Figures~\ref{Fig::obs-comparison-Sigmagas} and \ref{Fig::obs-comparison-pressures}.

\section*{Software}
The code used to analyze the simulations in this work is available at \url{https://github.com/sjeffreson/pressure_regulated_SF_analysis}.


\bibliography{bibliography}{}

\begin{thebibliography}{}
\expandafter\ifx\csname natexlab\endcsname\relax\def\natexlab#1{#1}\fi
\providecommand{\url}[1]{\href{#1}{#1}}
\providecommand{\dodoi}[1]{doi:~\href{http://doi.org/#1}{\nolinkurl{#1}}}
\providecommand{\doeprint}[1]{\href{http://ascl.net/#1}{\nolinkurl{http://ascl.net/#1}}}
\providecommand{\doarXiv}[1]{\href{https://arxiv.org/abs/#1}{\nolinkurl{https://arxiv.org/abs/#1}}}

\bibitem[{{Agertz} {et~al.}(2013){Agertz}, {Kravtsov}, {Leitner}, \&
  {Gnedin}}]{Agertz13}
{Agertz}, O., {Kravtsov}, A.~V., {Leitner}, S.~N., \& {Gnedin}, N.~Y. 2013,
  \apj, 770, 25, \dodoi{10.1088/0004-637X/770/1/25}

\bibitem[{{Benincasa} {et~al.}(2016){Benincasa}, {Wadsley}, {Couchman}, \&
  {Keller}}]{Benincasa16}
{Benincasa}, S.~M., {Wadsley}, J., {Couchman}, H.~M.~P., \& {Keller}, B.~W.
  2016, \mnras, 462, 3053, \dodoi{10.1093/mnras/stw1741}

\bibitem[{{Bertoldi} \& {McKee}(1992)}]{BertoldiMcKee1992}
{Bertoldi}, F., \& {McKee}, C.~F. 1992, \apj, 395, 140, \dodoi{10.1086/171638}

\bibitem[{{Bigiel} {et~al.}(2008){Bigiel}, {Leroy}, {Walter}, {Brinks}, {de
  Blok}, {Madore}, \& {Thornley}}]{Bigiel08}
{Bigiel}, F., {Leroy}, A., {Walter}, F., {et~al.} 2008, \aj, 136, 2846,
  \dodoi{10.1088/0004-6256/136/6/2846}

\bibitem[{{Bigiel} {et~al.}(2011){Bigiel}, {Leroy}, {Walter}, {Brinks}, {de
  Blok}, {Kramer}, {Rix}, {Schruba}, {Schuster}, {Usero}, \&
  {Wiesemeyer}}]{Bigiel11}
{Bigiel}, F., {Leroy}, A.~K., {Walter}, F., {et~al.} 2011, \apjl, 730, L13,
  \dodoi{10.1088/2041-8205/730/2/L13}

\bibitem[{{Bolatto} {et~al.}(2013){Bolatto}, {Wolfire}, \& {Leroy}}]{Bolatto13}
{Bolatto}, A.~D., {Wolfire}, M., \& {Leroy}, A.~K. 2013, \araa, 51, 207,
  \dodoi{10.1146/annurev-astro-082812-140944}

\bibitem[{{Cappellari} {et~al.}(2011){Cappellari}, {Emsellem}, {Krajnovi{\'c}},
  {McDermid}, {Scott}, {Verdoes Kleijn}, {Young}, {Alatalo}, {Bacon}, {Blitz},
  {Bois}, {Bournaud}, {Bureau}, {Davies}, {Davis}, {de Zeeuw}, {Duc},
  {Khochfar}, {Kuntschner}, {Lablanche}, {Morganti}, {Naab}, {Oosterloo},
  {Sarzi}, {Serra}, \& {Weijmans}}]{2011MNRAS.413..813C}
{Cappellari}, M., {Emsellem}, E., {Krajnovi{\'c}}, D., {et~al.} 2011, \mnras,
  413, 813, \dodoi{10.1111/j.1365-2966.2010.18174.x}

\bibitem[{{Cappellari} {et~al.}(2013){Cappellari}, {Scott}, {Alatalo}, {Blitz},
  {Bois}, {Bournaud}, {Bureau}, {Crocker}, {Davies}, {Davis}, {de Zeeuw},
  {Duc}, {Emsellem}, {Khochfar}, {Krajnovi{\'c}}, {Kuntschner}, {McDermid},
  {Morganti}, {Naab}, {Oosterloo}, {Sarzi}, {Serra}, {Weijmans}, \&
  {Young}}]{2013MNRAS.432.1709C}
{Cappellari}, M., {Scott}, N., {Alatalo}, K., {et~al.} 2013, \mnras, 432, 1709,
  \dodoi{10.1093/mnras/stt562}

\bibitem[{{Catinella} {et~al.}(2018){Catinella}, {Saintonge}, {Janowiecki},
  {Cortese}, {Dav{\'e}}, {Lemonias}, {Cooper}, {Schiminovich}, {Hummels},
  {Fabello}, {Ger{\'e}b}, {Kilborn}, \& {Wang}}]{2018MNRAS.476..875C}
{Catinella}, B., {Saintonge}, A., {Janowiecki}, S., {et~al.} 2018, \mnras, 476,
  875, \dodoi{10.1093/mnras/sty089}

\bibitem[{{Chabrier}(2003)}]{Chabrier03}
{Chabrier}, G. 2003, \pasp, 115, 763, \dodoi{10.1086/376392}

\bibitem[{{Chen} {et~al.}(2016){Chen}, {Bryan}, \&
  {Salem}}]{2016MNRAS.460.3335C}
{Chen}, J., {Bryan}, G.~L., \& {Salem}, M. 2016, \mnras, 460, 3335,
  \dodoi{10.1093/mnras/stw1197}

\bibitem[{{Clark} {et~al.}(2012){Clark}, {Glover}, \& {Klessen}}]{Clark12}
{Clark}, P.~C., {Glover}, S.~C.~O., \& {Klessen}, R.~S. 2012, \mnras, 420, 745,
  \dodoi{10.1111/j.1365-2966.2011.20087.x}

\bibitem[{{Colombo} {et~al.}(2020){Colombo}, {Sanchez}, {Bolatto}, {Kalinova},
  {Wei{\ss}}, {Wong}, {Rosolowsky}, {Vogel}, {Barrera-Ballesteros},
  {Dannerbauer}, {Cao}, {Levy}, {Utomo}, \& {Blitz}}]{2020A&A...644A..97C}
{Colombo}, D., {Sanchez}, S.~F., {Bolatto}, A.~D., {et~al.} 2020, \aap, 644,
  A97, \dodoi{10.1051/0004-6361/202039005}

\bibitem[{{Combes} {et~al.}(2007){Combes}, {Young}, \&
  {Bureau}}]{2007MNRAS.377.1795C}
{Combes}, F., {Young}, L.~M., \& {Bureau}, M. 2007, \mnras, 377, 1795,
  \dodoi{10.1111/j.1365-2966.2007.11759.x}

\bibitem[{{da Silva} {et~al.}(2014){da Silva}, {Fumagalli}, \&
  {Krumholz}}]{daSilva14}
{da Silva}, R.~L., {Fumagalli}, M., \& {Krumholz}, M.~R. 2014, \mnras, 444,
  3275, \dodoi{10.1093/mnras/stu1688}

\bibitem[{{Dav{\'e}} {et~al.}(2019){Dav{\'e}}, {Angl{\'e}s-Alc{\'a}zar},
  {Narayanan}, {Li}, {Rafieferantsoa}, \& {Appleby}}]{2019MNRAS.486.2827D}
{Dav{\'e}}, R., {Angl{\'e}s-Alc{\'a}zar}, D., {Narayanan}, D., {et~al.} 2019,
  \mnras, 486, 2827, \dodoi{10.1093/mnras/stz937}

\bibitem[{{Dav{\'e}} {et~al.}(2016){Dav{\'e}}, {Thompson}, \&
  {Hopkins}}]{2016MNRAS.462.3265D}
{Dav{\'e}}, R., {Thompson}, R., \& {Hopkins}, P.~F. 2016, \mnras, 462, 3265,
  \dodoi{10.1093/mnras/stw1862}

\bibitem[{{Davis} {et~al.}(2019){Davis}, {Greene}, {Ma}, {Blakeslee}, {Dawson},
  {Pandya}, {Veale}, \& {Zabel}}]{2019MNRAS.486.1404D}
{Davis}, T.~A., {Greene}, J.~E., {Ma}, C.-P., {et~al.} 2019, \mnras, 486, 1404,
  \dodoi{10.1093/mnras/stz871}

\bibitem[{{Davis} {et~al.}(2013){Davis}, {Alatalo}, {Bureau}, {Cappellari},
  {Scott}, {Young}, {Blitz}, {Crocker}, {Bayet}, {Bois}, {Bournaud}, {Davies},
  {de Zeeuw}, {Duc}, {Emsellem}, {Khochfar}, {Krajnovi{\'c}}, {Kuntschner},
  {Lablanche}, {McDermid}, {Morganti}, {Naab}, {Oosterloo}, {Sarzi}, {Serra},
  \& {Weijmans}}]{2013MNRAS.429..534D}
{Davis}, T.~A., {Alatalo}, K., {Bureau}, M., {et~al.} 2013, \mnras, 429, 534,
  \dodoi{10.1093/mnras/sts353}

\bibitem[{{Davis} {et~al.}(2014){Davis}, {Young}, {Crocker}, {Bureau}, {Blitz},
  {Alatalo}, {Emsellem}, {Naab}, {Bayet}, {Bois}, {Bournaud}, {Cappellari},
  {Davies}, {de Zeeuw}, {Duc}, {Khochfar}, {Krajnovi{\'c}}, {Kuntschner},
  {McDermid}, {Morganti}, {Oosterloo}, {Sarzi}, {Scott}, {Serra}, \&
  {Weijmans}}]{2014MNRAS.444.3427D}
{Davis}, T.~A., {Young}, L.~M., {Crocker}, A.~F., {et~al.} 2014, \mnras, 444,
  3427, \dodoi{10.1093/mnras/stu570}

\bibitem[{{Davis} {et~al.}(2022){Davis}, {Gensior}, {Bureau}, {Cappellari},
  {Choi}, {Elford}, {Kruijssen}, {Lelli}, {Liang}, {Liu}, {Ruffa}, {Saito},
  {Sarzi}, {Schruba}, \& {Williams}}]{Davis2022}
{Davis}, T.~A., {Gensior}, J., {Bureau}, M., {et~al.} 2022, \mnras, 512, 1522,
  \dodoi{10.1093/mnras/stac600}

\bibitem[{{den Brok} {et~al.}(2023){den Brok}, {Bigiel}, {Chastenet},
  {Sandstrom}, {Leroy}, {Usero}, {Schinnerer}, {Rosolowsky}, {Koch}, {Chiang},
  {Barnes}, {Puschnig}, {Saito}, {Be{\v{s}}li{\'c}}, {Chevance}, {Dale},
  {Eibensteiner}, {Glover}, {Jim{\'e}nez-Donaire}, {Teng}, \&
  {Williams}}]{2023A&A...676A..93D}
{den Brok}, J.~S., {Bigiel}, F., {Chastenet}, J., {et~al.} 2023, \aap, 676,
  A93, \dodoi{10.1051/0004-6361/202245718}

\bibitem[{{Donnari} {et~al.}(2021){Donnari}, {Pillepich}, {Nelson},
  {Marinacci}, {Vogelsberger}, \& {Hernquist}}]{2021MNRAS.506.4760D}
{Donnari}, M., {Pillepich}, A., {Nelson}, D., {et~al.} 2021, \mnras, 506, 4760,
  \dodoi{10.1093/mnras/stab1950}

\bibitem[{{Donnari} {et~al.}(2019){Donnari}, {Pillepich}, {Nelson},
  {Vogelsberger}, {Genel}, {Weinberger}, {Marinacci}, {Springel}, \&
  {Hernquist}}]{2019MNRAS.485.4817D}
---. 2019, \mnras, 485, 4817, \dodoi{10.1093/mnras/stz712}

\bibitem[{{Dutton} \& {Macci{\`o}}(2014)}]{2014MNRAS.441.3359D}
{Dutton}, A.~A., \& {Macci{\`o}}, A.~V. 2014, \mnras, 441, 3359,
  \dodoi{10.1093/mnras/stu742}

\bibitem[{{Elmegreen}(2011)}]{2011ApJ...737...10E}
{Elmegreen}, B.~G. 2011, \apj, 737, 10, \dodoi{10.1088/0004-637X/737/1/10}

\bibitem[{{Fabian}(2012)}]{2012ARA&A..50..455F}
{Fabian}, A.~C. 2012, \araa, 50, 455,
  \dodoi{10.1146/annurev-astro-081811-125521}

\bibitem[{{Faucher-Gigu{\`e}re} {et~al.}(2013){Faucher-Gigu{\`e}re},
  {Quataert}, \& {Hopkins}}]{2013MNRAS.433.1970F}
{Faucher-Gigu{\`e}re}, C.-A., {Quataert}, E., \& {Hopkins}, P.~F. 2013, \mnras,
  433, 1970, \dodoi{10.1093/mnras/stt866}

\bibitem[{{Fielding} {et~al.}(2018){Fielding}, {Quataert}, \&
  {Martizzi}}]{2018MNRAS.481.3325F}
{Fielding}, D., {Quataert}, E., \& {Martizzi}, D. 2018, \mnras, 481, 3325,
  \dodoi{10.1093/mnras/sty2466}

\bibitem[{{Gallagher} {et~al.}(2018){Gallagher}, {Leroy}, {Bigiel}, {Cormier},
  {Jim{\'e}nez-Donaire}, {Ostriker}, {Usero}, {Bolatto}, {Garc{\'\i}a-Burillo},
  {Hughes}, {Kepley}, {Krumholz}, {Meidt}, {Meier}, {Murphy}, {Pety},
  {Rosolowsky}, {Schinnerer}, {Schruba}, \& {Walter}}]{2018ApJ...858...90G}
{Gallagher}, M.~J., {Leroy}, A.~K., {Bigiel}, F., {et~al.} 2018, \apj, 858, 90,
  \dodoi{10.3847/1538-4357/aabad8}

\bibitem[{{Gensior} {et~al.}(2023){Gensior}, {Davis}, {Bureau}, {Kruijssen},
  {Cappellari}, {Ruffa}, \& {Williams}}]{Gensior2023}
{Gensior}, J., {Davis}, T.~A., {Bureau}, M., {et~al.} 2023, \mnras, 526, 5590,
  \dodoi{10.1093/mnras/stad3127}

\bibitem[{{Gensior} \& {Kruijssen}(2021)}]{2021MNRAS.500.2000G}
{Gensior}, J., \& {Kruijssen}, J.~M.~D. 2021, \mnras, 500, 2000,
  \dodoi{10.1093/mnras/staa3453}

\bibitem[{{Gensior} {et~al.}(2020){Gensior}, {Kruijssen}, \&
  {Keller}}]{2020MNRAS.495..199G}
{Gensior}, J., {Kruijssen}, J.~M.~D., \& {Keller}, B.~W. 2020, \mnras, 495,
  199, \dodoi{10.1093/mnras/staa1184}

\bibitem[{{Gentry} {et~al.}(2017){Gentry}, {Krumholz}, {Dekel}, \&
  {Madau}}]{Gentry17}
{Gentry}, E.~S., {Krumholz}, M.~R., {Dekel}, A., \& {Madau}, P. 2017, \mnras,
  465, 2471, \dodoi{10.1093/mnras/stw2746}

\bibitem[{{Glover} \& {Clark}(2012)}]{Glover&Clark12}
{Glover}, S. C.~O., \& {Clark}, P.~C. 2012, \mnras, 421, 116,
  \dodoi{10.1111/j.1365-2966.2011.20260.x}

\bibitem[{{Glover} {et~al.}(2010){Glover}, {Federrath}, {Mac Low}, \&
  {Klessen}}]{Glover10}
{Glover}, S.~C.~O., {Federrath}, C., {Mac Low}, M.~M., \& {Klessen}, R.~S.
  2010, \mnras, 404, 2, \dodoi{10.1111/j.1365-2966.2009.15718.x}

\bibitem[{{Glover} \& {Mac Low}(2007{\natexlab{a}})}]{GloverMacLow07a}
{Glover}, S.~C.~O., \& {Mac Low}, M.-M. 2007{\natexlab{a}}, \apjs, 169, 239,
  \dodoi{10.1086/512238}

\bibitem[{{Glover} \& {Mac Low}(2007{\natexlab{b}})}]{GloverMacLow07b}
---. 2007{\natexlab{b}}, \apj, 659, 1317, \dodoi{10.1086/512227}

\bibitem[{{Goldbaum} {et~al.}(2015){Goldbaum}, {Krumholz}, \&
  {Forbes}}]{Goldbaum15}
{Goldbaum}, N.~J., {Krumholz}, M.~R., \& {Forbes}, J.~C. 2015, \apj, 814, 131,
  \dodoi{10.1088/0004-637X/814/2/131}

\bibitem[{{Gong} {et~al.}(2017){Gong}, {Ostriker}, \& {Wolfire}}]{Gong17}
{Gong}, M., {Ostriker}, E.~C., \& {Wolfire}, M.~G. 2017, \apj, 843, 38,
  \dodoi{10.3847/1538-4357/aa7561}

\bibitem[{{Grand} {et~al.}(2017){Grand}, {G{\'o}mez}, {Marinacci}, {Pakmor},
  {Springel}, {Campbell}, {Frenk}, {Jenkins}, \& {White}}]{2017MNRAS.467..179G}
{Grand}, R. J.~J., {G{\'o}mez}, F.~A., {Marinacci}, F., {et~al.} 2017, \mnras,
  467, 179, \dodoi{10.1093/mnras/stx071}

\bibitem[{{Grudi{\'c}} {et~al.}(2018){Grudi{\'c}}, {Hopkins},
  {Faucher-Gigu{\`e}re}, {Quataert}, {Murray}, \&
  {Kere{\v{s}}}}]{2018MNRAS.475.3511G}
{Grudi{\'c}}, M.~Y., {Hopkins}, P.~F., {Faucher-Gigu{\`e}re}, C.-A., {et~al.}
  2018, \mnras, 475, 3511, \dodoi{10.1093/mnras/sty035}

\bibitem[{{Gurvich} {et~al.}(2020){Gurvich}, {Faucher-Gigu{\`e}re}, {Richings},
  {Hopkins}, {Grudi{\'c}}, {Hafen}, {Wellons}, {Stern}, {Quataert}, {Chan},
  {Orr}, {Kere{\v{s}}}, {Wetzel}, {Hayward}, {Loebman}, \&
  {Murray}}]{2020MNRAS.498.3664G}
{Gurvich}, A.~B., {Faucher-Gigu{\`e}re}, C.-A., {Richings}, A.~J., {et~al.}
  2020, \mnras, 498, 3664, \dodoi{10.1093/mnras/staa2578}

\bibitem[{{Habing}(1968)}]{Habing68}
{Habing}, H.~J. 1968, \bain, 19, 421

\bibitem[{{Hassan} {et~al.}(2023){Hassan}, {Ostriker}, {Kim}, {Bryan},
  {Fielding}, {Genel}, {Jeffreson}, {Motwani}, {Smith}, {Somerville}, \&
  {Steinwandel}}]{Hassan2023}
{Hassan}, S., {Ostriker}, E.~C., {Kim}, C.-G., {et~al.} 2023,
  \mnras~to~be~submitted

\bibitem[{{Hayward} \& {Hopkins}(2017)}]{2017MNRAS.465.1682H}
{Hayward}, C.~C., \& {Hopkins}, P.~F. 2017, \mnras, 465, 1682,
  \dodoi{10.1093/mnras/stw2888}

\bibitem[{{Hernandez} {et~al.}(2007){Hernandez}, {Park}, {Cervantes-Sodi}, \&
  {Choi}}]{2007MNRAS.375..163H}
{Hernandez}, X., {Park}, C., {Cervantes-Sodi}, B., \& {Choi}, Y.-Y. 2007,
  \mnras, 375, 163, \dodoi{10.1111/j.1365-2966.2006.11274.x}

\bibitem[{{Hernquist}(1990)}]{Hernquist90}
{Hernquist}, L. 1990, \apj, 356, 359, \dodoi{10.1086/168845}

\bibitem[{Hines(2023)}]{RBFInterpolant}
Hines, T. 2023, Python package containing tools for radial basis function (RBF)
  applications.
\newblock \url{https://github.com/treverhines/RBF}

\bibitem[{{Hopkins} \& {Christiansen}(2013)}]{2013ApJ...776...48H}
{Hopkins}, P.~F., \& {Christiansen}, J.~L. 2013, \apj, 776, 48,
  \dodoi{10.1088/0004-637X/776/1/48}

\bibitem[{{Hopkins} {et~al.}(2011){Hopkins}, {Quataert}, \&
  {Murray}}]{Hopkins11}
{Hopkins}, P.~F., {Quataert}, E., \& {Murray}, N. 2011, \mnras, 417, 950,
  \dodoi{10.1111/j.1365-2966.2011.19306.x}

\bibitem[{{Hopkins} {et~al.}(2018){Hopkins}, {Wetzel}, {Kere{\v s}},
  {Faucher-Gigu{\`e}re}, {Quataert}, {Boylan-Kolchin}, {Murray}, {Hayward},
  {Garrison-Kimmel}, {Hummels}, {Feldmann}, {Torrey}, {Ma},
  {Angl{\'e}s-Alc{\'a}zar}, {Su}, {Orr}, {Schmitz}, {Escala}, {Sanderson},
  {Grudi{\'c}}, {Hafen}, {Kim}, {Fitts}, {Bullock}, {Wheeler}, {Chan},
  {Elbert}, \& {Narayanan}}]{Hopkins18}
{Hopkins}, P.~F., {Wetzel}, A., {Kere{\v s}}, D., {et~al.} 2018, \mnras, 480,
  800, \dodoi{10.1093/mnras/sty1690}

\bibitem[{{Indriolo} \& {McCall}(2012)}]{Indriolo&McCall12}
{Indriolo}, N., \& {McCall}, B.~J. 2012, \apj, 745, 91,
  \dodoi{10.1088/0004-637X/745/1/91}

\bibitem[{{Jeffreson} {et~al.}(2021){Jeffreson}, {Krumholz}, {Fujimoto},
  {Armillotta}, {Keller}, {Chevance}, \& {Kruijssen}}]{2021MNRAS.505.3470J}
{Jeffreson}, S. M.~R., {Krumholz}, M.~R., {Fujimoto}, Y., {et~al.} 2021,
  \mnras, 505, 3470, \dodoi{10.1093/mnras/stab1536}

\bibitem[{{Jeffreson} {et~al.}(2024){Jeffreson}, {Semenov}, \&
  {Krumholz}}]{2024MNRAS.527.7093J}
{Jeffreson}, S. M.~R., {Semenov}, V.~A., \& {Krumholz}, M.~R. 2024, \mnras,
  527, 7093, \dodoi{10.1093/mnras/stad3550}

\bibitem[{{Kennicutt}(1998)}]{Kennicutt1998}
{Kennicutt}, Jr., R.~C. 1998, \apj, 498, 541, \dodoi{10.1086/305588}

\bibitem[{{Kim} {et~al.}(2023{\natexlab{a}}){Kim}, {Kim}, {Gong}, \&
  {Ostriker}}]{2023ApJ...946....3K}
{Kim}, C.-G., {Kim}, J.-G., {Gong}, M., \& {Ostriker}, E.~C.
  2023{\natexlab{a}}, \apj, 946, 3, \dodoi{10.3847/1538-4357/acbd3a}

\bibitem[{{Kim} {et~al.}(2011){Kim}, {Kim}, \&
  {Ostriker}}]{2011ApJ...743...25K}
{Kim}, C.-G., {Kim}, W.-T., \& {Ostriker}, E.~C. 2011, \apj, 743, 25,
  \dodoi{10.1088/0004-637X/743/1/25}

\bibitem[{{Kim} \& {Ostriker}(2015)}]{KimCG&Ostriker15b}
{Kim}, C.-G., \& {Ostriker}, E.~C. 2015, \apj, 815, 67,
  \dodoi{10.1088/0004-637X/815/1/67}

\bibitem[{{Kim} \& {Ostriker}(2017)}]{KimCG&Ostriker17}
---. 2017, \apj, 846, 133, \dodoi{10.3847/1538-4357/aa8599}

\bibitem[{{Kim} {et~al.}(2013){Kim}, {Ostriker}, \&
  {Kim}}]{2013ApJ...776....1K}
{Kim}, C.-G., {Ostriker}, E.~C., \& {Kim}, W.-T. 2013, \apj, 776, 1,
  \dodoi{10.1088/0004-637X/776/1/1}

\bibitem[{{Kim} {et~al.}(2017){Kim}, {Ostriker}, \&
  {Raileanu}}]{2017ApJ...834...25K}
{Kim}, C.-G., {Ostriker}, E.~C., \& {Raileanu}, R. 2017, \apj, 834, 25,
  \dodoi{10.3847/1538-4357/834/1/25}

\bibitem[{{Kim} {et~al.}(2020){Kim}, {Ostriker}, {Somerville}, {Bryan},
  {Fielding}, {Forbes}, {Hayward}, {Hernquist}, \&
  {Pandya}}]{2020ApJ...900...61K}
{Kim}, C.-G., {Ostriker}, E.~C., {Somerville}, R.~S., {et~al.} 2020, \apj, 900,
  61, \dodoi{10.3847/1538-4357/aba962}

\bibitem[{{Kim} {et~al.}(2023{\natexlab{b}}){Kim}, {Gong}, {Kim}, \&
  {Ostriker}}]{2023ApJS..264...10K}
{Kim}, J.-G., {Gong}, M., {Kim}, C.-G., \& {Ostriker}, E.~C.
  2023{\natexlab{b}}, \apjs, 264, 10, \dodoi{10.3847/1538-4365/ac9b1d}

\bibitem[{{Kim} {et~al.}(2014){Kim}, {Abel}, {Agertz}, {Bryan}, {Ceverino},
  {Christensen}, {Conroy}, {Dekel}, {Gnedin}, \& {Goldbaum}}]{Kim14}
{Kim}, J.-h., {Abel}, T., {Agertz}, O., {et~al.} 2014, \apjs, 210, 14,
  \dodoi{10.1088/0067-0049/210/1/14}

\bibitem[{{Kim} \& {Ostriker}(2001)}]{2001ApJ...559...70K}
{Kim}, W.-T., \& {Ostriker}, E.~C. 2001, \apj, 559, 70, \dodoi{10.1086/322330}

\bibitem[{{Kim} {et~al.}(2002){Kim}, {Ostriker}, \&
  {Stone}}]{2002ApJ...581.1080K}
{Kim}, W.-T., {Ostriker}, E.~C., \& {Stone}, J.~M. 2002, \apj, 581, 1080,
  \dodoi{10.1086/344367}

\bibitem[{{Koch} {et~al.}(2018){Koch}, {Rosolowsky}, {Lockman}, {Kepley},
  {Leroy}, {Schruba}, {Braine}, {Dalcanton}, {Johnson}, \&
  {Stanimirovi{\'c}}}]{2018MNRAS.479.2505K}
{Koch}, E.~W., {Rosolowsky}, E.~W., {Lockman}, F.~J., {et~al.} 2018, \mnras,
  479, 2505, \dodoi{10.1093/mnras/sty1674}

\bibitem[{{Krajnovi{\'c}} {et~al.}(2013){Krajnovi{\'c}}, {Alatalo}, {Blitz},
  {Bois}, {Bournaud}, {Bureau}, {Cappellari}, {Davies}, {Davis}, {de Zeeuw},
  {Duc}, {Emsellem}, {Khochfar}, {Kuntschner}, {McDermid}, {Morganti}, {Naab},
  {Oosterloo}, {Sarzi}, {Scott}, {Serra}, {Weijmans}, \&
  {Young}}]{2013MNRAS.432.1768K}
{Krajnovi{\'c}}, D., {Alatalo}, K., {Blitz}, L., {et~al.} 2013, \mnras, 432,
  1768, \dodoi{10.1093/mnras/sts315}

\bibitem[{{Kruijssen} {et~al.}(2019){Kruijssen}, {Schruba}, {Chevance},
  {Longmore}, {Hygate}, {Haydon}, {McLeod}, {Dalcanton}, {Tacconi}, \& {van
  Dishoeck}}]{Kruijssen2019}
{Kruijssen}, J.~M.~D., {Schruba}, A., {Chevance}, M., {et~al.} 2019, \nat, 569,
  519, \dodoi{10.1038/s41586-019-1194-3}

\bibitem[{{Krumholz} \& {Burkert}(2010)}]{2010ApJ...724..895K}
{Krumholz}, M., \& {Burkert}, A. 2010, \apj, 724, 895,
  \dodoi{10.1088/0004-637X/724/2/895}

\bibitem[{{Krumholz}(2013)}]{Krumholz13a}
{Krumholz}, M.~R. 2013, {DESPOTIC: Derive the Energetics and SPectra of
  Optically Thick Interstellar Clouds}, Astrophysics Source Code Library.
\newblock \doeprint{1304.007}

\bibitem[{{Krumholz}(2014)}]{Krumholz14}
---. 2014, \mnras, 437, 1662, \dodoi{10.1093/mnras/stt2000}

\bibitem[{{Krumholz} {et~al.}(2018){Krumholz}, {Burkhart}, {Forbes}, \&
  {Crocker}}]{Krumholz18b}
{Krumholz}, M.~R., {Burkhart}, B., {Forbes}, J.~C., \& {Crocker}, R.~M. 2018,
  \mnras, 477, 2716, \dodoi{10.1093/mnras/sty852}

\bibitem[{{Krumholz} {et~al.}(2015){Krumholz}, {Fumagalli}, {da Silva},
  {Rendahl}, \& {Parra}}]{Krumholz15}
{Krumholz}, M.~R., {Fumagalli}, M., {da Silva}, R.~L., {Rendahl}, T., \&
  {Parra}, J. 2015, \mnras, 452, 1447, \dodoi{10.1093/mnras/stv1374}

\bibitem[{{Krumholz} \& {Gnedin}(2011)}]{2011ApJ...729...36K}
{Krumholz}, M.~R., \& {Gnedin}, N.~Y. 2011, \apj, 729, 36,
  \dodoi{10.1088/0004-637X/729/1/36}

\bibitem[{{Krumholz} \& {Matzner}(2009)}]{KrumholzMatzner09}
{Krumholz}, M.~R., \& {Matzner}, C.~D. 2009, \apj, 703, 1352,
  \dodoi{10.1088/0004-637X/703/2/1352}

\bibitem[{{Leroy} {et~al.}(2021){Leroy}, {Schinnerer}, {Hughes}, {Rosolowsky},
  {Pety}, {Schruba}, {Usero}, {Blanc}, {Chevance}, {Emsellem}, {Faesi},
  {Herrera}, {Liu}, {Meidt}, {Querejeta}, {Saito}, {Sandstrom}, {Sun},
  {Williams}, {Anand}, {Barnes}, {Behrens}, {Belfiore}, {Benincasa},
  {Be{\v{s}}li{\'c}}, {Bigiel}, {Bolatto}, {den Brok}, {Cao}, {Chandar},
  {Chastenet}, {Chiang}, {Congiu}, {Dale}, {Deger}, {Eibensteiner}, {Egorov},
  {Garc{\'\i}a-Rodr{\'\i}guez}, {Glover}, {Grasha}, {Henshaw}, {Ho}, {Kepley},
  {Kim}, {Klessen}, {Kreckel}, {Koch}, {Kruijssen}, {Larson}, {Lee}, {Lopez},
  {Machado}, {Mayker}, {McElroy}, {Murphy}, {Ostriker}, {Pan}, {Pessa},
  {Puschnig}, {Razza}, {S{\'a}nchez-Bl{\'a}zquez}, {Santoro}, {Sardone},
  {Scheuermann}, {Sliwa}, {Sormani}, {Stuber}, {Thilker}, {Turner}, {Utomo},
  {Watkins}, \& {Whitmore}}]{Leroy2021a}
{Leroy}, A.~K., {Schinnerer}, E., {Hughes}, A., {et~al.} 2021, \apjs, 257, 43,
  \dodoi{10.3847/1538-4365/ac17f3}

\bibitem[{{Liu} {et~al.}(2021{\natexlab{a}}){Liu}, {Bureau}, {Blitz}, {Davis},
  {Onishi}, {Smith}, {North}, \& {Iguchi}}]{Liu2021}
{Liu}, L., {Bureau}, M., {Blitz}, L., {et~al.} 2021{\natexlab{a}}, \mnras, 505,
  4048, \dodoi{10.1093/mnras/stab1537}

\bibitem[{{Liu} {et~al.}(2021{\natexlab{b}}){Liu}, {Bureau}, {Blitz}, {Davis},
  {Onishi}, {Smith}, {North}, \& {Iguchi}}]{2021MNRAS.505.4048L}
---. 2021{\natexlab{b}}, \mnras, 505, 4048, \dodoi{10.1093/mnras/stab1537}

\bibitem[{{Lu} {et~al.}(2024){Lu}, {Haggard}, {Bureau}, {Gensior}, {Jeffreson},
  {Robert}, {Williams}, {Liang}, {Choi}, {Davis}, {Babic}, {Boyce}, {Cheung},
  {Drissen}, {Elford}, {Liu}, {Martin}, {Rhea}, {Rosseau-Nepton}, \&
  {Ruffa}}]{2024MNRAS.subm}
{Lu}, A., {Haggard}, D., {Bureau}, M., {et~al.} 2024, \mnras~submitted

\bibitem[{{Ma} {et~al.}(2014){Ma}, {Greene}, {McConnell}, {Janish},
  {Blakeslee}, {Thomas}, \& {Murphy}}]{2014ApJ.795.158M}
{Ma}, C.-P., {Greene}, J.~E., {McConnell}, N., {et~al.} 2014, \apj, 795, 158,
  \dodoi{10.1088/0004-637X/795/2/158}

\bibitem[{{MacLaren} {et~al.}(1988){MacLaren}, {Richardson}, \&
  {Wolfendale}}]{MacLaren1988}
{MacLaren}, I., {Richardson}, K.~M., \& {Wolfendale}, A.~W. 1988, \apj, 333,
  821, \dodoi{10.1086/166791}

\bibitem[{{Martig} {et~al.}(2009){Martig}, {Bournaud}, {Teyssier}, \&
  {Dekel}}]{2009ApJ...707..250M}
{Martig}, M., {Bournaud}, F., {Teyssier}, R., \& {Dekel}, A. 2009, \apj, 707,
  250, \dodoi{10.1088/0004-637X/707/1/250}

\bibitem[{{Martig} {et~al.}(2013){Martig}, {Crocker}, {Bournaud}, {Emsellem},
  {Gabor}, {Alatalo}, {Blitz}, {Bois}, {Bureau}, {Cappellari}, {Davies},
  {Davis}, {Dekel}, {de Zeeuw}, {Duc}, {Falc{\'o}n-Barroso}, {Khochfar},
  {Krajnovi{\'c}}, {Kuntschner}, {Morganti}, {McDermid}, {Naab}, {Oosterloo},
  {Sarzi}, {Scott}, {Serra}, {Griffin}, {Teyssier}, {Weijmans}, \&
  {Young}}]{Martig2013}
{Martig}, M., {Crocker}, A.~F., {Bournaud}, F., {et~al.} 2013, \mnras, 432,
  1914, \dodoi{10.1093/mnras/sts594}

\bibitem[{{Martizzi} {et~al.}(2015){Martizzi}, {Faucher-Gigu{\`e}re}, \&
  {Quataert}}]{Martizzi15}
{Martizzi}, D., {Faucher-Gigu{\`e}re}, C.-A., \& {Quataert}, E. 2015, \mnras,
  450, 504, \dodoi{10.1093/mnras/stv562}

\bibitem[{{Mathis} {et~al.}(1983){Mathis}, {Mezger}, \& {Panagia}}]{Mathis83}
{Mathis}, J.~S., {Mezger}, P.~G., \& {Panagia}, N. 1983, \aap, 500, 259

\bibitem[{{Matzner}(2002)}]{Matzner02}
{Matzner}, C.~D. 2002, \apj, 566, 302, \dodoi{10.1086/338030}

\bibitem[{{Murray} {et~al.}(2010){Murray}, {Quataert}, \&
  {Thompson}}]{Murray10}
{Murray}, N., {Quataert}, E., \& {Thompson}, T.~A. 2010, \apj, 709, 191,
  \dodoi{10.1088/0004-637X/709/1/191}

\bibitem[{{Murray} \& {Rahman}(2010)}]{Murray&Rahman10}
{Murray}, N., \& {Rahman}, M. 2010, \apj, 709, 424,
  \dodoi{10.1088/0004-637X/709/1/424}

\bibitem[{{Navarro} {et~al.}(1997){Navarro}, {Frenk}, \& {White}}]{Navarro97}
{Navarro}, J.~F., {Frenk}, C.~S., \& {White}, S. D.~M. 1997, \apj, 490, 493,
  \dodoi{10.1086/304888}

\bibitem[{{Nelson} {et~al.}(2018){Nelson}, {Pillepich}, {Springel},
  {Weinberger}, {Hernquist}, {Pakmor}, {Genel}, {Torrey}, {Vogelsberger},
  {Kauffmann}, {Marinacci}, \& {Naiman}}]{2018MNRAS.475..624N}
{Nelson}, D., {Pillepich}, A., {Springel}, V., {et~al.} 2018, \mnras, 475, 624,
  \dodoi{10.1093/mnras/stx3040}

\bibitem[{{Nelson} \& {Langer}(1997)}]{NelsonLanger97}
{Nelson}, R.~P., \& {Langer}, W.~D. 1997, \apj, 482, 796,
  \dodoi{10.1086/304167}

\bibitem[{{Orr} {et~al.}(2022){Orr}, {Fielding}, {Hayward}, \&
  {Burkhart}}]{2022ApJ...932...88O}
{Orr}, M.~E., {Fielding}, D.~B., {Hayward}, C.~C., \& {Burkhart}, B. 2022,
  \apj, 932, 88, \dodoi{10.3847/1538-4357/ac6c26}

\bibitem[{{Ostriker} \& {Kim}(2022)}]{2022ApJ...936..137O}
{Ostriker}, E.~C., \& {Kim}, C.-G. 2022, \apj, 936, 137,
  \dodoi{10.3847/1538-4357/ac7de2}

\bibitem[{{Ostriker} {et~al.}(2010){Ostriker}, {McKee}, \&
  {Leroy}}]{Ostriker+10}
{Ostriker}, E.~C., {McKee}, C.~F., \& {Leroy}, A.~K. 2010, \apj, 721, 975,
  \dodoi{10.1088/0004-637X/721/2/975}

\bibitem[{{Ostriker} \& {Shetty}(2011)}]{OstrikerShetty2011}
{Ostriker}, E.~C., \& {Shetty}, R. 2011, \apj, 731, 41,
  \dodoi{10.1088/0004-637X/731/1/41}

\bibitem[{{O'Sullivan} {et~al.}(2018){O'Sullivan}, {Combes}, {Salom{\'e}},
  {David}, {Babul}, {Vrtilek}, {Lim}, {Olivares}, {Raychaudhury}, \&
  {Schellenberger}}]{2018A&A...618A.126O}
{O'Sullivan}, E., {Combes}, F., {Salom{\'e}}, P., {et~al.} 2018, \aap, 618,
  A126, \dodoi{10.1051/0004-6361/201833580}

\bibitem[{{Padoan} {et~al.}(2017){Padoan}, {Haugb{\o}lle}, {Nordlund}, \&
  {Frimann}}]{Padoan17}
{Padoan}, P., {Haugb{\o}lle}, T., {Nordlund}, {\r{A}}., \& {Frimann}, S. 2017,
  \apj, 840, 48, \dodoi{10.3847/1538-4357/aa6afa}

\bibitem[{{Phillips} {et~al.}(1987){Phillips}, {Ellison}, {Keene}, {Leighton},
  {Howard}, {Masson}, {Sanders}, {Veidt}, \& {Young}}]{1987ApJ...322L..73P}
{Phillips}, T.~G., {Ellison}, B.~N., {Keene}, J.~B., {et~al.} 1987, \apjl, 322,
  L73, \dodoi{10.1086/185039}

\bibitem[{{Piotrowska} {et~al.}(2022){Piotrowska}, {Bluck}, {Maiolino}, \&
  {Peng}}]{2022MNRAS.512.1052P}
{Piotrowska}, J.~M., {Bluck}, A. F.~L., {Maiolino}, R., \& {Peng}, Y. 2022,
  \mnras, 512, 1052, \dodoi{10.1093/mnras/stab3673}

\bibitem[{{Plummer}(1911)}]{Plummer1911}
{Plummer}, H.~C. 1911, \mnras, 71, 460, \dodoi{10.1093/mnras/71.5.460}

\bibitem[{{Power} {et~al.}(2003){Power}, {Navarro}, {Jenkins}, {Frenk},
  {White}, {Springel}, {Stadel}, \& {Quinn}}]{2003MNRAS.338...14P}
{Power}, C., {Navarro}, J.~F., {Jenkins}, A., {et~al.} 2003, \mnras, 338, 14,
  \dodoi{10.1046/j.1365-8711.2003.05925.x}

\bibitem[{{Romeo} {et~al.}(2010){Romeo}, {Burkert}, \&
  {Agertz}}]{2010MNRAS.407.1223R}
{Romeo}, A.~B., {Burkert}, A., \& {Agertz}, O. 2010, \mnras, 407, 1223,
  \dodoi{10.1111/j.1365-2966.2010.16975.x}

\bibitem[{{Romeo} \& {Falstad}(2013)}]{2013MNRAS.433.1389R}
{Romeo}, A.~B., \& {Falstad}, N. 2013, \mnras, 433, 1389,
  \dodoi{10.1093/mnras/stt809}

\bibitem[{{Romeo} \& {Wiegert}(2011)}]{2011MNRAS.416.1191R}
{Romeo}, A.~B., \& {Wiegert}, J. 2011, \mnras, 416, 1191,
  \dodoi{10.1111/j.1365-2966.2011.19120.x}

\bibitem[{{Russell} {et~al.}(2016){Russell}, {McNamara}, {Fabian}, {Nulsen},
  {Edge}, {Combes}, {Murray}, {Parrish}, {Salom{\'e}}, {Sanders}, {Baum},
  {Donahue}, {Main}, {O'Connell}, {O'Dea}, {Oonk}, {Tremblay}, {Vantyghem}, \&
  {Voit}}]{2016MNRAS.458.3134R}
{Russell}, H.~R., {McNamara}, B.~R., {Fabian}, A.~C., {et~al.} 2016, \mnras,
  458, 3134, \dodoi{10.1093/mnras/stw409}

\bibitem[{{Russell} {et~al.}(2019){Russell}, {McNamara}, {Fabian}, {Nulsen},
  {Combes}, {Edge}, {Madar}, {Olivares}, {Salom{\'e}}, \&
  {Vantyghem}}]{2019MNRAS.490.3025R}
---. 2019, \mnras, 490, 3025, \dodoi{10.1093/mnras/stz2719}

\bibitem[{{Safranek-Shrader} {et~al.}(2017){Safranek-Shrader}, {Krumholz},
  {Kim}, {Ostriker}, {Klein}, {Li}, {McKee}, \& {Stone}}]{Safranek-Shrader+17}
{Safranek-Shrader}, C., {Krumholz}, M.~R., {Kim}, C.-G., {et~al.} 2017, \mnras,
  465, 885, \dodoi{10.1093/mnras/stw2647}

\bibitem[{{Saintonge} {et~al.}(2012){Saintonge}, {Tacconi}, {Fabello}, {Wang},
  {Catinella}, {Genzel}, {Graci{\'a}-Carpio}, {Kramer}, {Moran}, {Heckman},
  {Schiminovich}, {Schuster}, \& {Wuyts}}]{Saintonge2012}
{Saintonge}, A., {Tacconi}, L.~J., {Fabello}, S., {et~al.} 2012, \apj, 758, 73,
  \dodoi{10.1088/0004-637X/758/2/73}

\bibitem[{{Saintonge} {et~al.}(2017){Saintonge}, {Catinella}, {Tacconi},
  {Kauffmann}, {Genzel}, {Cortese}, {Dav{\'e}}, {Fletcher},
  {Graci{\'a}-Carpio}, {Kramer}, {Heckman}, {Janowiecki}, {Lutz}, {Rosario},
  {Schiminovich}, {Schuster}, {Wang}, {Wuyts}, {Borthakur}, {Lamperti}, \&
  {Roberts-Borsani}}]{2017ApJS..233...22S}
{Saintonge}, A., {Catinella}, B., {Tacconi}, L.~J., {et~al.} 2017, \apjs, 233,
  22, \dodoi{10.3847/1538-4365/aa97e0}

\bibitem[{{Salom{\'e}} \& {Combes}(2003)}]{2003A&A...412..657S}
{Salom{\'e}}, P., \& {Combes}, F. 2003, \aap, 412, 657,
  \dodoi{10.1051/0004-6361:20031438}

\bibitem[{{Schaye} {et~al.}(2015){Schaye}, {Crain}, {Bower}, {Furlong},
  {Schaller}, {Theuns}, {Dalla Vecchia}, {Frenk}, {McCarthy}, {Helly},
  {Jenkins}, {Rosas-Guevara}, {White}, {Baes}, {Booth}, {Camps}, {Navarro},
  {Qu}, {Rahmati}, {Sawala}, {Thomas}, \& {Trayford}}]{2015MNRAS.446..521S}
{Schaye}, J., {Crain}, R.~A., {Bower}, R.~G., {et~al.} 2015, \mnras, 446, 521,
  \dodoi{10.1093/mnras/stu2058}

\bibitem[{{Smith} {et~al.}(2020){Smith}, {Bryan}, {Somerville}, {Hu},
  {Teyssier}, {Burkhart}, \& {Hernquist}}]{2020arXiv200911309S}
{Smith}, M.~C., {Bryan}, G.~L., {Somerville}, R.~S., {et~al.} 2020, arXiv
  e-prints, arXiv:2009.11309.
\newblock \doarXiv{2009.11309}

\bibitem[{{Smith} {et~al.}(2023){Smith}, {Fielding}, {Bryan}, {Kim},
  {Ostriker}, {Somerville}, {Stern}, {Su}, {Weinberger}, {Hu}, {Forbes},
  {Hernquist}, {Burkhart}, \& {Li}}]{2023arXiv230107116S}
{Smith}, M.~C., {Fielding}, D.~B., {Bryan}, G.~L., {et~al.} 2023, arXiv
  e-prints, arXiv:2301.07116, \dodoi{10.48550/arXiv.2301.07116}

\bibitem[{{Sobolev}(1960)}]{1960mes..book.....S}
{Sobolev}, V.~V. 1960, {Moving Envelopes of Stars},
  \dodoi{10.4159/harvard.9780674864658}

\bibitem[{{Solomon} \& {Vanden Bout}(2005)}]{SolomonVandenBout05}
{Solomon}, P.~M., \& {Vanden Bout}, P.~A. 2005, \araa, 43, 677,
  \dodoi{10.1146/annurev.astro.43.051804.102221}

\bibitem[{{Springel}(2010)}]{Springel10}
{Springel}, V. 2010, \mnras, 401, 791, \dodoi{10.1111/j.1365-2966.2009.15715.x}

\bibitem[{{Springel} {et~al.}(2005){Springel}, {Di Matteo}, \&
  {Hernquist}}]{Springel05}
{Springel}, V., {Di Matteo}, T., \& {Hernquist}, L. 2005, \mnras, 361, 776,
  \dodoi{10.1111/j.1365-2966.2005.09238.x}

\bibitem[{{Springel} \& {Hernquist}(2003)}]{Springel03}
{Springel}, V., \& {Hernquist}, L. 2003, \mnras, 339, 289,
  \dodoi{10.1046/j.1365-8711.2003.06206.x}

\bibitem[{{Stone} \& {Gardiner}(2009)}]{2009NewA...14..139S}
{Stone}, J.~M., \& {Gardiner}, T. 2009, \na, 14, 139,
  \dodoi{10.1016/j.newast.2008.06.003}

\bibitem[{{Stone} {et~al.}(2008){Stone}, {Gardiner}, {Teuben}, {Hawley}, \&
  {Simon}}]{2008ApJS..178..137S}
{Stone}, J.~M., {Gardiner}, T.~A., {Teuben}, P., {Hawley}, J.~F., \& {Simon},
  J.~B. 2008, \apjs, 178, 137, \dodoi{10.1086/588755}

\bibitem[{{Sun} {et~al.}(2020){Sun}, {Leroy}, {Ostriker}, {Hughes},
  {Rosolowsky}, {Schruba}, {Schinnerer}, {Blanc}, {Faesi}, {Kruijssen},
  {Meidt}, {Utomo}, {Bigiel}, {Bolatto}, {Chevance}, {Chiang}, {Dale},
  {Emsellem}, {Glover}, {Grasha}, {Henshaw}, {Herrera}, {Jimenez-Donaire},
  {Lee}, {Pety}, {Querejeta}, {Saito}, {Sandstrom}, \& {Usero}}]{Sun2020}
{Sun}, J., {Leroy}, A.~K., {Ostriker}, E.~C., {et~al.} 2020, arXiv e-prints,
  arXiv:2002.08964.
\newblock \doarXiv{2002.08964}

\bibitem[{{Sun} {et~al.}(2023){Sun}, {Leroy}, {Ostriker}, {Meidt},
  {Rosolowsky}, {Schinnerer}, {Wilson}, {Utomo}, {Belfiore}, {Blanc},
  {Emsellem}, {Faesi}, {Groves}, {Hughes}, {Koch}, {Kreckel}, {Liu}, {Pan},
  {Pety}, {Querejeta}, {Razza}, {Saito}, {Sardone}, {Usero}, {Williams},
  {Bigiel}, {Bolatto}, {Chevance}, {Dale}, {Gensior}, {Glover}, {Grasha},
  {Henshaw}, {Jim{\'e}nez-Donaire}, {Klessen}, {Kruijssen}, {Murphy},
  {Neumann}, {Teng}, \& {Thilker}}]{2023ApJ...945L..19S}
---. 2023, \apjl, 945, L19, \dodoi{10.3847/2041-8213/acbd9c}

\bibitem[{{Tacconi} {et~al.}(2018){Tacconi}, {Genzel}, {Saintonge}, {Combes},
  {Garc{\'\i}a-Burillo}, {Neri}, {Bolatto}, {Contini}, {F{\"o}rster Schreiber},
  {Lilly}, {Lutz}, {Wuyts}, {Accurso}, {Boissier}, {Boone}, {Bouch{\'e}},
  {Bournaud}, {Burkert}, {Carollo}, {Cooper}, {Cox}, {Feruglio}, {Freundlich},
  {Herrera-Camus}, {Juneau}, {Lippa}, {Naab}, {Renzini}, {Salome}, {Sternberg},
  {Tadaki}, {{\"U}bler}, {Walter}, {Weiner}, \& {Weiss}}]{2018ApJ...853..179T}
{Tacconi}, L.~J., {Genzel}, R., {Saintonge}, A., {et~al.} 2018, \apj, 853, 179,
  \dodoi{10.3847/1538-4357/aaa4b4}

\bibitem[{{Teng} {et~al.}(2023){Teng}, {Sandstrom}, {Sun}, {Gong}, {Bolatto},
  {Chiang}, {Leroy}, {Usero}, {Glover}, {Klessen}, {Liu}, {Querejeta},
  {Schinnerer}, {Bigiel}, {Cao}, {Chevance}, {Eibensteiner}, {Grasha},
  {Israel}, {Murphy}, {Neumann}, {Pan}, {Pinna}, {Sormani}, {Smith}, {Walter},
  \& {Williams}}]{2023ApJ...950..119T}
{Teng}, Y.-H., {Sandstrom}, K.~M., {Sun}, J., {et~al.} 2023, \apj, 950, 119,
  \dodoi{10.3847/1538-4357/accb86}

\bibitem[{{Thompson} {et~al.}(2005){Thompson}, {Quataert}, \&
  {Murray}}]{2005ApJ...630..167T}
{Thompson}, T.~A., {Quataert}, E., \& {Murray}, N. 2005, \apj, 630, 167,
  \dodoi{10.1086/431923}

\bibitem[{{Utomo} {et~al.}(2015){Utomo}, {Blitz}, {Davis}, {Rosolowsky},
  {Bureau}, {Cappellari}, \& {Sarzi}}]{Utomo2015}
{Utomo}, D., {Blitz}, L., {Davis}, T., {et~al.} 2015, \apj, 803, 16,
  \dodoi{10.1088/0004-637X/803/1/16}

\bibitem[{{Veale} {et~al.}(2018){Veale}, {Ma}, {Greene}, {Thomas}, {Blakeslee},
  {Walsh}, \& {Ito}}]{2018MNRAS.473.5446V}
{Veale}, M., {Ma}, C.-P., {Greene}, J.~E., {et~al.} 2018, \mnras, 473, 5446,
  \dodoi{10.1093/mnras/stx2717}

\bibitem[{{Vijayan} {et~al.}(2020){Vijayan}, {Kim}, {Armillotta}, {Ostriker},
  \& {Li}}]{2020ApJ...894...12V}
{Vijayan}, A., {Kim}, C.-G., {Armillotta}, L., {Ostriker}, E.~C., \& {Li}, M.
  2020, \apj, 894, 12, \dodoi{10.3847/1538-4357/ab8474}

\bibitem[{{Vogelsberger} {et~al.}(2013){Vogelsberger}, {Genel}, {Sijacki},
  {Torrey}, {Springel}, \& {Hernquist}}]{2013MNRAS.436.3031V}
{Vogelsberger}, M., {Genel}, S., {Sijacki}, D., {et~al.} 2013, \mnras, 436,
  3031, \dodoi{10.1093/mnras/stt1789}

\bibitem[{{Vogelsberger} {et~al.}(2014){Vogelsberger}, {Genel}, {Springel},
  {Torrey}, {Sijacki}, {Xu}, {Snyder}, {Bird}, {Nelson}, \&
  {Hernquist}}]{2014Natur.509..177V}
{Vogelsberger}, M., {Genel}, S., {Springel}, V., {et~al.} 2014, \nat, 509, 177,
  \dodoi{10.1038/nature13316}

\bibitem[{{Welch} \& {Sage}(2003)}]{2003ApJ...584..260W}
{Welch}, G.~A., \& {Sage}, L.~J. 2003, \apj, 584, 260, \dodoi{10.1086/345537}

\bibitem[{{Westmeier} {et~al.}(2011){Westmeier}, {Braun}, \&
  {Koribalski}}]{2011MNRAS.410.2217W}
{Westmeier}, T., {Braun}, R., \& {Koribalski}, B.~S. 2011, \mnras, 410, 2217,
  \dodoi{10.1111/j.1365-2966.2010.17596.x}

\bibitem[{{Wiklind} \& {Rydbeck}(1986)}]{1986A&A...164L..22W}
{Wiklind}, T., \& {Rydbeck}, G. 1986, \aap, 164, L22

\bibitem[{{Williams} {et~al.}(2023){Williams}, {Bureau}, {Davis}, {Cappellari},
  {Choi}, {Elford}, {Iguchi}, {Gensior}, {Liang}, {Lu}, {Ruffa}, \&
  {Zhang}}]{2023arXiv230805146W}
{Williams}, T.~G., {Bureau}, M., {Davis}, T.~A., {et~al.} 2023, arXiv e-prints,
  arXiv:2308.05146.
\newblock \doarXiv{2308.05146}

\bibitem[{{Young} {et~al.}(2011){Young}, {Bureau}, {Davis}, {Combes},
  {McDermid}, {Alatalo}, {Blitz}, {Bois}, {Bournaud}, {Cappellari}, {Davies},
  {de Zeeuw}, {Emsellem}, {Khochfar}, {Krajnovi{\'c}}, {Kuntschner},
  {Lablanche}, {Morganti}, {Naab}, {Oosterloo}, {Sarzi}, {Scott}, {Serra}, \&
  {Weijmans}}]{2011MNRAS.414..940Y}
{Young}, L.~M., {Bureau}, M., {Davis}, T.~A., {et~al.} 2011, \mnras, 414, 940,
  \dodoi{10.1111/j.1365-2966.2011.18561.x}

\end{thebibliography}
\bibliographystyle{aasjournal}



\appendix

\section{Chemical post-processing} \label{App::chem-postproc}
As noted in Section~\ref{Sec::ICs}, the CO-luminous gas fraction in our simulations is calculated in post-processing using the {\sc Despotic} model for astrochemistry and radiative transfer~\citep{Krumholz13a}. The self- and dust-shielding of CO molecules from the ambient UV radiation field cannot be accurately computed during run-time at the mass resolution of our simulation. Within {\sc Despotic}, the escape probability formalism is applied to compute the CO line emission from each gas cell according to its hydrogen atom number density $n_{\rm H}$, column density $N_{\rm H}$ and virial parameter $\alpha_{\rm vir}$, assuming that the cells are approximately spherical. In practice, the line luminosity varies smoothly with the variables $n_{\rm H}$, $N_{\rm H}$, and $\alpha_{\rm vir}$. We therefore interpolate over a grid of pre-calculated models at regularly-spaced logarithmic intervals in these variables to reduce computational cost. The hydrogen column density is estimated via the local approximation of~\cite{Safranek-Shrader+17} as $N_{\rm H}=\lambda_{\rm J} n_{\rm H}$, where $\lambda_{\rm J}=(\pi c_s^2/G\rho)^{1/2}$ is the Jeans length, with an upper limit of $T=40~{\rm K}$ on the gas cell temperature. The virial parameter is calculated from the turbulent velocity dispersion of each gas cell according to~\cite{MacLaren1988,BertoldiMcKee1992}. The line emission is self-consistently coupled to the chemical and thermal evolution of the gas, including carbon and oxygen chemistry~\citep{Gong17}, gas heating by cosmic rays and the grain photo-electric effect, line cooling due to ${\rm C}^+$, ${\rm C}$, ${\rm O}$ and ${\rm CO}$ and thermal exchange between dust and gas. We match the ISRF strength and cosmic ionization rate to the values used in our live chemistry.

Having calculated values of the CO line luminosity for each simulated gas cell, we compute the CO-bright molecular hydrogen surface density as
\begin{equation} \label{Eqn::DESPOTIC}
\begin{split}
\Sigma_{\rm H_2, CO}[{\rm M}_\odot{\rm pc}^{-2}] = &\frac{2.3 \times 10^{-29}{\rm M}_\odot({\rm erg~s}^{-1})^{-1}}{m_{\rm H}[{\rm M}_\odot]} \\
&\times \int^{\infty}_{-\infty}{\dd z^\prime \rho_{\rm g}(z^\prime) L_{\rm CO}[{\rm erg~s}^{-1}~{\rm H~atom}^{-1}]},
\end{split}
\end{equation}
where $\rho_{\rm g}(z)$ is the total gas volume density in ${\rm M}_\odot~{\rm pc}^{-3}$ at a distance $z$ (in ${\rm pc}$) from the galactic mid-plane. The factor of $2.3 \times 10^{-29}~{\rm M}_\odot~({\rm erg~s}^{-1})^{-1}$ combines the mass-to-luminosity conversion factor $\alpha_{\rm CO}=4.3~{\rm M}_\odot{\rm pc}^{-2}({\rm K}~{\rm kms}^{-1})^{-1}$ of~\cite{Bolatto13} with the line-luminosity conversion factor $5.31 \times 10^{-30}({\rm K~kms}^{-1}{\rm pc}^2)/({\rm erg~s}^{-1})$ for the CO $J=1\rightarrow 0$ transition at redshift $z=0$~\citep{SolomonVandenBout05}.

We note that our assumption of a constant ${\rm H_2}$-to-CO conversion factor may introduce an over-estimate of the CO-luminous molecular gas surface density at high gas surface densities. Additionally, for high column-density regions in which the CO $J=1$-$0$ line becomes optically-thick, Equation~(\ref{Eqn::DESPOTIC}) may overestimate the integrated emission for this particular line. Our CO-luminous molecular gas surface density is therefore necessarily an upper limit to the CO-luminous ${\rm H}_2$ column.

We emphasize that this post-processing calculation of CO-bright ${\rm H}_2$ emission is used only for the comparison of molecular half-mass radii and average surface densities in the creation of initial conditions to match the observational samples shown in Figure~\ref{Fig::ICs}, for which Equation~(\ref{Eqn::DESPOTIC}) is sufficient.  All other ${\rm H_2}$ column densities are computed via the chemical network and shielding prescription outlined in Section~\ref{Sec::chem-SF-feedback}, independently of the CO luminosity.

\begin{figure}
\begin{centering}
  \includegraphics[width=.5\linewidth]{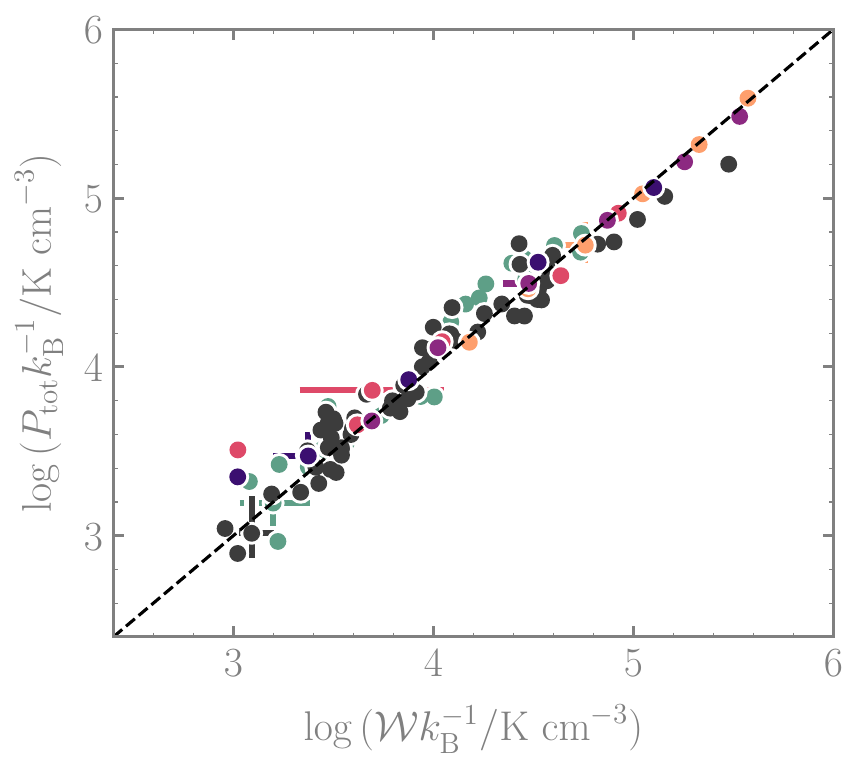}
  \caption{Total mid-plane gas pressure $P_{\rm tot}$ as a function of the interstellar medium weight $\mathcal{W}$ for our six simulated galaxies. The dashed line represents the 1-1 relationship expected in dynamical equilibrium. Filled data points represent median values over time for each simulated galaxy, measured within overlapping radial annuli of width $500$~pc. The Milky Way-like galaxy is represented by grey data points, the NGC~300-like galaxy by green data points, and the ETGs by the purple, pink and orange data points. For visual clarity, the corresponding interquartile ranges are shown at just one representative radius in each galaxy.}
  \label{Fig::pressures-vs-weight}
\end{centering}
\end{figure}

\section{Calculation of interstellar medium weights, pressures, velocity dispersions and star formation rate surface densities} \label{App::P-and-W}
\subsection{Values of weight, pressure and star formation rate from simulated data}
In Figures~\ref{Fig::eEoS},~\ref{Fig::obs-comparison-pressures}, and~\ref{Fig::rotcurves} we show the total mid-plane pressure $P_{\rm tot}$, the mid-plane volume density $n_{\rm H}$, and the gravitational weight $\mathcal{W}$ of the interstellar medium across our simulation suite. These quantities are computed on a cylindrical three-dimensional grid in galactocentric radius $R$, azimuthal angle $\theta$ and vertical distance $z$ from the galactic mid-plane. The $R$-bins have a width of $500$~pc and a separation of $200$~pc, while the $z$-bins have a width of $10$~pc and a separation of $10$~pc. Twelve $\theta$ bins with $\theta \in [0, 2\pi]$ are used in every case.

In the calculation of all gas properties below, we exclude gas that is gravitationally-bound with $\alpha_{\rm vir}<2$, or that is in the hot, feedback-heated phase with $T>2 \times 10^4$~K. That is, we include only the cool-warm gas phase; this gas is assumed to be in a state of vertical dynamical equilibrium in the theory of~\cite{Ostriker+10,2022ApJ...936..137O}, and this equilibrium is directly demonstrated  for a range of $z$ within multiphase simulations in~\citet{KimCG&Ostriker15b,2020ApJ...894...12V,2020MNRAS.498.3664G}.

The interstellar medium weight is computed over the set of gas cells within each $(R, \theta, z)$ bin, such that
\begin{equation} \label{Eqn::ISM-weight}
\mathcal{W} = \Big|\min{\Big[\sum^{z_{\rm max}}_{z_{\Phi_{\rm min}}}{\rho(z) \frac{\partial \Phi}{\partial z}}(z), \sum^{z_{\Phi_{\rm min}}}_{-z_{\rm max}}{\rho(z) \frac{\partial \Phi}{\partial z}}(z)\Big]}\Big| \Delta z, 
\end{equation}
where $\rho$ is the gas volume density, $z_{\rm max} = 300$~pc for the ETG simulations and $z_{\rm max} = 1.5$~kpc for the Milky Way-like and NGC~300-like simulations. The potential $\Phi$ in each bin is given by interpolating the gravitational potential values across the 150 particle centroids (stellar, dark matter or gas particles of any phase) nearest to the center of the bin, using radial basis function interpolation~\citep{RBFInterpolant}. The gradient $\partial{\Phi}/\partial{z}$ is then taken as the difference between adjacent bins along the $z$-axis.

In Equation (\ref{Eqn::ISM-weight}), the quantity $z_{\rm \Phi_{\rm min}}$ represents the $z$-bin in each $(R, \phi)$ column for which $\rho \Phi$ is minimized. We treat this as the mid-plane of the gas disk. The minimum of the two sums on either side of $z_{\Phi_{\rm min}}$ then provides a measure of the compressive force per unit area that acts on the gas disk (the interstellar medium weight). The difference in the absolute values of these sums provides a force per unit area that pushes the gas disk in one direction, rather than compressing it, and is therefore excluded.

Correspondingly, the volume-weighted mid-plane pressure is calculated for each column of $(R, \phi)$ at $z=z_{\Phi_{\rm min}}$, such that $P_{\rm tot} = P_{\rm th} + P_{\rm turb}$, with
\begin{equation}
P_{\rm th} = \rho c_s^2 |_{z=z_{\Phi_{\rm min}}},
\end{equation}
and
\begin{equation}
P_{\rm turb} = \rho (v_z - \langle v_z \rangle)^2 |_{z=z_{\Phi_{\rm min}}}.
\end{equation}
The volume density $\rho = n_{\rm H} \mu m_{\rm p}$ of the gas cells is simply the sum of the gas cell masses divided by the bin volume (equivalent to a volume-weighted average of the gas cell densities). The angled brackets denote mass-weighted averages over each $(R, \phi)$ column (note that the volume-weighted pressure is given by the product of the volume-weighted gas density and the mass-weighted velocity dispersion, as shown explicitly in~\cite{Ostriker+10}). The gas velocity perpendicular to the galactic mid-plane is given by $v_z$, and $c_s$ is the gas sound-speed.  Figure~\ref{Fig::rotcurves} shows that total midplane pressure does indeed balance the ISM disk's vertical weight in our simulations.

Finally, we calculate the star formation rate surface density $\Sigma_{\rm SFR}$ in each $(R, \phi)$ column by simply summing the instantaneous star formation rates of the gas cells in each column, and dividing by its surface area. We note that Figures~\ref{Fig::gal-main-sequence} and~\ref{Fig::KS-relations}, by contrast, use star formation rates calculated as averages over star particles with ages $\leq 5$~Myr, similar to the values that would be observed in ${\rm H}\alpha$ emission. We find good agreement between the the star formation rates computed via these two methods.

The values of $\mathcal{W}$, $P_{\rm tot}$ and $\Sigma_{\rm SFR}$ shown in Figures~\ref{Fig::eEoS}-\ref{Fig::rotcurves} are median values over time and azimuthal angle. Because the turbulent velocity dispersion used to compute $P_{\rm turb}$ is a statistical quantity (standard deviation of vertical velocities) and the star formation rate is computed stochastically for the gas cells in our simulation, we take medians and interquartile ranges only over the set of voxels with $\geq 100$ gas cells.

\subsection{Calculation of the radial velocity dispersions, $\sigma_{R}$ and $\sigma_{R, *}$}
In Figure~\ref{Fig::Romeo-Qs}, we show the radial component of the velocity dispersion for the gas and stellar components of the galactic disk, which provides support against the gravitational collapse of gas, as encapsulated in the Toomre $Q$ parameter. We calculate these quantities as
\begin{equation}
\begin{split}
\sigma_R^2 &= \langle (v_R - \langle v_R \rangle)^2 \rangle \\
\sigma_{R,*}^2 &= \langle (v_{R, *} - \langle v_{R, *} \rangle)^2 \rangle,
\end{split}
\end{equation}
where angled brackets again denote mass-weighted averages over the gas cells/stellar particles in each $(R, \phi)$ column.

\subsection{Estimated values for the ATLAS$^{\rm 3D}$ sample}
On the right-hand side of Figure~\ref{Fig::obs-comparison-pressures}, we estimate the positions of the ATLAS$^{\rm 3D}$ galaxies in the plane of ISM weight $\mathcal{W}$ vs. the star formation rate surface density $\Sigma_{\rm SFR}$ (grey crosses). We have approximated $\mathcal{W}$ for the ATLAS$^{\rm 3D}$ sample by making a number of geometrical approximations regarding the gas disk, stellar bulge, and dark matter halo. The median disk-to-bulge ratio in the galaxies is zero, such that
\begin{equation}
\mathcal{W} = \mathcal{W}_{\rm g} + \mathcal{W}_{*, {\rm b}} + \mathcal{W}_{\rm dm},
\end{equation}
where $\mathcal{W}_{\rm g}$ is the weight of the gas due to its own gravitational potential, $\mathcal{W}_{*, {\rm b}}$ is the weight due to the potential associated with the stellar bulge, and $\mathcal{W}_{\rm dm}$ is the weight due to the potential associated with the dark matter halo. Assuming a plane-parallel geometry for the gas,
\begin{equation}
\mathcal{W}_{\rm g} = \frac{\pi G \Sigma_{\rm g}^2}{2},
\end{equation}
where $\Sigma_{\rm g}$ is the gas surface density. Both the stellar bulge and dark matter components have spherical distributions, such that their combined weight can be approximated as
\begin{equation}
\mathcal{W}_{\rm *, b} + \mathcal{W}_{\rm dm} = \zeta \Sigma_{\rm g}(\Omega_{*, \rm b}^2 + \Omega_{\rm dm}^2) h_{\rm g},
\end{equation}
where $h_{\rm g}$ is the gas disk scale-height, and we have assumed that $h_{\rm g}$ is much smaller than the scale-lengths of both the bulge and the halo, with $\zeta \sim 1/3$~\citep[see][]{OstrikerShetty2011}. For the ATLAS$^{\rm 3D}$ galaxies, we assume a Plummer profile for the bulge and an NFW profile for the halo, as in our simulations, and calculate $\Omega_{\rm *, b}$ and $\Omega_{\rm dm}$ from the measured values of the stellar half-light radius $R_{*, 1/2}$ and the virial halo mass $M_{200}$, as shown in Figure~\ref{Fig::ICs}. We set a gas disk scale-height of $h_{\rm g} = 25$~pc---the median scale-height within our ETG simulations.

\end{document}